\pdfoutput=1
\documentclass[a4paper,12pt]{article}
\usepackage{amsfonts}
\usepackage{mathrsfs}
\usepackage{amsmath}
\usepackage{amssymb}
\usepackage{framed}
\usepackage[medium]{titlesec}
\usepackage{bm}
\usepackage{cite}
\usepackage[normalem]{ulem}
\usepackage{extarrows}
\usepackage{slashed}
\usepackage{isodateo}
\usepackage{graphicx}
\usepackage{xcolor}
\usepackage[bookmarksnumbered=true,bookmarksopen=true]{hyperref}
 \hypersetup{colorlinks,%
             linkcolor=[rgb]{0,0.3,0.6}, %
             citecolor=[rgb]{0,0.3,0.6}, %
             urlcolor=[rgb]{0,0.3,0.6}}
\usepackage[hmargin=.7in,vmargin=1.1in]{geometry}
\usepackage{indentfirst}
\usepackage{booktabs}

\newcommand{\FR}[2]{\displaystyle\frac{\,{#1}\,}{#2}}
\newcommand{\fr}[2]{\mbox{$\frac{\,{#1}\,}{#2}$}}
\newcommand{\n}{\nonumber}

\graphicspath{{fig/}}

\def\bge{\begin{equation}}
\def\ede{\end{equation}}
\def\bga{\begin{aligned}}
\def\eda{\end{aligned}}
\def\bgp{\begin{pmatrix}}
\def\edp{\end{pmatrix}}
\def\bgm{\begin{matrix}}
\def\edm{\end{matrix}}
\def\bgs{\begin{subequations}}
\def\eds{\end{subequations}}

\def\di{{\mathrm{d}}}

\def\mb{\mathbf}

\def\pd{\partial}
\def\ld{{\mathscr{L}}}

\def\la{\langle}\def\ra{\rangle}

\def\to{\rightarrow}

\def\ii{\mathrm{i}}

\def\al{\alpha}
\def\be{\beta}
\def\ga{\gamma}
\def\de{\delta}
\def\ep{\epsilon}
\def\ka{\kappa}
\def\lam{\lambda}

\def\si{\sigma}

\def\Wh{\mathrm{W}}

\def\aa{\mathsf{a}}
\def\bb{\mathsf{b}}

\usepackage{mdframed}

\newmdenv[skipabove=0pt,%
          skipbelow=5pt,%
          leftmargin=0pt,%
          rightmargin=0pt,%
          innertopmargin=-5pt,%
          innerbottommargin=7pt,%
          innerleftmargin=2pt,%
          innerrightmargin=2pt,%
          splittopskip=0pt,%
          splitbottomskip=0pt,%
          linewidth=0pt,%
          nobreak=true]%
          {keyeqn2}

\newmdenv[backgroundcolor=gray!15,%
          skipabove=0pt,%
          skipbelow=5pt,%
          leftmargin=0pt,%
          rightmargin=0pt,%
          innertopmargin=-5pt,%
          innerbottommargin=7pt,%
          innerleftmargin=2pt,%
          innerrightmargin=2pt,%
          splittopskip=0pt,%
          splitbottomskip=0pt,%
          linewidth=0pt,%
          nobreak=true]%
          {keyeqn}

\usepackage{titlesec}          
\titleformat{\section}
{\normalfont\fontsize{15}{20}\bfseries}{\thesection}{1em}{}

\newcommand{\ob}[1]{\mkern 2mu \overline{\mkern -2mu #1 \mkern -2mu}\mkern 2mu}
\newcommand{\wt}[1]{\mkern 2mu \widetilde{\mkern -2mu #1 \mkern -2mu}\mkern 2mu}
\newcommand{\wh}[1]{\mkern 2mu \widehat{\mkern-2mu#1\mkern-2mu}\mkern 2mu}

\newcommand{\fnemail}[1]{\footnote{Email: \href{mailto:#1}{\nolinkurl{#1}}}}

\begin{document}


\title{\Large\textbf{Helical Inflation Correlators: \\[2mm]
Partial Mellin-Barnes and Bootstrap Equations}\\[2mm]}

\author{Zhehan Qin\fnemail{qzh21@mails.tsinghua.edu.cn}~~~~~ and ~~~~~Zhong-Zhi Xianyu\fnemail{zxianyu@tsinghua.edu.cn}\\[5mm]
\normalsize{\emph{Department of Physics, Tsinghua University, Beijing 100084, China}}}

\date{}
\maketitle

\vspace{20mm}

\begin{abstract}
\vspace{10mm}

Massive spinning particles acquire helicity-dependent chemical potentials during the inflation from axion-type couplings. Such spinning fields can mediate sizable inflaton correlators which we call the helical inflation correlators. Helical inflaton correlators are approximately scale invariant, dS boost breaking, parity violating, and are promising observables of cosmological collider physics. In this work, we present complete and analytical results for 4-point helical inflation correlators with tree-level exchanges of massive spinning particles, including both the smooth background and the oscillatory signals. We compute the bulk Schwinger-Keldysh integrals in two independent ways, including the partial Mellin-Barnes representation and solving bootstrap equations. We also present new closed-form analytical results for 3-point functions with massive scalar or helical spinning exchanges. The analytical results allow us to concretely and efficiently explore the phenomenological consequences of helicity-dependent chemical potentials. In particular, we show that the chemical potential can exponentially enhance oscillatory signals of both local and nonlocal types, but only affects the background in a rather mild way. Our results extend the de Sitter bootstrap program to include nonperturbative breaking of de Sitter boosts. Our results also explicitly verify the recently proposed cutting rule for cosmological collider signals.

\end{abstract}

\newpage
\tableofcontents

\newpage
\section{Introduction}

The matter distribution of our universe at large scales features an essentially uniform background with small but important fluctuations. According to the widely accepted inflation paradigm, the large-scale inhomogeneities are originated from the primordial quantum fluctuations of spacetime generated at very high-energy (small-distance) scales during the inflation. Therefore, measurements of large-scale inhomogeneities including cosmic microwave background (CMB) observations and large-scale structure (LSS) surveys provide us unique windows to microscopic physical processes happening during the inflation \cite{Achucarro:2022qrl}. 

Mathematically, information about the primordial fluctuations is encoded in their correlation functions, which we call \emph{inflation correlators} for simplicity. We can think of inflation correlators as statistically measured correlation function $\la\varphi(\mb x_1)\cdots\varphi(\mb x_n)\ra$ where $\varphi(\mb x)$ is the inflaton fluctuation measured at the future boundary of the inflation spacetime. Nontrivial correlators start from $n=2$, which is more often called the power spectrum, while $n$-point correlators with $n\geq 3$ are conventionally called the non-Gaussianities \cite{Maldacena:2002vr,Chen:2010xka,Wang:2013eqj,Meerburg:2019qqi}. The power spectrum of the primordial scalar fluctuation has been well measured at CMB scales, providing us a wealth of information about the inflationary universe \cite{Planck:2018jri}. The non-Gaussianities, on the other hand, are yet to be discovered. The current CMB measurements only provide limits on their sizes \cite{Planck:2019kim}. However, we expect non-Gaussianities to be present, since they are signals of interactions, and interactions must be present during the inflation. 

In recent years, there have been increasing interests in the study of inflation correlators, and significant progresses have been made in various directions. The motivations behind this revived interest in the inflation correlators are at least three-fold.

First, observational progresses are expected to be made in the near and far future, allowing us to probe inflation correlators with unprecedented precision. The expected sensitivity to 3-point functions with near-future LSS surveys can be one order of magnitude higher than the current CMB constraints \cite{Achucarro:2022qrl,Ferraro:2022cmj}. More futuristic 21cm tomography could further improve this sensitivity by orders of magnitudes \cite{Munoz:2015eqa,Liu:2022iyy}. These future observations would open up entirely new parameter space with rich physics, calling for more careful and more systematic theoretical studies of inflation correlators. 

Second, the inflation correlators encode the information about the physics around the inflation scale, which could be as high as $10^{14}$GeV \cite{Planck:2018jri,BICEP:2021xfz}. This is certainly among the highest energy scale that  we can ever have access to in nature. Therefore, probing the inflation correlators provides us a unique chance to learn potential new physics at an extremely high energy scale. This idea was already realized in the early studies of quasi-single-field inflations \cite{Chen:2009we,Chen:2009zp,Baumann:2011nk,Chen:2012ge,Pi:2012gf,Noumi:2012vr,Gong:2013sma}, and was emphasized more recently in \cite{Arkani-Hamed:2015bza} under the name of cosmological collider (CC) physics. In \cite{Chen:2009we,Chen:2009zp,Baumann:2011nk,Chen:2012ge,Pi:2012gf,Noumi:2012vr,Gong:2013sma,Arkani-Hamed:2015bza}, it was shown that the fast expansion of the inflation background can trigger spontaneous production of heavy particles with mass up to the inflation Hubble scale $H$. Along with their subsequent evolution, these heavy particles can imprint characteristic oscillatory signatures on the inflation correlators, of which a measurement could tell the mass and the spin of these heavy particles. This makes inflation correlators directly relevant to the study of fundamental particle physics.  

Last but not least, developing computation techniques for inflation correlators and understanding their analytical properties are interesting theoretical problems on their own.
 The inflation spacetime is close to the Poincaré patch of the de Sitter spacetime (dS), and the inflation correlators can be viewed as correlators of bulk quantum fields in dS, with their external points pinned to the future boundary. Inflation correlators are thus natural analogues of scattering amplitudes in Minkowski spacetime or boundary correlators in anti-de Sitter spacetime (AdS). However, compared to the highly developed territory of flat-space scattering amplitudes or AdS correlators, the study of dS correlators is still in its infancy. To give an example, the analytical result for the simplest CC process, namely the 4-point correlator with tree-level exchange of a massive scalar field (see Fig.\ \ref{fig_scalar_tree}), was found only quite recently \cite{Arkani-Hamed:2018kmz}. Full analytical results for CC processes at loop levels are still beyond our reach at the moment. As the readers would appreciate from the following sections, the  complications of computing inflation correlators come mainly from two sources. One is the space-time asymmetric nature of the problem, which renders a fully covariant 4-momentum representation inapplicable. Also, while some inflation correlators are covariant under full dS isometries, phenomenologically more interesting correlators typically break dS boosts. As often happens in physics, less symmetry implies increased difficulty. The other complication is that the mode functions of fields are often special functions due to the distortion from the spacetime background. One has to deal with time-ordered integrals of Hankel functions (for ordinary massive fields) or Whittaker functions (for fields with helical chemical potentials) instead of plane waves in flat space. 
 
Despite of the difficulties, progresses have been made in past few years. In a recently developed cosmological bootstrap program \cite{Baumann:2022jpr}, full analytical results have been found for dS covariant 4-point and 3-point functions with tree-level exchange of particles of arbitrary mass and (integer) spin \cite{Arkani-Hamed:2018kmz,Baumann:2019oyu,Baumann:2020dch}. Similar results were also obtained with an AdS-inspired Mellin space approach \cite{Sleight:2019mgd,Sleight:2019hfp,Sleight:2020obc,Sleight:2021iix,Sleight:2021plv}. The bootstrap program was further extended to include dS boost-breaking effects such as non-unit sound speed and nonconvariant couplings, first to correlators of purely massless fields \cite{Pajer:2020wnj,Pajer:2020wxk,Cabass:2021fnw}, and more recently to correlators containing massive fields (and thus CC signals) \cite{Pimentel:2022fsc,Jazayeri:2022kjy}. 
At the loop level, precision computations of 1-loop correlators of both dS covariant type and boost-breaking type have been done in \cite{Wang:2021qez} with a full numerical approach. Analytically, an approached called partial Mellin-Barnes (MB) representation was recently introduced in \cite{Qin:2022lva}, with which the complete analytical results for the nonlocal CC signals at 1-loop level were obtained, for both dS covariant case and boost-breaking case. See also \cite{Goodhew:2020hob,Jazayeri:2021fvk,Melville:2021lst,Goodhew:2021oqg,DiPietro:2021sjt,Tong:2021wai,Bonifacio:2021azc,Hogervorst:2021uvp,Meltzer:2021zin,Heckelbacher:2022hbq,Gomez:2021qfd,Gomez:2021ujt,Baumann:2021fxj} for recent works on the analytical properties of inflation correlators or  wavefunction coefficients, and see \cite{Baumann:2022jpr} for a review. 

From particle physics perspective, the rich CC phenomenology has also been explored in recent years, including particle physics within and beyond the Standard Model \cite{Chen:2015lza,Chen:2016nrs,Chen:2016uwp,Chen:2016hrz,Lee:2016vti,An:2017hlx,An:2017rwo,Iyer:2017qzw,Kumar:2017ecc,Chen:2017ryl,Tong:2018tqf,Chen:2018sce,Chen:2018xck,Chen:2018cgg,Chua:2018dqh,Domenech:2018bnf,Wu:2018lmx,Saito:2018omt,Li:2019ves,Lu:2019tjj,Liu:2019fag,Hook:2019zxa,Hook:2019vcn,Kumar:2018jxz,Kumar:2019ebj,Alexander:2019vtb,Wang:2019gbi,Wang:2019gok,Wang:2020uic,Li:2020xwr,Wang:2020ioa,Fan:2020xgh,Aoki:2020zbj,Bodas:2020yho,Maru:2021ezc,Lu:2021gso,Sou:2021juh,Lu:2021wxu,Pinol:2021aun,Cui:2021iie,Tong:2022cdz,Reece:2022soh,Chen:2022vzh}, the observation strategies and forecasts \cite{Meerburg:2016zdz,MoradinezhadDizgah:2017szk,MoradinezhadDizgah:2018ssw,Kogai:2020vzz}. Among the many CC models that have been considered, a scenario with helicity-dependent chemical potentials has been emphasized owing to its distinct and promising phenomenology, as well as its simplicity and generality in model buildings \cite{Chen:2018xck,Hook:2019zxa,Hook:2019vcn,Wang:2019gbi,Wang:2020ioa,Wang:2021qez,Tong:2022cdz,Qin:2022lva}. Such chemical potentials arise naturally from axion-type couplings $\pd_\mu J_\text{A}^\mu/\Lambda$ between the inflaton $\phi$ and axial currents $J^\mu_\text{A}$ formed by massive spinning particles, and thus are naturally present in various axion inflation models. Examples include axial current $J_\text{A}^\mu=\ob{\Psi}\ga^\mu\ga^5\Psi$ for spin-$1/2$ fermions \cite{Chen:2018xck,Wang:2019gbi} and Chern-Simons current $J_\text{A}^\mu=\ep^{\mu\nu\rho\si}A_\nu F_{\rho\si}$ for spin-1 bosons \cite{Wang:2020ioa}. Generalizations to higher spins are also possible \cite{Tong:2022cdz}. With the rolling inflaton background $\dot\phi_0\neq 0$, these operators turn into the number density operators weighted by the helicity of the particle, namely $(\dot\phi_0/\Lambda) J_\text{A}^0$ and thus the combination $\mu\equiv \dot\phi_0/\Lambda$ can be viewed as a helicity-dependent chemical potential, which we shall call \emph{helical chemical potential}. As a result, the spontaneous production of one helicity state will be boosted while the other being suppressed.\footnote{The chemical potential can also be introduced to ordinary number density operator without the helicity dependence. However, as was shown in \cite{Wang:2019gbi}, only the helicity-dependent chemical potential can lead to nontrivial enhancement of CC signals. See \cite{Sou:2021juh} for more discussions. In \cite{Bodas:2020yho} a helicity-independent chemical potential was introduced for scalar fields. But this chemical potential enhance the signal from the vertex rather than from the propagator, and thus is quite different from the chemical potential in our current context.} Cosmological consequences of helicity-dependent chemical potential have also been extensively explored in a wider context beyond the CC physics; See, e.g., \cite{Anber:2006xt, Durrer:2010mq,Barnaby:2010vf,Peloso:2016gqs,Bugaev:2013fya,Anber:2009ua,Anber:2012du,Adshead:2015kza,Adshead:2015pva,Adshead:2015jza,Agrawal:2018vin,Berghaus:2019whh}.

The chemical potential operator $(\pd_\mu\phi)J_\text{A}^\mu$ is the lowest dimension operator that couples massive spinning particle and the inflaton field and respects the shift symmetry of the inflaton $\phi\to\phi+\text{const.}$, and thus is rather generic from the effective field theory (EFT) viewpoint. A chemical potential comparable to or greater than the particle mass can natually alleviate the Boltzmann suppression factor $e^{-\pi m/H}$ that affects most of CC signals when the mass $m$ is significantly greater than the Hubble scale $H$. In particular, in parameter space that can be easily realized in particle models ($\mu\sim m\gg H$ for spin-1 and $\mu\gg m\gg H$ for spin-1/2), the resulting CC signals possess large amplitudes and large frequencies, while the smooth background contribution is expected to be negligible \cite{Wang:2019gbi,Wang:2020ioa}. It is clear that the helical chemical potential necessarily breaks the dS boosts as well as the spatial parity. The resulting correlators possess characteristic angular dependence which reflects the helicity asymmetric nature of the underlying dynamics \cite{Liu:2019fag,Tong:2022cdz}. We will call these objects \emph{helical inflation correlators} to highlight their characteristic dependences on the helicity of the intermediate states. 

For reliable studies of CC physics involving helical chemical potential, it is clearly essential to develop techniques for precise and efficient computation of helical inflation correlators. Ideally, we would like to gain full analytical control over these objects. However, the presence of the chemical potential makes the computation more involved. The primary reason is that the chemical potential comes from the inflaton rolling $\dot\phi_0\neq 0$, and thus breaks the dS boost. In particular, when the chemical potential is large, the boost-breaking operator becomes \emph{nonperturbative}. (See Fig.\ \ref{fig_vecboot} and related discussions in Sec.\ \ref{sec_boot}.) Therefore one must treat the chemical potential insertions nonperturbatively, resumming all such insertions to get the \emph{dressed} propagators for the massive spinning particle. This nonperturbative boost breaking makes it hard to use the previously developed tools that rely on full dS isometries. The result from recently developed boostless bootstrap was not immediately applicable to this case either, as we shall further explain in Sec.\ \ref{sec_boot}. 
 
On the other hand, it is possible to complete the resummation of chemical-potential insertions by directly solving the equation of motion, resulting in a solution expressed by the Whittaker W function, which is technically more complicated than the more familiar Hankel functions. (In comparison, there is another type of strong boost-breaking effects, namely the non-unit sound speed $c_s\neq 1$, for which the mode function  remains to be the Hankel functions, only with a shift in the momentum $k\to c_s k$.) Therefore, in order to find analytical results for inflation correlators with helical chemical potentials, it is essential to develop appropriate tools to deal with Whittaker functions. 

In this work, we set out to find full analytical results for a class of helical inflation correlators at the tree level. We shall attack the problem from two independent approaches. 

In the first approach, we use the method of partial Mellin-Barnes (MB) representation introduced in our previous work \cite{Qin:2022lva}. In this approach, the complicated special functions in the intermediate massive propagators are recast into MB integrals over power functions, which trivialize the time integral. We then finish the MB integrals by properly closing the contours and applying the residue theorem. In this work, we shall show how to compute both the signal and the background part of the correlators using this method. We will further explore the freedom of MB representation for Whittaker functions to give two distinct derivations of the same result. 

In the second approach, we derive differential equations satisfied by the helical correlators and then solve them with appropriate boundary conditions. We call them \emph{bootstrap equations}, since this approach is more in line with the cosmological bootstrap program. Thus our bootstrap equations can be viewed as an extension of the cosmological bootstrap program to the theories with helical chemical potential, which is a nonperturbative source of dS boost breaking. We also present a new way to take folded limit of the bootstrapped series, which allows us to write down closed-form analytical results for various 3-point functions of massive mediations without any series expansion. 

The central object of this paper is a \emph{vector seed integral} (\ref{eq_vecIab}) defined in Sec.\ \ref{sec_HCP}, from which a large number of helical correlators can be built. For illustrating the method and for completeness, we also present a similar computation for the simpler massive scalar exchange, by defining and computing a \emph{scalar seed integral}. By computing these objects with both partial MB and bootstrap equations, we cross-check the results from the two very different methods, and find nice agreements between them. 

With full analytical results for these seed integrals, we can build more general correlators with helical massive spinning exchanges for all helicity states. These results also allow us to explore more reliably and more efficiently the phenomenological consequences of these correlators. In particular, we confirm a previous expectation that the CC signals from these helical correlators are parametrically larger than background even for mildly large chemical potentials. Also, we show that both the local and nonlocal CC signals can be enhanced exponentially by the chemical potential, but the detailed parameter dependence shows interesting difference between these two types of signals. The fast numerical implementation of our results makes it easier for parameter scanning in model building and also for generating templates when confronting the model with data. Therefore, we hope that the results presented in the work is not only of theory interest, but also of practical use.

\paragraph{Outline of this work.} Since this work is mainly about analytical computation of helical inflation correlators, a large portion of the main text will be rather technical.  In order to make this work more accessible, below we will give a brief summary for each of the following sections.

In Sec.\ \ref{sec_HCP}, after making some general remarks about the helical chemical potentials, we present in Sec.\ \ref{sec_HeliSpin1} a detailed treatment of quantizing a massive spin-1 field with chemical potential in dS (henceforth helical spin-1 fields), which leads to a set of \emph{mode functions} for the helical spin-1 fields, summarized in (\ref{eq_BmodePM})-(\ref{eq_BmodeL}). Following a standard procedure in the Schwinger-Keldysh (SK) formalism \cite{Chen:2017ryl}, we can construct propagators, and eventually Feynman diagrams, from these mode functions. Then in Sec.\ \ref{sec_VSI} we discuss the most general couplings that give rise to the 4-point functions in Fig.\ \ref{fd_vector}. Then we show that the computation of this class of 4-point correlators can be reduced to that of a vector seed integral, defined in (\ref{eq_vecIab}).

In Sec.\ \ref{sec_PMB}, we introduce our main tool for computing the SK integrals, namely the partial MB representation. For illustrative purpose and also for completeness, we use the partial MB to compute a simpler process, namely the 4-point function with tree-level exchange of a massive scalar particle. To this end, we define a scalar seed integral in (\ref{eq_ScalarSeedInt}), and then present all the details of computation. Readers uninterested in the technical details can directly go to Sec.\ \ref{sec_scalar_summary} for a summary of the result.

In Sec.\ \ref{sec_vec}, we present the full analytical results for the vector seed integral defined in Sec.\ \ref{sec_HCP}. It turns out that the correlators mediated by the two transverse states and by the longitudinal state require separate treatments. Therefore, we present the computation of the transverse part in Sec.\ \ref{sec_vec_trans}, and the longitudinal part in Sec.\ \ref{sec_vec_long}. In Sec.\ \ref{sec_vec_trans}, we discuss the flexibility in choosing the MB representation for the Whittaker function. In the same subsection, we provide the full computation using the so-called \emph{partially resolved MB representation}. The computation of the same integral with an alternative MB representation, which we call the \emph{completely resolved MB representation}, will be presented in App.\ \ref{app_AlterMB}. Again, readers uninterested in the details can directly jump to Sec.\ \ref{sec_vec_summary} for a summary of the results in this section. In Sec.\ \ref{sec_vec_summary}, we also comment on the relation between our results and a recently proposed tree-level cutting rule for CC signals \cite{Tong:2021wai}. Our results explicitly verified the cutting rule for both the scalar exchanges and the helical spinning exchanges. Therefore it is possible to use our method to give a more rigorous proof of the cutting rule stated in \cite{Tong:2021wai}.

In Sec.\ \ref{sec_boot}, we turn to the method of bootstrap equations. We will first make some general remarks about the bootstrap method, and then derive the scalar and vector bootstrap equations in the three subsections. We then provide the particular solutions to the inhomogeneous equations in each case in terms of Taylor series of momentum ratios, and compare these results with the ones obtained from partial MB representation. It turns out that the series solution can be found for transverse correlators only if we make a suitable change of variables, as suggested by the results from partial MB. We then solve the homogeneous equations to get the signal part of the correlator. Different from previous works, we impose the boundary conditions from the squeezed configurations, which correspond to the late-time limit of the bulk integral. Given the more free parameters in the homogeneous solutions, it turns out that imposing the boundary condition from the late-time limit is the easiest choice. We then check our results by sending them to the folded limit, where all superficial divergences should cancel, as required by the Bunch-Davies initial condition \cite{Arkani-Hamed:2015bza,Arkani-Hamed:2018kmz}. We will also show how an appropriate change of variables can help us to find neat and closed expressions for the 4-point function in the folded limit.

In Sec.\ \ref{sec_var}, we demonstrate how to extend our results to more general helical correlators, by working out a particular example, namely a mixed 3-point function with two scalar modes and one tensor mode, where the tensor mode is mixed with a massive spin-2 field with helical chemical potential. This is a process that generalizes the CC physics to tensor modes with large oscillatory signals, and has been considered in \cite{Tong:2022cdz}. In this section, we provide full analytical result for this process, including the signal and the background, in terms of the vector seed integral computed in Sec.\ \ref{sec_vec}. For completeness, we will also present a corresponding closed-form expression for the scalar 3-point function with massive scalar exchange, which seems new to us.

In Sec.\ \ref{sec_pheno}, we discuss the phenomenology of the helical 4-point correlators. In particular, we show how the CC signals and the background depend on the mass and the chemical potential of the intermediate particles. We also provide representative plots for the signals and the backgrounds in Fig.\ \ref{fig_NLsignal}, and also plots for typical full correlators in Fig.\ \ref{fig_signal}.

Further discussions and outlooks are presented in Sec.\ \ref{sec_concl}. Some useful formulae are collected in App.\ \ref{app_formulae}. In App.\ \ref{app_general} we give the results of the 4-point spin-1 correlator with the most general effective couplings in terms of the vector seed integral. In App.\ \ref{app_AlterMB} we present the result for the vector seed integral using the completely resolved MB representation for the Whittaker function. 

\paragraph{Notations and conventions.} Most of special notations used in this work will be defined in a self-contained way when they appear. We use the mostly-minus spacetime metric. For the most part, we use conformal time $\tau$ and comoving spatial coordinates $\mb x$, with which the dS metric reads $\di s^2=a^2(\tau)(-\di\tau^2+\di\mb x^2)$, $a(\tau)=-1/(H\tau)$ is the scale factor and $H$ is the inflation Hubble parameter which we will always take to be a constant. In most part of this paper, we also take $H=1$ for notational simplicity. We occasionally use the comoving time $t$, with which the metric reads $\di s^2=-\di t^2+a^2(t)\di\mb x^2$ and $a(t)=e^{Ht}$. Our notations and conventions for describing the 4-point inflation correlators are mostly in line with \cite{Qin:2022lva}, in which the readers can find detailed discussion about the kinematics and the symmetry of the 4-point functions. In particular, we will often use notations like $k_{12}\equiv k_1+k_2$ (for momenta; magnitude sum rather than vector sum), $s_{12}=s_1+s_2$ (for Mellin variables), $n_{12}=n_1+n_2$ (for summation variables), etc. Our diagrammatic notations and computations follow the treatment of SK formalism presented in \cite{Chen:2017ryl}, in which the detailed diagrammatic rules about SK integrals for given diagrams have been spelled out. We refer readers unfamiliar with the SK formalism to \cite{Chen:2017ryl} for a pedagogic review. 

\section{Spinning Fields with Helical Chemical Potential}
\label{sec_HCP}

The inflationary universe is very close to the Poincar\'e patch of the de Sitter space, which has the largest possible number of isometries, including the space translations and rotations, the dilatation, and also the three dS boosts. The observed scalar power spectrum suggests that the inflation correlators should be covariant under the space translations and rotations. The dilatation symmetry is also realized approximately in the form of scale invariance. On the contrary, we have no observational evidence of the three dS boosts. From a theoretical viewpoint, the rolling background of the inflaton itself is a source of boost breaking. Therefore, when constructing inflation correlators, it is appropriate to consider theories respecting all dS isometries expect the three boosts. 

There are two distinct ways to introduce boost breaking dynamics to a single propagating species in dS. One is to modify the relative size between the kinetic term and the gradient term, resulting a non-unit sound speed $c_s$. This is a well-known possibility and has been explored in the context of CC physics recently in \cite{Pimentel:2022fsc,Jazayeri:2022kjy}. The other possibility, which is realized for spinning particles only, is the helical chemical potential and is the main focus of this work. Usually, these two boost breaking effects can both be obtained by evaluating some Lorentz invariant operators in an underlying theory with the rolling inflaton background. In this respect, the chemical potential term appears first from the dimension-5 operators, as mentioned in the Introduction. The non-unit sound speed, on the other hand, normally requires higher dimensional operators. Therefore, from an effective field theory point of view, it seems that the chemical potential is more relevant than the non-unit sound speed whenever it is present. For this reason, we shall neglect any non-unit sound speed in the following discussions.

\subsection{Helical spin-1 field}
\label{sec_HeliSpin1}

Helical chemical potentials can in principle be introduced to massive particles of arbitrary nonzero spin. In this work we mainly focus on the case of massive spin-1 particles. Generalizations to higher spins or half-integer spins are similar, and we shall show an example of massive spin-2 in Sec.\ \ref{sec_var}. For completeness, we shall present the details of canonical quantization from the classical Lagrangian to the mode functions and finally to the SK propagators. Experts may wish to skip this subsection.

We can work with effective field theory of inflation to keep things general, but we find it convenient enough to start from a (local) Lorentz invariant Lagrangian. Therefore we introduce a real background scalar field $\phi$ which can be either the inflaton field or a spectator field or a mixture of the both. In either case, we require that the scalar field has a homogeneous and isotropic background:
\bge
\label{eq_RollingScalarBg}
\phi_0(t)=\la\phi(t,\mb x)\ra, 
\ede
which is independent of space coordinates $\mb x$ but can have nontrivial time dependence. For a wide range of phenomenological applications, we can take the time dependence of $\phi_0(t)$ to be $\phi_0(t)=\dot\phi_0 t+\text{const.}$ where $\dot\phi_0$ is a \emph{time-independent} constant speed. Such form of $\phi_0(t)$ is usually a good approximation for either a rolling inflaton along its flat potential, or a light spectator field with the mass smaller than the Hubble scale $H$. We also note that such form of time-dependence leaves the inflation correlators scale invariant with appropriate couplings. 

\paragraph{The action and the equations of motion.} Now we can write down an action for the spin-1 field $A_\mu$ of mass $m$, with a fixed spacetime background $g_{\mu\nu}$ and the scalar background introduced above:
\bge
\label{eq_ActionVec}
  S=\int\di^4x\,\bigg[\sqrt{-g}\Big(-\FR{1}{4}g^{\mu\rho}g^{\nu\si}F_{\mu\nu}F_{\rho\si}-\FR{1}{2}m^2g^{\mu\nu}A_\mu A_\nu\Big)+\FR{\phi}{4\Lambda}\ep^{\mu\nu\rho\si}F_{\mu\nu}F_{\rho\si}\bigg].
\ede
Here $\ep^{\mu\nu\rho\si}$ is the total antisymmetric tensor with $\ep^{0123}=1$. 
We do not assume a specific origin for the mass $m$ of the spin-1 field $A_\mu$. When it is originated from a Higgs mechanism, we assume that the Higgs degree is heavy enough so that it can be decoupled from our problem, and the above action can be regarded as written in the unitary gauge. 

The dimension-5 coupling $\phi F\wt F$ is a rather generic coupling between a CP odd scalar field $\phi$ and a spin-1 field $A_\mu$. It is also the lowest-dimension coupling between the spin-1 field $A_\mu$ to the scalar $\phi$ that respects both the shift symmetry of the scalar field $\phi\to \phi+\text{const.}$ and the (albeit broken) gauge symmetry of the spin-1 field. As mentioned in the introductory section and will be detailed now, this coupling is also the origin of the helical chemical potential of the massive spin-1 field once the scalar field $\phi$ acquires its rolling background (\ref{eq_RollingScalarBg}). 

Specifically, we take the dS limit of the inflation background $\di s^2=a^2(\tau)(-\di\tau^2+\di\mb x^2)$ with the scale factor $a(\tau)=-1/(H\tau)$, and also the scalar background (\ref{eq_RollingScalarBg}). Then the action (\ref{eq_ActionVec}) can be rewritten as:
\bge
\label{eq_ActionVecInfBg}
  S=\int\di\tau\di^3\mb x\,\ld,~~~~~\ld= \Big(-\FR{1}{4}F_{\mu\nu}F^{\mu\nu}-\FR{1}{2}a^2m^2 A_\mu A^\mu\Big)+\FR{\dot\phi_0 t}{4\Lambda}\ep^{\mu\nu\rho\si}F_{\mu\nu}F_{\rho\si} .
\ede
From this equation onward, all the spacetime indices are lowered and raised by the Minkowski metric $\eta_{\mu\nu}$ and its inverse, unless otherwise stated. Also, in (\ref{eq_ActionVecInfBg}) we have thrown away a constant in the scalar background $\phi_0(t)$ which contributes only to a boundary term that is irrelevant to our study. We stress that it is the physical-time derivative of the inflaton background $\dot\phi_0=\di\phi_0/\di t$ that is nearly the constant. The quantity $\phi_0'\equiv\di\phi_0/\di\tau= a\dot\phi_0=e^{Ht}\dot\phi_0$ actually has strong time dependence. Also, the appearance of explicit time dependence in (\ref{eq_ActionVecInfBg}) may look a bit weird, but it can be easily removed with integration by parts, leaving a term that has an uncontracted  temporal index. Such bare temporal indices are direct consequences of broken time diffeomorphism by the rolling scalar background \cite{Cheung:2007st}. 

From the action (\ref{eq_ActionVecInfBg}) it is straightforward to get the equation of motion for the spin-1 field $A_\mu$:
\begin{align}
\label{eq_VecEoM}
  \pd_\mu F^{\mu\nu}-a^2m^2A^\nu- 2a\mu\ep^{0\nu\rho\si}F_{\rho\si}=0.
\end{align} 
Here the dimension-1 parameter $\mu\equiv \dot\phi_0/\Lambda$ is nothing but the helical chemical potential. 
Acting $\pd_\nu$ on the equation of motion (\ref{eq_VecEoM}), we get a constraint: 
\bge
\label{eq_Aconstraint}
  2aH A^0+ \pd_\nu A^\nu  =0,
\ede
which can be written in a more covariant form $g^{\mu\nu}\nabla_\mu A_\nu=0$ although we prefer not to do so.\footnote{Our treatment here also applies to a Higgsed $U(1)$ gauge theory so long as one chooses (\ref{eq_Aconstraint}) as the gauge condition. However, in the case of a Higgsed gauge theory, other choices of gauge conditions are also possible, which may lead to expressions superficially different from what we are presenting here.}
Substituting the constraint (\ref{eq_Aconstraint}) back into the equation (\ref{eq_VecEoM}), we get two equations for the temporal and spatial components of $A_\mu=(A_0,A_i)$, respectively:
\begin{align}
  \label{eq_AEoMTemporal}
  &A_0''-\pd_j^2 A_0+(2aHA_0)'+a^2m^2A_0=0,\\
  \label{eq_AEoMSpatial}
  &A_i''-\pd_j^2 A_i+2aH\pd_i A_0+a^2m^2A_i+2a\mu\ep_{ijk}\pd^j A^k=0.
\end{align}
Although we have 4 equations here, they are not independent due to the constraint (\ref{eq_Aconstraint}). As is well known, a massive spin-1 field as discussed here gives rise to 3 independent propagating degrees of freedom.\footnote{This applies to all mass $0<m<H/2$ and $m>H/2$. Although the spin-1 representations of dS isometry group belong to different categories for $0<m<H/2$ and $m>H/2$, this peculiarity of dS representation does not affect the degree counting. In particular, the possibility of partially massless states kicks in only for spin $s> 2$ \cite{Baumann:2017jvh}. } It is convenient to choose helicity eigenstates with $h=0,\pm1$ to describe them. The helicity $h=\mb s\cdot\mb k/|\mb k|$ is the angular momentum of the mode along the direction of propagation. In a Lorentz invariant theory, the helicity is frame dependent, and thus is not an intrinsic property for massive spinning particles. More explicitly, one can always boost a helicity eigenstate to flip the direction of propagation while keeping the direction of angular momentum invariant, and thereby flip the sign of the helicity. However, in our case, the dS boosts (the dS counterparts of the Lorentz boosts) are spontaneously broken by the scalar background $\phi_0(t)$. As a result, the scalar background no longer takes the space-independent form $\phi_0(t)$ for an observer boosted relative to the local comoving frame. Therefore, the background scalar $\phi_0(t)$ unambiguously picks out a comoving frame in which the universe looks isotropic. It can be easily seen that the helicity is invariant under all other unbroken symmetries, including the space translations and rotations, and the dilatation. The dilatation is of course broken by the scalar background $\dot\phi_0$ as well. However, this is inconsequential so long as other fields couple only derivatively to the scalar and so long as $\dot\phi_0$ is constant.

\paragraph{Canonical quantization.}  Now we review the canonical quantization of the massive spin-1 field in dS in the presence of a helical chemical potential. Given the symmetry of the problem, it is convenient to go to the Fourier space for the 3-dimensional space and consider a specific Fourier mode with comoving momentum $\mb k$. Then, the Fourier modes $A_{\mu,\mb k}(\tau)$ can be written as linear superposition of annihilation and creation operators, $a_{\mb k}^{(h)}$ and $a_{\mb k}^{(h)\dag}$, for each helicity eigenstate:
\begin{align}
\label{eq_VecFourierMode}
  &A_\mu(\tau,\mb x)=\int\FR{\di^3\mb k}{(2\pi)^3}e^{\ii\mb k\cdot\mb x} A_{\mu,\mb k}(\tau) ,
  &&A_{\mu,\mb k}(\tau)=\sum_{h=-1}^{+1}\Big[ A^{(h)}_{\mu,\mb k}(\tau)a_{\mb k}^{(h)}+A^{(h)*}_{\mu,-\mb k}(\tau)a_{-\mb k}^{(h)\dag}\Big],
\end{align}
where $A_{\mu,\mb k}^{(h)}(\tau)$ are the helicity eigenstates, with $h=0$ denoting the longitudinal mode and $h=\pm 1$ denoting the two transverse modes. In the flat spacetime, each helicity eigenstate can be further written as a product of a time-dependent mode function $e^{-\ii E_{\mb k}t}$ and a time-independent polarization vector $e_{\mb k}^{(h)}$. Things are a little more complicated in dS for the longitudinal mode: As a consequence of nontrivial time dependence in the mode function, together with the constraint equation (\ref{eq_Aconstraint}), the longitudinal mode is no longer a simple product of a time-dependent mode function and a time-independent polarization vector. To take care of this complication, we take $\mb k=(0,0,k)^T$ without loss of generality. Then we can define four polarization vectors $e^{(\lam)}$ with $\lam=T,+,-,L$ as below (Note that $\lam$ is \emph{not} a label of helicity):
\begin{align}
\label{eq_PolarVec}
  &e_{\mu,\mb k}^{(T)}=\bgp 1 \\ 0 \\0 \\0 \edp,
  &&e_{\mu,\mb k}^{(\pm)}=\FR{1}{\sqrt2}\bgp 0 \\ 1 \\ \pm\ii \\0 \edp,
  &&e_{\mu,\mb k}^{(L)}= \bgp 0 \\ 0 \\ 0 \\ 1 \edp,
  \text{~~~with~~~}\mb k=\bgp 0\\0\\k\edp.
\end{align}
For general $\mb k$ not in the 3-direction, the corresponding polarization vectors can be obtained by the usual rule of 3-dimensional rotations. That is, $e^{(T)}_\mu$ will remain intact as a scalar under 3-rotations, while $e_{\mu,\mb k}^{(L)}=\wh{\mb k}$ and $e_{\mu,\mb k}^{(\pm)}=(\wh{\mb{u}}\pm\ii\wh{\mb{v}})/\sqrt2$ up to arbitrary complex phase factors, where $(\wh{\mb u},\wh{\mb v},\wh{\mb k})$ form a right-handed frame of the 3-dimensional space. We note in particular that these expressions are valid only in the comoving frame, and we are not allowed to perform dS boost due to the boost breaking nature of the problem. We will not need expressions for these polarization vectors in boosted frames. 

With the aid of the (physically redundant) set of polarization basis (\ref{eq_PolarVec}), we can write the helicity eigenstates $A_{\mu,\mb k}^{(h)}(\tau)$ in terms of a set of mode functions $B^{(\lam)}$ as below:
\begin{align}
\label{eq_Bpm}
  &A_{\mu,\mb k}^{(\pm1)}(\tau)=B^{(\pm)}(k,\tau)e_{\mu,\mb k}^{(\pm)},\\
\label{eq_Blong}
  &A_{\mu,\mb k}^{(0)}(\tau)=B^{(T)}(k,\tau)e_{\mu,\mb k}^{(T)}+B^{(L)}(k,\tau)e_{\mu,\mb k}^{(L)},
\end{align}
The equations of motion for the field $A_\mu$ then give rise to a set of equations for the mode functions. The temporal equation (\ref{eq_AEoMTemporal}) gives:
\begin{align}
\label{eq_ModeEqTemp}
  B^{(T)}{}''+2aHB^{(T)}{}'+\big[k^2+a^2(m^2+2H^2)\big]B^{(T)}=0,
\end{align}
while the spatial equation (\ref{eq_AEoMSpatial}) mixes different mode functions. To decouple the spatial equation, we contract it with the polarization vectors in (\ref{eq_PolarVec}). 
Contracting (\ref{eq_AEoMSpatial}) with $e_{\mu,\mb k}^{(L)}$ gives the longitudinal equation:
\begin{align}
  B^{(L)}{}''+(k^2+a^2m^2)B^{(L)}+2\ii aHk B^{(T)}=0,
\end{align}
and contracting (\ref{eq_AEoMSpatial}) with $e_{\mu,\mb k}^{(\mp)}$ gives the transverse equations:
\begin{align}
\label{eq_ModeEqTrans}
  B^{(\pm)}{}''+(k^2\pm 2a\mu k+a^2m^2)B^{(\pm)}=0.
\end{align}
It can be readily checked that the longitudinal equation and the temporal equation are compatible with the constraint (\ref{eq_Aconstraint}) and therefore are not independent. We can conveniently choose (\ref{eq_ModeEqTemp}) and (\ref{eq_ModeEqTrans}) as an independent and complete set of equations, and the longitudinal mode function $B^{(L)}$ can be found from $B^{(T)}$ by solving the constraint (\ref{eq_Aconstraint}):
\bge
\label{eq_BLsol}
  B^{(L)}=\FR{1}{\ii k}\big(B^{(T)}{}'+2aHB^{(T)}\big).
\ede
Now it is straightforward to solve the mode equations up to overall normalization constants:
\begin{align}
  B^{(\pm)}(k,\tau)=&~\mathcal{N}^{(\pm)}\FR{e^{\mp\pi\wt\mu/2}}{\sqrt{2k}}\mathrm{W}_{\pm\ii\wt\mu,\ii\wt\nu}(2\ii k\tau), \\
  B^{(T)}(k,\tau)=&~\mathcal{N}^{(T)}\FR{\sqrt\pi}{2}e^{-\pi\wt\nu/2}H(-\tau)^{3/2}\mathrm{H}^{(1)}_{\ii\wt\nu}(-k\tau).
\end{align}
Here $\wt\mu\equiv\mu/H$ and $\wt\nu\equiv\sqrt{(m/H)^2-1/4}$, $\mathrm{W}_{\ka,\mu}(z)$ is the Whittaker W function and $\mathrm{H}_{\mu}^{(1)}(z)$ is the Hankel function of the first kind. We shall call $\wt\nu$ the mass parameter and we add a tilde to distinguish it from the more conventionally defined parameter $\nu=\sqrt{1/4-m^2/H^2}$. In this paper, we shall exclusively consider massive fields in the principal series, meaning that $\nu$ is purely imaginary, and $\wt\nu$ is positive real. Note also that the relation between $\wt\nu$ and $m$ depends on the spin: For spin-$s$ $(s\neq 0)$ fields, $\wt\nu=\sqrt{(m/H)^2-(s-1/2)^2}$, and for scalars $(s=0)$, $\wt\nu=\sqrt{(m/H)^2-9/4}$.

With the application of CC physics in mind, we always assume $m>H/2$, the so-called principal series. The generalization to complementary series $0<m<H/2$ is straightforward. 

The normalization constants $\mathcal{N}^{(\pm)}$ and $\mathcal{N}^{(T)}$ are determined by the canonical commutators. On the one hand, we have the canonical equal-time commutator of the field variable $A_i(\tau,\mb x)$ with its canonical conjugate momentum $\Pi^i(\tau,\mb x)\equiv\pd\ld/\pd A_i'$ where $\ld$ is given in (\ref{eq_ActionVecInfBg}):
\begin{align}
  [A_{i}(\tau,\mb x),\Pi^j(\tau,\mb y)]=\ii\,\de_{i}^j\de^{(3)}(\mb x-\mb y),
\end{align}
and all other brackets vanish. On the other hand, we have the canonical commutators of annihilation and creation operators $[a_{\mb k}^{(h)},a_{\mb q}^{(h')\dag}]=(2\pi)^3\de^{hh'}\de^{(3)}(\mb k-\mb q)$ and all other commutators vanish. These two sets of commutators, together with the mode decomposition, fix the normalization of the mode functions. More explicitly, we can again take $\mb k=(0,0,k)^T$ without loss of generality. Then, the transverse field commutators $[A_i(\tau,\mb x),A_j'(\tau,\mb y)]=\ii\de_{ij}\de^{(3)}(\mb x-\mb y)$ with $i=1,2$ lead to the normalization condition for the transverse mode functions:
\bge
  \ii=\mathcal{W}\Big(B^{(\pm)},B^{(\pm)*}\Big)=\big|\mathcal{N}^{(\pm)}\big|^2.
\ede
where $\mathcal{W}(f,g)=fg'-f'g$ is the Wronskian of two functions $f$ and $g$. Thus the coefficients of the transverse modes are fixed to be $\mathcal{N}^{(\pm)}=1$ up to irrelevant phases.

For the longitudinal mode $A_{\mu,\mb k}^{(0)}$, we have $\Pi_3=A_3'-\pd_3 A_0$, and thus the field commutator reads $\big[A_{3}(\tau,\mb x),A_{3}'(\tau,\mb y)-\pd_3 A_{0}(\tau,\mb y)\big]=\ii \de^{(3)}(\mb  x-\mb y)$. Together with $[a_{\mb p}^{(0)},a_{-\mb q}^{(0)\dag}]=(2\pi)^3\de^{(3)}(\mb p+\mb q)$, it gives rise to the following condition:
\begin{align}
  &\Big[B^{(L)}(k,\tau)B^{(L)*}{}'(k,\tau)-B^{(L)*}(k,\tau)B^{(L)}{}'(k,\tau)\Big]\n\\
  &-\ii k\Big[B^{(L)}(k,\tau)B^{(T)*}(k,\tau)-B^{(L)*}(k,\tau)B^{(T)}(k,\tau)\Big]=\ii
\end{align}
Using (\ref{eq_BLsol}) to remove $B^{(L)}$ in the second line, we get a somewhat unusual Wronskian condition:
\bge
  \ii=\mathcal{W}\Big(B^{(L)}(k,\tau),B^{(L)*}(k,\tau)\Big)-\mathcal{W}\Big(B^{(T)}(k,\tau),B^{(T)*}(k,\tau)\Big)=\ii \big|\mathcal{N}^{(T)}\big|^2\FR{m^2}{k^2}.
\ede 
This fixes the normalization coefficient $\mathcal{N}^{(T)}=k/m$ up to an irrelevant phase. 

To summarize, we have obtained the mode functions for the massive spin-1 fields in dS with helical chemical potential as follows:
\begin{keyeqn}
\begin{align}
\label{eq_BmodePM}
  B^{(\pm)}(k,\tau)=&~ \FR{e^{\mp\pi\wt\mu/2}}{\sqrt{2k}}\mathrm{W}_{\pm\ii\wt\mu,\ii\wt\nu}(2\ii k\tau), \\
\label{eq_BmodeT}
  B^{(T)}(k,\tau)=&~ \FR{\sqrt\pi k}{2m}e^{-\pi\wt\nu/2}H(-\tau)^{3/2}\mathrm{H}^{(1)}_{\ii\wt\nu}(-k\tau), \\
\label{eq_BmodeL}
  B^{(L)}(k,\tau)=&- \FR{\ii\sqrt\pi}{4m}e^{-\pi\wt\nu/2}H (-\tau)^{1/2} \Big[\mathrm{H}^{(1)}_{\ii\wt\nu}(-k\tau)-k\tau\Big(\mathrm{H}^{(1)}_{1+\ii\wt\nu}(-k\tau)-\mathrm{H}^{(1)}_{-1+\ii\wt\nu}(-k\tau)\Big)\Big].
\end{align}
\end{keyeqn}
 
\paragraph{SK propagators.} With the mode functions obtained above, it is straightforward to construct the propagators of the massive spinning fields. Owing to the aforementioned complication of the longitudinal mode, namely the temporal component $B^{(T)}$ not proportional to the longitudinal component $B^{(L)}$, we find it more useful to define the propagators with respect to the redundant polarization basis $e_{\mu,\mb k}^{(\lam)}$ given in (\ref{eq_PolarVec}). Following the standard SK path integral formalism \cite{Chen:2017ryl}, the tree-level propagators can be obtained as:
\begin{align}
  D_{++}^{(\lam)}(k;\tau_1,\tau_2)=&~e_{\mb k}^{(\lam)\mu*}e_{-\mb k}^{(\lam)\nu*}\big\la 0|\text{T}\{A_{\mu,\mb k}(\tau_1)A_{\nu,-\mb k}(\tau_2)\} |0\big\ra', \\
  D_{--}^{(\lam)}(k;\tau_1,\tau_2)=&~e_{\mb k}^{(\lam)\mu*}e_{-\mb k}^{(\lam)\nu*}\big\la 0|\ob{\text{T}}\{A_{\mu,\mb k}(\tau_1)A_{\nu,-\mb k}(\tau_2)\} |0\big\ra', \\
  D_{+-}^{(\lam)}(k;\tau_1,\tau_2)=&~e_{\mb k}^{(\lam)\mu*}e_{-\mb k}^{(\lam)\nu*}\big\la 0| A_{\nu,-\mb k}(\tau_2)A_{\mu,\mb k}(\tau_1)  |0\big\ra',\\
  D_{-+}^{(\lam)}(k;\tau_1,\tau_2)=&~e_{\mb k}^{(\lam)\mu*}e_{-\mb k}^{(\lam)\nu*}\big\la 0| A_{\mu,\mb k}(\tau_1)A_{\nu,-\mb k}(\tau_2) |0\big\ra',
\end{align}
where $|0\ra$ is the initial state of the problem which we will take as the usual Bunch-Davies vacuum, $\text{T}$ and $\ob{\text{T}}$ denote the time ordering and anti-time ordering of operators, respectively, and $\la\cdots\ra'$ denotes a correlator with momentum-conserving $\de$-function removed. Then, with the equations (\ref{eq_VecFourierMode}), (\ref{eq_Bpm}), (\ref{eq_Blong}), and (\ref{eq_BmodePM})-(\ref{eq_BmodeL}), one can find the explicit expressions of the above propagators in terms of mode functions. All these propagators can be expressed in terms of the Wightman functions, namely $D_>^{(\lam)}(k;\tau_1,\tau_2)$ and its complex conjugate $D_<^{(\lam)}(k;\tau_1,\tau_2)\equiv D_>^{(\lam)*}(k;\tau_1,\tau_2)$:
\bge
\label{eq_DVecGreater}
  D_>^{(\lam)}(k;\tau_1,\tau_2)\equiv B^{(\lam)}(k,\tau_1)B^{(\lam)*}(k,\tau_2).
\ede
Then, we have:
\begin{align}
\label{eq_DVecPP}
  &D_{++}^{(\lam)}(k;\tau_1,\tau_2)=D_>^{(\lam)}(k;\tau_1,\tau_2)\theta(\tau_1-\tau_2)+D_<^{(\lam)}(k;\tau_1,\tau_2)\theta(\tau_2-\tau_1),\\ 
\label{eq_DVecMM}
  &D_{--}^{(\lam)}(k;\tau_1,\tau_2)=D_<^{(\lam)}(k;\tau_1,\tau_2)\theta(\tau_1-\tau_2)+D_<^{(\lam)}(k;\tau_1,\tau_2)\theta(\tau_2-\tau_1),\\ 
\label{eq_DVecPM}
  &D_{+-}^{(\lam)}(k;\tau_1,\tau_2)=D_<^{(\lam)}(k;\tau_1,\tau_2), \\
\label{eq_DVecMP}
  &D_{-+}^{(\lam)}(k;\tau_1,\tau_2)=D_>^{(\lam)}(k;\tau_1,\tau_2), 
\end{align}
One can show, as we will do in Sec.\ \ref{sec_boot}, that the two opposite-sign propagators $D_{\pm\mp}^{(\lam)}$ satisfies the sourceless (and thus homogenous) equation of motion for the vector field $A_\mu$, while the same-sign propagators satisfy the same equation but with a nonzero source (and thus inhomogeneous). For this reason we will also call $D^{(\lam)}_{\pm\pm}$ the inhomogeneous propagators and $D^{(\lam)}_{\pm\mp}$ the homogeneous propagators. 

For later use, we also present the explicit expressions for the Wightman function $D_>^{(\lam)}$ for various polarizations: 
\begin{align}
  \label{eq_VecPropPM}
  D^{(\pm)}_>(k;\tau_1,\tau_2)=&~B^{(\pm)}(k,\tau_1)B^{(\pm)*}(k,\tau_2)=\FR{e^{\mp\pi\wt\mu}}{2k}\mathrm{W}_{\pm\ii\wt\mu,\ii\wt\nu}(2\ii k\tau_1)\mathrm{W}_{\mp\ii\wt\mu,\ii\wt\nu}(-2\ii k\tau_2),\\
  \label{eq_VecPropT}
  D^{(T)}_>(k;\tau_1,\tau_2)=&~B^{(T)}(k,\tau_1)B^{(T)*}(k,\tau_2)=\FR{\pi H^2k^2}{4m^2}e^{-\pi\wt\nu}(\tau_1\tau_2)^{3/2}\mathrm{H}^{(1)}_{\ii\wt\nu}(-k\tau_1)\mathrm{H}^{(2)}_{-\ii\wt\nu}(-k\tau_2),\\
  \label{eq_VecPropL}
  D^{(L)}_>(k;\tau_1,\tau_2)=&~\FR{1}{k^2}\bigg(\pd_{\tau_1}-\FR{2}{\tau_1}\bigg)\bigg(\pd_{\tau_2}-\FR{2}{\tau_2}\bigg)D^{(T)}_>(k;\tau_1,\tau_2).
\end{align}

One caveat is that the relation between the longitudinal propagator and the temporal propagator as written in (\ref{eq_VecPropL}) applies only to homogeneous type propagators, namely $D^{(L)}_\gtrless$ or $D^{(L)}_{\pm\mp}$. Owing to the presence of Heaviside $\theta$-functions, the two inhomogeneous propagators $D_{\pm\pm}^{(L)}$ and $D_{\pm\pm}^{(T)}$ are not related by simply acting the differential operators $\pd_{\tau_{1,2}}-2/\tau_{1,2}$ as in (\ref{eq_VecPropL}). There will be additional contact term generated involved in such a relation. This point will be relevant for our computation of longitudinal correlators below, especially when using the bootstrap equations in Sec.\ \ref{sec_boot}.

\subsection{Correlators and the vector seed integral}
\label{sec_VSI}

From this point onward, we shall take $H=1$ to simplify expressions. 

With the massive spin-1 fields properly quantized and their propagators obtained, we are now ready to construct helical inflation correlators. In this work, we focus on the tree-level exchange of massive spinning particles mainly in 4-point correlators, namely the trispectrum. The 3-point correlators, namely the bispectrum, will be considered in Sec.\ \ref{sec_var} as soft limits of 4-point correlators. 

\begin{figure}
\centering
  \parbox{0.32\textwidth}{\includegraphics[width=0.32\textwidth]{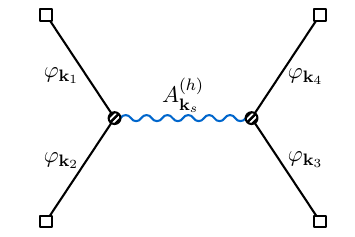}}
  \parbox{0.32\textwidth}{\includegraphics[width=0.32\textwidth]{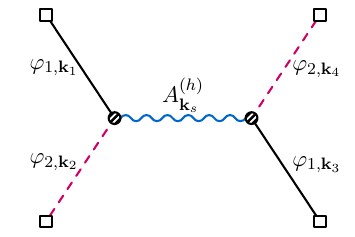}}
  \caption{The 4-point correlators mediated by a massive spin-1 field in the $s$-channel at the tree level. The left diagram shows the correlator of four identical external scalars $\la\varphi_{\mb k_1}\varphi_{\mb k_2}\varphi_{\mb k_3}\varphi_{\mb k_4}\ra'$, which corresponds to the amplitude in (\ref{eq_T1}). The right diagram corresponds to correlators with two distinct external scalars $\la\varphi_{1,\mb k_1}\varphi_{2,\mb k_2}\varphi_{1,\mb k_3}\varphi_{2,\mb k_4}\ra'$, which corresponds to the amplitude in (\ref{eq_T2}).}
  \label{fd_vector}
\end{figure}

The 4-point correlators we will be considering are shown in Fig.\ \ref{fd_vector}. In this figure, we only draw the $s$-channel exchange. There are two other similar processes with $t$-channel and $u$-channel exchanges that can be obtained by simple permutations $\mb k_1\leftrightarrow \mb k_4$ and $\mb k_1\leftrightarrow \mb k_3$, respectively. To form a correlator as in Fig.\ \ref{fd_vector}, we need a trilinear interaction vertex with two external scalar fields and one internal spinning field, to be introduced below.

\paragraph{Trilinear couplings.} From the previous subsection, we see that the helical chemical potential affects only the transverse components. Therefore, the trilinear coupling must see the spatial components $A_i$ of the spin-1 field in order to see the effect of the chemical potential. For this reason, we shall only consider couplings involving $A_i$ and neglect couplings to $A_0$. Also, for convenience, we shall consider a process with two distinct external massless scalars $\varphi_1$ and $\varphi_2$, shown by the right diagram in Fig.\ \ref{fd_vector}. (The result for four identical external scalars, namely the left diagram of Fig.\ \ref{fd_vector}, can be obtained by permuting the momenta. See below.) With the symmetry of the problem given, we can write down the following most general trilinear interaction vertex among $\varphi_1$, $\varphi_2$, and $A_i$:
\bge
\label{eq_CoupGeneral}
  \mathcal{O}_{\mathbb{P}}= \lam_{\mathbb{P}}a^{2-(J+K)-2(M+N+R)} \Big[\pd_{\tau}^{J} \pd_{i_1}\cdots\pd_{i_{M}}(\pd_j\pd^j)^{N}\varphi_1\Big] \Big[\pd_{\tau}^{K}\pd^i\pd^{i_1}\cdots\pd^{i_{M}}(\pd_k\pd^k)^{R}\varphi_2\Big] A_i.
\ede
Here the subscript $\mathbb{P}=(J,K,M,N,R)$ is a set of nonnegative integers specifying the number of various temporal and spatial derivatives, and it also serves as a label for the operator $\mathcal{O}_{\mathbb{P}}$ and the coupling strength $\lam_{\mathbb{P}}$. The power of the scale factor $a$ is uniquely fixed by the scale invariance. It is of course possible to act temporal and spatial derivatives on $A_i$, but these derivatives can always be moved to $\varphi_1$ and $\varphi_2$ using integration by parts. Therefore, any operators involving derivatives of $A_i$ can be written as linear combinations of operators of the above form up to irrelevant boundary terms. 

It might be useful to find the result for the correlator with couplings written in the above very general form. However, it is clearly rather cumbersome to stay in the most general case throughout our presentation. Also, the final result will be rather complicated and uninspiring. Therefore we shall present the result for the 4-point correlator with the most general coupling (\ref{eq_CoupGeneral}) in App.\ \ref{app_general}. Here, for the sake of clarity, we will set $J=1$ and $K=M=N=R=0$ in the operator $\mathcal{O}_{\mathbb{P}}$ in (\ref{eq_CoupGeneral}), and we simply call the resulting operator $\mathcal{O}$ :
\begin{equation}
\label{eq_CoupSimp}
    \mathcal{O} = \lam a \varphi_1' (\pd^i \varphi_2) A_i.
\end{equation}
This could well be the simplest choice for the trilinear coupling, and is the lowest dimension coupling one can write down for the three fields $\varphi_1$, $\varphi_2$, and $A_i$ that respects the shift symmetry of the scalar fields. Note in particular that the shift symmetry requires us to retain at least one derivative to each of the two scalar fields.

\paragraph{SK integral.} Now we are ready to write down an expression for the 4-point correlator shown in the right diagram of Fig.\ \ref{fd_vector}, with the two vertices given by the simple operator $\mathcal{O}$ in (\ref{eq_CoupSimp}), following the diagrammatic rule summarized in \cite{Chen:2017ryl}. More explicitly, we use $G_\aa(k,\tau)$ to denote the bulk-to-boundary propagator for the external scalar fields:
\bge
\label{eq_BtoBprop}
  G_\aa(k,\tau)=\FR{1}{2k^3}(1-\ii\aa k\tau)e^{\ii\aa k\tau}.
\ede
Since we are assuming both $\varphi_1$ and $\varphi_2$ massless, we do not use different letters to distinguish their propagators. Then, we can write down the amplitude mediated by each of the four polarizations of the spin-1 field:
\begin{align}
\label{eq_T2}
    \mathcal{T}_{\text{2}}^{(\lam)}
    =&-\lam^2(\ii\mb k_2\cdot \mb e_{\mb k_s}^{(\lam)})(\ii\mb k_4\cdot \mb e_{\mb k_s}^{(\lam)*})\sum_{\mathsf{a},\mathsf{b}=\pm}\mathsf{ab}\int_{-\infty}^0 \FR{\di\tau_1}{-\tau_1}\FR{\di\tau_2}{-\tau_2}\n\\
    & \times 
     \big[\pd_{\tau_1} G_\mathsf{a} (k_1,\tau_1)\big]  G_\mathsf{a}(k_2,\tau_1) 
     \big[\pd_{\tau_2} G_\mathsf{b}(k_3,\tau_2)\big] G_\mathsf{b}(k_4,\tau_2)
    D_{\mathsf{ab}}^{(\lam)}(k_s;\tau_1,\tau_2). 
\end{align}
Then, use (\ref{eq_BtoBprop}), we find
\begin{align}
\label{eq_T2int2}
    \mathcal{T}_{\text{2}}^{(\lam)}
    =&~\lam^2(\mb k_2\cdot \mb e_{\mb k_s}^{(\lam)})(\mb k_4\cdot \mb e_{\mb k_s}^{(\lam)*})
     \FR{1}{16k_1k_2^3k_3k_4^3} \sum_{\aa,\bb=\pm}\aa\bb  \n\\
     &~\times\int_{-\infty}^0 \di\tau_1 \di\tau_2 (1-\ii\aa k_2\tau_1) (1-\ii\bb k_4\tau_2)e^{\ii \aa k_{12}\tau_1+\ii \bb k_{34}\tau_2 }D_{\aa\bb}^{(\lam)}(k_s;\tau_1,\tau_2).
\end{align}
The process with four identical scalars, namely the left diagram of Fig.\ \ref{fd_vector}, can be obtained directly by permutation of the four external momenta:
\begin{equation}
\label{eq_T1}
    \mathcal{T}_{\text{1}}^{(\lam)}(k_I) =\mathcal{T}_{\text{2}}^{(\lam)}(k_I)+(k_1\leftrightarrow k_2)+(k_3\leftrightarrow k_4)+\binom{k_1\leftrightarrow k_2}{k_3\leftrightarrow k_4}. 
\end{equation}
There are some cancellations in this permutation in the leading order CC signals in the squeezed limit. However, the CC signals are still present in the subleading terms, albeit with a different angular dependence. See \cite{Qin:2022lva} for more discussions. 

\paragraph{Helical Vector Seed Integral.}
One can well follow the standard procedure and substitute the explicit expressions of the spin-1 propagators (\ref{eq_VecPropPM}) and (\ref{eq_VecPropL}) into (\ref{eq_T2int2}). However, the computation of the two time integrals is not that trivial. Therefore, we will isolate this integral from the rest of the expression, by defining the following \emph{helical vector seed integral}:
\begin{keyeqn}
\begin{align} 
\label{eq_vecIab}
     \mathcal{I}^{(\lam)p_1p_2}_{\mathsf{ab}}  \equiv & -\mathsf{ab}\, k_s^{3+p_1+p_2}\int_{-\infty}^0 \di\tau_1\di\tau_2\,(-\tau_1)^{p_1}(-\tau_2)^{p_2}e^{\ii\mathsf a k_{12}\tau_1+\ii\mathsf b k_{34}\tau_2} D_{\mathsf{ab}}^{(\lam)}(k_s;\tau_1,\tau_2).
\end{align}
\end{keyeqn}

Let us make some remarks about this vector seed integral before going on to consider the 4-point correlator $\mathcal{T}_{\text{2}}^{(\lam)}$.

First, the minus sign and also the factor $\aa\bb$ in front of everything in (\ref{eq_vecIab}) are from the two vertices, which give $\ii\aa$ and $\ii\bb$ as the diagrammatic rule required. Second, we have introduced two arbitrary powers of time $(-\tau_1)^{p_1}$ and $(-\tau_2)^{p_2}$, and it turns out that this slight generalization suffices to accommodate a wide range of diagrams, including the diagrams with the most general coupling (\ref{eq_CoupGeneral}). It is for this reason that we call $\mathcal{I}^{(\lam)p_1p_2}_{\mathsf{ab}}$ a seed integral. Third, we have also introduced the factor $k_s^{3+p_1+p_2}$ to balance the powers of time inside the integral, so that the whole expression $\mathcal{I}^{(\lam)p_1p_2}_{\mathsf{ab}}$ is dimensionless and scale invariant, and must be expressible in terms of momentum ratios.\footnote{This is not the unique choice. We can also balance the $\tau$-powers by including a factor of $k_{12}^{3/2+p_1}k_{34}^{3/2+p_2}$. Our choice $k_s^{3+p_1+p_2}$ slightly simplifies some derivations and the final expressions.} It turns out that the integral (\ref{eq_vecIab}) depends on two independent momentum ratios, reminiscent of the behavior of a 4-point correlator in ordinary conformal field theories. As we shall see below, it is convenient to define the following momentum ratios:
\begin{align}
\label{eq_Kratios}
  &r_1\equiv\FR{k_s}{k_{12}},
 &&r_2\equiv\FR{k_s}{k_{34}}.
  &u_1\equiv\FR{2k_s}{k_{12}+k_s}=\FR{2r_1}{1+r_1},
 &&u_2\equiv\FR{2k_s}{k_{34}+k_s}=\FR{2r_2}{1+r_2}.
\end{align}
Clearly, the momentum conservation at each vertex, namely $\mb k_1+\mb k_2=\mb k_s=-(\mb k_3+\mb k_4)$, requires that $r_1,r_2,u_1,u_2\in[0,1]$. As we shall see in Sec.\ \ref{sec_vec}, it turns out more convenient to use $(r_1,r_2)$ as independent variables for the longitudinal integral $\mathcal{I}^{(L)p_1p_2}_{\mathsf{ab}}$ while $(u_1,u_2)$ is a better choice for the transverse integral $\mathcal{I}^{(\pm)p_1p_2}_{\mathsf{ab}}$.

\paragraph{From the seed integral to the correlator.} Now, in terms of the vector seed integral (\ref{eq_vecIab}), we can write the 4-point correlator $\mathcal{T}_{\text{2}}^{(\lam)}$ in two different ways. First, the correlator can be written as a linear combination of seed integrals with different SK indices: 
\begin{align}
\label{eq_T2inI}
    \mathcal{T}_{\text{2}}^{(\lam)} 
    =&~\FR{-\lam^2}{16 k_1k_2k_3k_4 k_s^3} 
     \FR{(\wh{\mb k}_2\cdot \mb e_{\mb k_s}^{(\lam)})(\wh{\mb k}_4\cdot \mb e_{\mb k_s}^{(\lam)*}) }{k_2k_4} \n\\
     &~\times\sum_{\aa,\bb=\pm} \Big[  \mathcal{I}_{\aa\bb}^{(\lam)0,0} 
      - \ii\aa \FR{k_2}{k_s} \mathcal{I}_{\aa\bb}^{(\lam)1,0} 
      - \ii\bb \FR{k_4}{k_s}\mathcal{I}_{\aa\bb}^{(\lam)0,1}
      -\aa\bb \FR{k_2k_4}{k_s^2}\mathcal{I}_{\aa\bb}^{(\lam)1,1}\Big]. 
\end{align}
The advantage of the representation (\ref{eq_T2inI}) is that one only need to form simple linear combination once the result for the vector seed integral is known. However, an unsatisfactory aspect of this representation is that the coefficients of some terms contain the SK indices $\aa,\bb$, which means that we have to know the result of $\mathcal{I}_{\aa\bb}^{p_1p_2}$ for each $\aa,\bb=\pm$. While there is no problem in obtaining the results for each individual integral, it is somewhat redundant to present the result for each $\aa,\bb$ separately. In fact, the results for different $\aa,\bb=\pm$ are intimately related, not least because the bulk propagators $D_{\aa\bb}$ can all be constructed from a single Wightman function $D_>$.

There is an alternative way to relate the vector seed integral and the correlator, which is more in line with the spirit of the recent bootstrap papers \cite{Arkani-Hamed:2018kmz,Baumann:2019oyu,Pimentel:2022fsc,Jazayeri:2022kjy}, where the correlators of more complicated processes are all constructed from a seed correlator by acting appropriate differential or integral operators. For our example of $\mathcal{T}_{\text{2}}^{(\lam)}$, one can easily verify the following relation:
\begin{align}
\label{eq_T2fromDI}
  \mathcal{T}_{\text{2}}^{(\lam)}
  =\FR{-\lam^2}{16 k_1k_2k_3k_4 k_s^3} 
     \FR{(\wh{\mb k}_2\cdot \mb e_{\mb k_s}^{(\lam)})(\wh{\mb k}_4\cdot \mb e_{\mb k_s}^{(\lam)*}) }{k_2k_4}\times(1-k_2\pd_{k_{12}})(1-k_4\pd_{k_{34}})\sum_{\aa,\bb=\pm}\mathcal{I}_{\aa\bb}^{(\lam)0,0} . 
\end{align}
In this representation, we only need the summed vector seed integral $\sum_{\aa,\bb}\mathcal{I}_{\aa\bb}^{(\lam)p_1p_2}$. However, we need to act differential operators to the summed integral, which is itself a very complicated expression. In practical calculations, we do not have to stick to one representation, but can freely exploit both representations in order to simplify the calculation. We will illustrate this point in the subsequent sections. Of course, whatever approach taken, the final answer should be the same, in part because acting differential or integral operators on the seed correlator is usually equivalent to changing some parameters in it.

To summarize, in this section, we have reduced the computation of the helical 4-point correlators to the computation of a vector seed integral (\ref{eq_vecIab}).
In Sec.\ \ref{sec_vec}, we will carry out this computation and present an exact and analytical expression for it. We will also provide independent derivations for the seed integrals for particular choices of $p_{1,2}$ in Sec.\ \ref{sec_boot}, using a bootstrap approach.

\section{Inflation Correlators with Partial Mellin-Barnes Representation}
\label{sec_PMB}

From the previous section, we see that the major task in computing the correlator is to evaluate the SK integrals, preferably in an exact and analytic form. For a wide range of applications involving helicity dependent chemical potentials, it suffices to compute a single vector seed integral, defined in (\ref{eq_vecIab}). The seed integral is a two-layer integral over products of special functions with time orderings, which make the computation difficult. 

The approach we shall take to compute these integrals is the so-called partial Mellin-Barnes representation that we introduced in \cite{Qin:2022lva}. Before applying this method  to the vector seed integral, however, we shall demonstrate the use of partial Mellin transformation with a simpler example, namely the 4-point correlator mediated by a massive scalar field. The mode function associated with massive scalars involves the Hankel functions, which is somewhat simpler than the Whittaker function in the spin-1 mode function. Also, the result obtained in this section can be recycled when we consider the longitudinal spin-1 exchange. Finally, we shall present a new expression for the ``background'' part of the correlator, which is expressed as a fast converging series of $k_s/k_{12}$ alone, while the dependence on $k_{12}/k_{34}$ is fully resummed into a hypergeometric function, as opposed to the double-series expressions given in previous works. A detailed comparison between these two types of results will be given in Sec.\ \ref{sec_boot}.

\paragraph{Mellin transformation.}
The Mellin transformation of a function $f(x)$ is defined by
\bge
  F(s)=\int_0^\infty\di x\, x^{s-1}f(x).
\ede
The inverse Mellin transformation, also known as the Mellin-Barnes (MB) representation, is given by
\bge
  f(x)=\int_{c-\ii\infty}^{c+\ii\infty}\FR{\di s}{2\pi\ii}\,x^{-s} F(s),
\ede
where $c\in (\al,\be)$ is a real number and $(\al,\be)$ is an open interval in the real axis that includes 0. Very often, we can simply take $c=0$.

Like the more familiar Fourier transformation, the Mellin transformation is a linear transformation of functions, but with power function $x^{s-1}$ instead of exponential function $e^{\ii k x}$, as the transformation kernel. In dS, the background geometry possesses a dilatation symmetry generated by $\tau\pd_\tau$, with power functions $(-\tau)^\Delta$ as eigenfunctions. (In AdS, it is the radial direction that has a similar symmetry.) Therefore, the Mellin transformation in dS is really a natural analogy of Fourier transformation of time in Minkowski space. One can thus imagine that the Feynman rules and computations for dS correlators can be drastically simplified if we Mellin transform the time direction, in addition to the usual spatial Fourier transformation. For instance, the difficult time integral for each interaction vertex is replaced by a $\de$-function of Mellin variables, much like a distorted form of energy conservation in curved space. Indeed, the advantages of Mellin amplitudes in the context of inflation correlators have been investigated in \cite{Sleight:2019mgd,Sleight:2019hfp,Sleight:2020obc,Sleight:2021iix,Sleight:2021plv}. 

On the other hand, the inflation correlators are boundary correlators of operators with fixed time $\tau=0$, and therefore, to obtain these correlators from the Mellin amplitudes, we need to perform inverse Mellin transformation separately at \emph{each} external points, which is likely not a trivial step, and this limits the direct use of Mellin amplitudes for practical study of inflation correlators. 

To take the advantage of Mellin transformation in a more practical way, we adopted the \emph{partial Mellin-Barnes representation} introduced in \cite{Qin:2022lva}. The essential idea is simple: We use MB representations only for all massive internal modes. The MB representation resolves various special functions associated with these massive modes into powers of time variable $\tau$, and thus trivializes the time integrals. Then, the MB integral can be carried out by closing the contour appropriately and picking up residues of enclosed singularities.

\subsection{Scalar exchange revisited}
Given that the partial MB representation has not been widely adopted in the study of inflation correlators, in this section, we shall walk the readers through the main steps of this method with a simpler example, namely the tree-level exchange of a massive real scalar field in 4-point correlator. The complete analytical result for this process has been worked out in \cite{Arkani-Hamed:2018kmz}. In addition to providing independent checks, our final expression features an improved series expression for the background piece of the correlator, which resums the powers of $r_1/r_2$ to all orders into tractable Gauss hypergeometric functions.

\begin{figure}[tbph]
\centering
  \parbox{0.35\textwidth}{\includegraphics[width=0.31\textwidth]{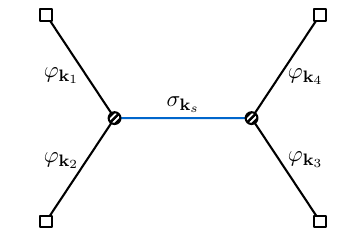}}
\caption{The 4-point inflation correlator with tree-level exchange of a massive scalar field $\si$ in the $s$-channel.}
\label{fig_scalar_tree}
\end{figure}

To be concrete, we consider the 4-point correlator shown in Fig.\;\ref{fig_scalar_tree}. The external legs represent bulk-to-boundary propagators of $\varphi$ field with momenta $\mb k_i$ ($i=1,\cdots,4$). In the study of realistic inflation correlators, $\varphi$ is usually taken to be the nearly massless inflaton fluctuation. Its bulk-to-boundary propagator is given in (\ref{eq_BtoBprop}).  

It is sometimes convenient to take the external legs to be conformal scalars $\phi_c$ with mass $m^2=2H^2$. These scalars do not propagate to the future boundary $\tau=0$ of dS; their mode functions are proportional to $-\tau$ and vanish when $\tau=0$. Therefore, to write their bulk-to-boundary propagator $C_\aa(k,\tau)$, we introduce a final-time cutoff $\tau=\tau_f$:
\bge
  C_\aa(k,\tau)=\FR{\tau\tau_f}{2k}e^{\aa\ii k\tau}.
\ede

The internal (blue) line in Fig.\;\ref{fig_scalar_tree} represents the bulk propagator $D_{\aa\bb}(k,\tau_1,\tau_2)$ of a massive scalar field $\si$. The four bulk propagators are again related to the Wightman functions $D_>$ and its complex conjugate $D_<$ in a way similar to the vector case shown in (\ref{eq_DVecPP})-(\ref{eq_DVecMP}). (One simply removes the superscript $(\lam)$ in those expressions.) For a massive scalar, the Wightman function $D_>$ is given by: 
\bge
\label{eq_ScalarDGreater}
  D_>(k;\tau_1,\tau_2)=\FR{\pi}{4}e^{-\pi\wt\nu}H^2(\tau_1\tau_2)^{3/2}\mathrm{H}_{\ii\wt\nu}^{(1)}(-k\tau_1)\mathrm{H}_{-\ii\wt\nu}^{(2)}(-k\tau_2),
\ede
where $\wt\nu\equiv\sqrt{m^2/H^2-9/4}$ is the mass parameter. 

There are many possible choices for the two vertices in Fig.\;\ref{fig_scalar_tree}. Since we are considering the scalar correlator only for illustrative purpose, we do not aim at the most general coupling. Instead, we will only mention several frequently encountered examples. 

In the case of the massless inflaton fluctuation being the external modes, the simplest choice is the following one:
\begin{equation}
\label{eq_ScalarTDCoup}
    \Delta\ld = \FR12 \lam a^2 \varphi'^2 \sigma.
\end{equation}
We choose the derivative coupling to keep the (approximate) shift symmetry of the inflaton field. 

For the conformal scalar $\phi_c$, one can simply choose a direct coupling:
\bge
\label{eq_ScalarConfCoup}
  \Delta\ld = \FR{1}{2}\lam_c a^4 \phi_c^2 \sigma.
\ede

With all the propagators and vertices given, it is straightforward to write down the expression for the correlator in Fig.\;\ref{fig_scalar_tree} following the standard procedure \cite{Chen:2017ryl}. For instance, the inflaton correlator $\la\varphi^4\ra$ with the interaction (\ref{eq_ScalarTDCoup}) is given by 
\begin{align}
    \mathcal T_\varphi \equiv &~ \la\varphi_{\mb k_1}\varphi_{\mb k_2}\varphi_{\mb k_3}\varphi_{\mb k_4}\ra_s'\n\\
  =&-\lam^2 \sum_{\mathsf{a,b}=\pm} \mathsf{ab} \int_{-\infty}^0 \FR{\di\tau_1}{(-\tau_1)^2}\FR{\di\tau_2}{(-\tau_2)^2}\,G_{\mathsf a}'(k_1,\tau_1)G_{\mathsf a}'(k_2,\tau_1)
    G_{\mathsf b}'(k_3,\tau_2)G_{\mathsf b}'(k_4,\tau_2)D_{\mathsf{ab}}(k_s;\tau_1,\tau_2)\n\\
  =&-\FR{\lam^2}{16k_1k_2k_3k_4} \sum_{\mathsf{a,b}=\pm} \mathsf{ab} \int_{-\infty}^0  \di\tau_1 \di\tau_2\, e^{\ii\mathsf{a}k_{12}\tau_1+\ii\mathsf{b}k_{34}\tau_2} D_{\mathsf{ab}}(k_s;\tau_1,\tau_2).
    \label{eq_sca4pt}
\end{align}
The conformal scalar correlator $\la\phi_c^4\ra$, on the other hand, is given by:
\begin{align}
  \mathcal{T}_{\phi_c}\equiv&~\la\phi_{c,\mb k_1}\phi_{c,\mb k_2}\phi_{c,\mb k_3}\phi_{c,\mb k_4}\ra_s'\n\\
  =&-\FR{\lam_c^2\tau_f^{4}}{16k_1k_2k_3k_4}\sum_{\mathsf{a,b}=\pm}\mathsf{ab}\int_{-\infty}^{0} \di\tau_1\di\tau_2\, (\tau_1\tau_2)^{-2}e^{\ii\mathsf{a}k_{12}\tau_1+\ii\mathsf{b}k_{34}\tau_2}D_{\mathsf{ab}}(k_s;\tau_1,\tau_2).
\end{align}

By examining these and other more general examples, it turns out useful to define the following \emph{scalar seed integral}: 
\begin{keyeqn}
\begin{align}
\label{eq_ScalarSeedInt}
  \mathcal{I}_{\aa\bb}^{p_1p_2}(r_1,r_2)\equiv -\mathsf{ab}\, k_s^{5+p_1+p_2}\int_{-\infty}^{0} \di\tau_1\di\tau_2\,(-\tau_1)^{p_1}(-\tau_2)^{p_2} e^{\ii\mathsf{a}k_{12}\tau_1+\ii\mathsf{b}k_{34}\tau_2}D_{\mathsf{ab}}(k_s;\tau_1,\tau_2).
\end{align}
\end{keyeqn} 
As in the vector seed integral, here we have kept the powers of $\tau_1$ and $\tau_2$ general. Also, the integral $\mathcal{I}_{\aa\bb}^{p_1p_2}(r_1,r_2)$ thus defined is dimensionless and scale invariant, and depends on various momenta only through two ratios $r_1= k_s/k_{12}$ and $r_2=k_s/k_{34}$.

It is now trivial to see that our previous two examples can be conveniently expressed as: 
\bge
  \mathcal{T}_\varphi=\FR{\lam^2}{16k_1k_2k_3k_4k_s^5}\sum_{\aa,\bb=\pm}\mathcal{I}^{00}_{\aa\bb}(r_1,r_2),
\ede 
and, 
\bge
\label{eq_Tvarphi}
  \mathcal{T}_{\phi_c}=\FR{\lam_c^2\tau_f^4}{16k_1k_2k_3k_4 k_s}\sum_{\aa,\bb=\pm}\mathcal{I}^{-2,-2}_{\aa\bb}(r_1,r_2).
\ede 
\begin{figure}
\centering
  \includegraphics[width=0.35\textwidth]{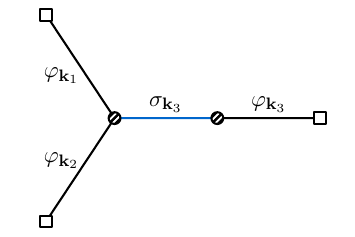}
  \caption{The 3-point correlator of the inflaton fluctuation $\varphi$ mediated by a massive scalar field $\si$ at the tree level.}
  \label{fig_fd_3ptScalar}
\end{figure}

The scalar seed integral thus defined can be used to compute more general inflation correlators. A phenomenologically important example is the 3-point inflaton correlator $\la\varphi^3\ra$, shown in Fig.\ \ref{fig_fd_3ptScalar}. In this figure, the left vertex is again given by (\ref{eq_ScalarTDCoup}), while the right two-point mixing vertex is given by the interaction $\Delta\ld=\lam_2a^3\varphi'\si$. Then the 3-point function can be expressed as 
\begin{align}
\label{eq_BispecToI}
  \mathcal{B}_\phi=\la\varphi_{\mb k_1}\varphi_{\mb k_2}\varphi_{\mb k_3}\ra_3'=\FR{\lam\lam_2}{8k_1k_2k_3^4}\lim_{r_2\to 1^-}\sum_{\aa,\bb=\pm}\mathcal{I}_{\aa\bb}^{0,-2}(r_1,r_2).
\end{align} 
Here we add a subscript 3 to highlight the factor that we are including only the figure with the massive propagator carrying momentum $\mb k_3$. The full result should also include two permutations, which we shall always neglect for simplicity.  
Incidentally, we note that setting directly $r_1=1$ or $r_2=1$ in $\mathcal{I}_{\aa\bb}^{p_1p_2}(r_1,r_2)$ may give rise to superficially divergent expressions, which are nevertheless canceled within each SK branch. We will come back to this point in Sec.\ \ref{sec_boot}. An explicit closed-form analytical expression for (\ref{eq_BispecToI}) will be given in (\ref{eq_Scalar3ptFullResult}).

Similar expressions can be obtained for more general examples including spatial derivative couplings. Therefore, the computation of scalar-exchange correlators has been largely reduced to the computation of scalar seed integral, which we shall elaborate in the rest of this section.

\subsection{Partial MB representation}

In this subsection we are going to use the partial MB representation to compute the scalar seed integral (\ref{eq_ScalarSeedInt}). This subsection is somewhat technical. Readers uninterested in the details can jump directly to Sec.\ \ref{sec_scalar_summary} for a summary of the result.

The main strategy is to use the MB representation to resolve the bulk massive propagators into powers of time. In the scalar seed integral, the bulk propagator (\ref{eq_ScalarDGreater}) includes a pair of Hankel functions. The MB representation of the Hankel function is given by \cite{nist:dlmf}:
\bge
  \text{H}_{\nu}^{(j)}(az)=\int_{-\ii\infty}^{\ii\infty}\FR{\di s}{2\pi\ii}\FR{(az/2)^{-2s}}{\pi}e^{(-1)^{j+1}(2s-\nu-1)\pi\ii/2}\Gamma\Big[s-\FR{\nu}{2},s+\FR{\nu}{2}\Big].~~~~(j=1,2)
\ede
It is then straightforward to get the MB representation for the four bulk propagators:
\begin{align}
\label{eq_DScalarMB1}
    D_{\pm\mp}(k;\tau_1,\tau_2) =&~ \FR{1}{4\pi}
    \int_{-\ii\infty}^{\ii\infty}
    \FR{\di s_1}{2\pi\ii}\FR{\di s_2}{2\pi\ii}\,
    e^{\mp\ii\pi(s_1-s_2)}\Big(\FR{k}2\Big)^{-2s_{12}}
    (-\tau_1)^{-2s_1+3/2}(-\tau_2)^{-2s_2+3/2}\n\\
    &\times \Gamma\Big[s_1-\FR{\ii\wt\nu}2,s_1+\FR{\ii\wt\nu}2,s_2-\FR{\ii\wt\nu}2,s_2+\FR{\ii\wt\nu}2\Big],\\
\label{eq_DScalarMB2}
    D_{\pm\pm}(k;\tau_1,\tau_2)=&~D_{\mp\pm}(k;\tau_1,\tau_2)\theta(\tau_1-\tau_2)+D_{\pm\mp}(k;\tau_1,\tau_2)\theta(\tau_2-\tau_1).
\end{align} 
Here and below, we use shorthand $s_{12}=s_1+s_2$.
Substituting the MB bulk propagators (\ref{eq_DScalarMB1}) and (\ref{eq_DScalarMB2}) into the scalar seed integral (\ref{eq_ScalarSeedInt}), we resolve all time dependences into powers, and thus simplify the time integral. This is particularly true for the opposite-sign integrals $\mathcal{I}_{\pm\mp}^{p_1p_2}(r_1,r_2)$, in which the time integrals yield new factors of Euler $\Gamma$ functions:  
\begin{align}
     \mathcal{I}_{\pm\mp}^{p_1p_2}(r_1,r_2) =&~ \FR{1}{4\pi}e^{\mp\ii\pi(p_1-p_2)/2}r_1^{5/2+p_1}r_2^{5/2+p_2}\int_{-\ii\infty}^{\ii\infty}\FR{\di s_1}{2\pi\ii}\FR{\di s_2}{2\pi\ii}\,
    \Big(\FR{r_1}2\Big)^{-2s_1}\Big(\FR{r_2}2\Big)^{-2s_2}\n\\
    &\times \Gamma\Big[p_1+\fr52-2s_1,p_2+\fr52-2s_2,s_1-\fr{\ii\wt\nu}2,s_1+\fr{\ii\wt\nu}2,s_2-\fr{\ii\wt\nu}2,s_2+\fr{\ii\wt\nu}2\Big].
\label{eq_scaIpmmpMB}
\end{align}  
The same-sign integrals, however, are a bit more complicated owing to the Heaviside $\theta$ functions for the (anti-)time-orderings. There are multiple ways to deal with this time ordering. It turns that different treatments will yield results with different convergent speeds, and the convergence speed also depends on whether $k_{12}>k_{34}$ or $k_{12}<k_{34}$. 

Assuming $k_{12}>k_{34}$ (and thus $r_1<r_2$) for the moment. It turns out useful to adopt the following rearrangement of the bulk propagator $D_{\pm\pm}$:
\begin{align}
  D_{\pm\pm}(k;\tau_1,\tau_2)
  =&~D_\gtrless(k;\tau_1,\tau_2)+\Big[D_\lessgtr(k;\tau_1,\tau_2)-D_\gtrless(k;\tau_1,\tau_2)\Big]\theta(\tau_2-\tau_1).
\end{align}
Correspondingly, we rewrite the seed integral $\mathcal{I}_{\pm\pm}^{p_1p_2}(r_1,r_2)$ as:
\bge
  \mathcal{I}_{\pm\pm}^{p_1p_2}(r_1,r_2)=\mathcal{I}_{\pm\pm,\text{F},>}^{p_1p_2}(r_1,r_2)+\mathcal{I}_{\pm\pm,\text{TO},>}^{p_1p_2}(r_1,r_2).~~~~(r_1<r_2)
\ede
Here the factorized (F) integral $\mathcal{I}_{\pm\pm,\text{F},>}^{p_1p_2}(r_1,r_2)$ and the time-ordered (TO) integral $\mathcal{I}_{\pm\pm,\text{TO},>}^{p_1p_2}(r_1,r_2)$ are respectively defined as: (The subscript $>$ associated with these integrals serves as a reminder that we are focusing on the region $k_{12}>k_{34}$.) 
\begin{align}
\label{eq_IpmpmF}
  \mathcal{I}_{\pm\pm,\text{F},>}^{p_1p_2}(r_1,r_2)
  \equiv & - k_s^{5+p_1+p_2}\int_{-\infty}^{0} \di\tau_1\di\tau_2\,(-\tau_1)^{p_1}(-\tau_2)^{p_2} e^{\pm \ii(k_{12}\tau_1+k_{34}\tau_2)}D_\gtrless(k_s;\tau_1,\tau_2),\\
\label{eq_IpmpmTO}
  \mathcal{I}_{\pm\pm,\text{TO},>}^{p_1p_2}(r_1,r_2)
  \equiv & - k_s^{5+p_1+p_2}\int_{-\infty}^{0} \di\tau_2\int_{-\infty}^{\tau_2}\di\tau_1\,(-\tau_1)^{p_1}(-\tau_2)^{p_2} e^{\pm \ii(k_{12}\tau_1+k_{34}\tau_2)}\n\\
  &~\times\Big[D_\lessgtr(k;\tau_1,\tau_2)-D_\gtrless(k;\tau_1,\tau_2)\Big].
\end{align}  
Then, the factorized time integral, very similar to the opposite-sign integrals (\ref{eq_scaIpmmpMB}), can be carried out directly: 
\begin{align}
{\mathcal I}_{\pm\pm,\text{F},>}^{p_1p_2}=&~\FR{1}{4\pi}e^{\mp\ii\pi(p_1+p_2)/2}r_1^{5/2+p_1}r_2^{5/2+p_2} \int_{-\ii\infty}^{\ii\infty}\FR{\di s_1}{2\pi\ii}\FR{\di s_2}{2\pi\ii}\,(\pm \ii e^{\pm 2\ii\pi s_1})
    \Big(\FR{r_1}2\Big)^{-2s_1}\Big(\FR{r_2}2\Big)^{-2s_2}\n\\
    &\times \Gamma\Big[p_1+\fr52-2s_1,p_2+\fr52-2s_2,s_1-\fr{\ii\wt\nu}2,s_1+\fr{\ii\wt\nu}2,s_2-\fr{\ii\wt\nu}2,s_2+\fr{\ii\wt\nu}2\Big],
\label{eq_scaIpmpmSnMB}
\end{align} 
The time-ordered integral can also be carried out with the aid of (\ref{eq_TOintFormula3}): 
\begin{align}
    {\mathcal I}_{\pm\pm,\text{TO},>}^{p_1p_2} 
    =&~ \FR{1}{4\pi }e^{\mp\ii\pi(p_1+p_2)/2}r_1^{5+p_1+p_2}
    \int_{-\ii\infty}^{\ii\infty}\FR{\di s_1}{2\pi\ii}\FR{\di s_2}{2\pi\ii}\,(\mp \ii e^{\pm 2\ii\pi s_1} \pm \ii e^{\pm 2\ii\pi s_2})
    \Big(\FR{r_1}2\Big)^{-2s_{12}}\n\\
    &\times \Gamma\Big[p_2+\fr52-2s_2,p_1+p_2+5-2s_{12},s_1-\fr{\ii\wt\nu}2,s_1+\fr{\ii\wt\nu}2,s_2-\fr{\ii\wt\nu}2,s_2+\fr{\ii\wt\nu}2\Big]\n\\
    &\times {}_2\wt{\mathrm{F}}_1\left[\bgm p_2+\fr52-2s_2,p_1+p_2+5-2s_{12}\\p_2+\fr72-2s_2\edm\middle|\,-\FR{r_1}{r_2}\right].
\label{eq_scaIpmpmBgMB}
\end{align} 
Here ${}_2\wt{\mathrm{F}}_1$ is the regularized hypergeometric function. See (\ref{eq_HyperGeoReg}) for the definition.

It turns out that the above results obtained with the assumption of $k_{12}>k_{34}$ are also valid when $k_{12}<k_{34}$. We suspect the reason is that our result for the background has resummed all powers of $k_{34}/k_{12}$ into a well behaved function which is analytic at $k_{12}=k_{34}$, so that the series of $r_1/r_2$ originally defined in the range of $k_{12}>k_{34}$ is analytically continued to the opposite side $k_{12}<k_{34}$. However, we also observe that the convergence of the $r_1$ series in (\ref{eq_scaIpmpmBgMB}) becomes slow as $r_1\to 1$. 

Alternatively, we can also perform the calculation again with the assumption $k_{12}<k_{34}$ from the very beginning. For instance, we can also define the factorized integral $\mathcal{I}_{\pm\pm,\text{F},<}^{p_1p_2}(r_1,r_2)$ and the time-ordered integral $\mathcal{I}_{\pm\pm,\text{TO},<}^{p_1p_2}(r_1,r_2)$. The results can be simply obtained by switching the variables $(r_1,p_1)\leftrightarrow (r_2,p_2)$. In practice, it turns out that the integrals expressed in the following way would have fast convergence speed for all physical range of parameters:
\begin{align}
  \mathcal{I}_{\pm\pm}^{p_1p_2}(r_1,r_2)
  =&~\mathcal{I}_{\pm\pm,\text{F}}^{p_1p_2}(r_1,r_2)+\mathcal{I}_{\pm\pm,\text{TO}}^{p_1p_2}(r_1,r_2),\\
  \mathcal{I}_{\pm\pm,\text{F}}^{p_1p_2}(r_1,r_2)
  =&~\mathcal{I}_{\pm\pm,\text{F},>}^{p_1p_2}(r_1,r_2)\theta(r_2-r_1)+\mathcal{I}_{\pm\pm,\text{F},>}^{{p_2p_1}}(r_2,r_1)\theta(r_1-r_2),\\
  \mathcal{I}_{\pm\pm,\text{TO}}^{p_1p_2}(r_1,r_2)
  =&~\mathcal{I}_{\pm\pm,\text{TO},>}^{p_1p_2}(r_1,r_2)\theta(r_2-r_1)+\mathcal{I}_{\pm\pm,\text{TO},>}^{{p_2p_1}}(r_2,r_1)\theta(r_1-r_2).
\end{align}

\paragraph{Signal and background.}
We are now ready to carry out the MB integrals in \eqref{eq_scaIpmmpMB}, \eqref{eq_scaIpmpmSnMB} and \eqref{eq_scaIpmpmBgMB}, by closing the contours and picking up the enclosed poles. Again let us focus on the case of $k_{12}>k_{34}$. In this region we have $0<r_1<r_2<1$, which shows that we should close the contour from left for each of the integrals. Then, all the enclosed poles are from the $\Gamma$ functions of the form $\Gamma(s_{1,2}\pm\ii\wt\nu/2)$, which are called ``left poles.'' These poles can be divided into two sets: 
\begin{align}
\label{eq_ScalarPole1}
    s_1 = -n_1 \mp \FR{\ii\wt\nu}2,\quad s_2 = -n_2 \pm \FR{\ii\wt\nu}2,\\
\label{eq_ScalarPole2}
    s_1 = -n_1 \mp \FR{\ii\wt\nu}2,\quad s_2 = -n_2 \mp \FR{\ii\wt\nu}2.
\end{align} 
Then using the residue theorem, we can complete the integrals.
The integrals \eqref{eq_scaIpmmpMB} and \eqref{eq_scaIpmpmSnMB} can be combined together:
\begin{align}
    {\mathcal I}_{\text{S},>}^{p_1p_2}(r_1,r_2) \equiv&~\mathcal{I}_{+-}^{p_1p_2}(r_1,r_2)+\mathcal{I}_{++,\text{F},>}^{p_1p_2}(r_1,r_2)+\text{c.c.}.
\end{align}
More explicitly,
\begin{align}
    &~{\mathcal I}_{\text{S},>}^{p_1p_2}(r_1,r_2)\n\\
    =&~ \FR{1}{4\pi} r_1^{5/2+p_1}r_2^{5/2+p_2}\int_{-\ii\infty}^{\ii\infty}\FR{\di s_1}{2\pi\ii}\FR{\di s_2}{2\pi\ii}\,
    \Big[e^{-\ii\pi(p_1-p_2)/2}+\ii e^{-\ii\pi(p_1+p_2)/2} e^{2\ii\pi s_1}\Big]
    \Big(\FR{r_1}2\Big)^{-2s_1}\Big(\FR{r_2}2\Big)^{-2s_2}\n\\
    &\times \Gamma\Big[p_1+\fr52-2s_1,p_2+\fr52-2s_2,s_1-\fr{\ii\wt\nu}2,s_1+\fr{\ii\wt\nu}2,s_2-\fr{\ii\wt\nu}2,s_2+\fr{\ii\wt\nu}2\Big]+\text{c.c.}.
\end{align} 
The result of this integral is obtained by summing over the residues of all the poles listed in (\ref{eq_ScalarPole1}) and (\ref{eq_ScalarPole2}). It is clear that the first set of poles (\ref{eq_ScalarPole1}) contribute to a result proportional to $(r_1/r_2)^{\pm\ii\wt\nu}$, which is the local (L) signal, while the second set (\ref{eq_ScalarPole2}) leads to the nonlocal (NL) signal of the form $(r_1r_2)^{\pm\ii\wt\nu}$. For positive real $\wt\nu$, it is clear that the local and nonlocal signals exhibit oscillatory behavior in $\log(r_1/r_2)$ and $\log(r_1r_2)$, respectively, and this is why they are called ``signals.'' For phenomenological applications, it would be useful to spell out these signals more clearly. Therefore, we make the following rearrangement:

\bge
  {\mathcal I}_{\text{S},>}^{p_1p_2}(r_1,r_2)={\mathcal I}_{\text{L},>}^{p_1p_2}(r_1,r_2)+{\mathcal I}_{\text{NL},>}^{p_1p_2}(r_1,r_2),
\ede
in which the local-signal integral is: 
\begin{align}
    &~{\mathcal I}_{\text{L},>}^{p_1p_2}(r_1,r_2) \n\\
    =& ~\FR{1}{4\pi} r_1^{5/2+p_1}r_2^{5/2+p_2}
    \sum_{n_1,n_2=0}^\infty\bigg\{\FR{(-1)^{n_{12}}}{n_1!n_2!}
    \Big[e^{-\ii\pi(p_1-p_2)/2}+\ii e^{-\ii\pi(p_1+p_2)/2} e^{\pi\wt\nu}\Big]
    \Big(\FR{r_1}{2}\Big)^{2n_1+\ii\wt\nu}\Big(\FR{r_1}{2}\Big)^{2n_2-\ii\wt\nu}\n\\
    &\times \Gamma\Big[2n_1+p_1+\fr52+\ii\wt\nu,2n_2+p_2+\fr52-\ii\wt\nu,-n_1-\ii\wt\nu,-n_2+\ii\wt\nu\Big]+(\wt\nu\to-\wt\nu)\bigg\}+\text{c.c.}\n\\
    =&~\mathcal{C}_{\wt\nu}^{p_1p_2}\Big(\FR{r_1}{r_2}\Big)^{\ii\wt\nu} 
    \mb F_{\wt\nu}^{p_1}(r_1) \mb F_{-\wt\nu}^{p_2}(r_2) + \text{c.c.}.
    \label{eq_scaLSn}
\end{align} 
Here we have defined 
\begin{equation}
\label{eq_Fbold}
    \mb F_{\wt\nu}^p(r)\equiv r^{5/2+p}\Gamma\Big[\FR52+p+\ii\wt\nu,-\ii\wt\nu\Big]
    {}_2\mathrm{F}_1\left[\bgm \fr54+\fr{p}2+\fr{\ii\wt\nu}2,\fr74+\fr{p}2+\fr{\ii\wt\nu}2\\1+\ii\wt\nu\edm\middle|\,r^2\right],
\end{equation} 
and
\bge
\label{eq_Ccoeff}
  \mathcal{C}_{\wt\nu}^{p_1p_2}\equiv\FR{1}{2\pi}\Big[\cos\big(\fr{\pi}{2}(p_1-p_2)\big)+\sin\big(\fr{\pi}{2}(p_1+p_2+2\ii\wt\nu)\big)\Big].
\ede
Similarly, the nonlocal signal can be worked out as
\begin{align}
    {\mathcal I}_{\text{NL},>}^{p_1p_2}
    =&~\mathcal{C}_{\wt\nu}^{p_1p_2} \Big(\FR{r_1r_2}4\Big)^{\ii\wt\nu} 
    \mb F_{\wt\nu}^{p_1}(r_1) \mb F_{\wt\nu}^{p_2}(r_2) + \text{c.c.}.
    \label{eq_scaNLSn}
\end{align}
It then remains to compute the time-ordered integral \eqref{eq_scaIpmpmBgMB}. It can be seen that the second set of poles (\ref{eq_ScalarPole2}) do not contribute to the result, so that we can consider the first set of poles (\ref{eq_ScalarPole1}) only. (The regularized hypergeometric function is regular with respect to all the three orders and thus contributes no poles.) The result is thus analytic in $r_1$ and $r_2$ at $r_1=r_2=0$, and corresponds to the ``EFT'' part, or the ``background'' (BG) part, of the correlator. (Again, see \cite{Qin:2022lva} for explanations of the background part.) Therefore, we can write:
\begin{align}
  {\mathcal I}_{\text{BG},>}^{p_1p_2}(r_1,r_2) \equiv {\mathcal I}_{++,\text{TO},>}^{p_1p_2}(r_1,r_2)+{\mathcal I}_{--,\text{TO},>}^{p_1p_2}(r_1,r_2), 
\end{align}
and more explicitly,
\begin{align}
    {\mathcal I}_{\text{BG},>}^{p_1p_2}(r_1,r_2) 
    =&~ \FR{1}{\ii\wt\nu}\cos\big[\fr{\pi}{2}(p_1+p_2)\big]r_1^{5+p_1+p_2}\sum_{n_1,n_2=0}^{\infty}
    \FR{(-1)^{n_{12}}}{n_1!n_2!}\Big(\FR {r_1}2 \Big)^{2n_{12}}(-\ii\wt\nu)_{-n_1}(\ii\wt\nu)_{-n_2}\n\\    &\times {\mathcal{F}}\left[\bgm 2n_2+p_2+\fr52-\ii\wt\nu,2n_{12}+p_1+p_2+5\\2n_2+p_2+\fr72-\ii\wt\nu\edm\middle|\,-\FR{r_1}{r_2}\right]+\text{c.c.},
    \label{eq_scaBg}
\end{align} 
where we have used the Pochhammer symbol $(z)_n\equiv \Gamma(z+n)/\Gamma(z)$ and the ``dressed'' hypergeometric function $\mathcal{F}$. See (\ref{eq_HyperGeoDressed}) for the definition. 
Thus we have finished the computation of the scalar seed integral $\mathcal{I}_{\aa\bb}^{p_1p_2}(r_1,r_2)$ for $r_1<r_2$. The result for $r_1>r_2$ is obtained by switching $(r_1,p_1)\leftrightarrow (r_2,p_2)$ in all the results above. We note that the nonlocal signal \eqref{eq_scaNLSn} is symmetric with respect to $(r_1,p_1)\leftrightarrow (r_2,p_2)$, so that we can simply drop the subscript ``$>$'':
\begin{equation}
	{\mathcal I}_{\text{NL}}^{p_1p_2}(r_1,r_2) = 	{\mathcal I}_{\text{NL},>}^{p_1p_2}(r_1,r_2).
\end{equation}
On the other hand, the local signal \eqref{eq_scaLSn} and the background \eqref{eq_scaBg} are asymmetric. 
 
\subsection{Summary}
\label{sec_scalar_summary}

Now we summarize the result for the scalar seed integral obtained above. As we have seen, the scalar seed integral can be expressed as a function of two independent momentum ratios $r_1=k_s/k_{12}$ and $r_2=k_s/k_{34}$. The following expressions apply to the case of $r_1<r_2$. The results for $r_1>r_2$ can be obtained from the following expressions simply by switching the variables $(r_1,p_1)\leftrightarrow (r_2,p_2)$.

For convenience, we only present the scalar seed integral with all SK indices summed. According to the analytic properties at $r_1=0$ and $r_2=0$, the summed integral can be put into the following three terms, respectively called the background (BG), the local signal (L), and the nonlocal signal (NL):
\begin{align}
\label{eq_scalarSeedSummed}
  \sum_{\aa,\bb=\pm}\mathcal{I}_{\aa\bb}^{p_1p_2}(r_1,r_2)=\mathcal{I}_{\text{BG},>}^{p_1p_2}(r_1,r_2)+\mathcal{I}_{\text{L},>}^{p_1p_2}(r_1,r_2)+\mathcal{I}_\text{NL}^{p_1p_2}(r_1,r_2)
\end{align}
The background piece $\mathcal{I}_{\text{BG},>}^{p_1p_2}$ is analytic in both $r_1$ and $r_2$ as $r_1,r_2\to 0$, and therefore possesses a power-series representation. The local signal $\mathcal{I}_{\text{L},>}^{p_1p_2}$ is nonanalytic in $r_1/r_2$ as the ratio reaches 0, but is analytic in $k_s$ as $k_s\to 0$. Finally, the nonlocal signal $\mathcal{I}_{\text{NL},>}^{p_1p_2}$ is nonanalytic in $r_1r_2$ as $r_1r_2\to 0$, and thus is also nonanalytic in $k_s$ as $k_s\to 0$. The explicit expressions for these integrals are summarized as follows: 
\begin{keyeqn}
\begin{align}
\label{eq_ScalarSeedBG}
  \mathcal{I}_{\text{BG},>}^{p_1p_2}(r_1,r_2)
  =&~\FR{1}{\ii\wt\nu}\cos\big[\fr{\pi}{2}(p_1+p_2)\big]r_1^{5+p_1+p_2}
  \sum_{n_1,n_2=0}^{\infty} \FR{(-1)^{n_{12}}}{n_1!n_2!}\Big(\FR{r_1}{2} \Big)^{2n_{12}}(-\ii\wt\nu)_{-n_1}(\ii\wt\nu)_{-n_2}\n\\    &\times {\mathcal{F}}\left[\bgm 2n_2+p_2+\fr52-\ii\wt\nu,2n_{12}+p_1+p_2+5\\2n_2+p_2+\fr72-\ii\wt\nu\edm\middle|\,-\FR{r_1}{r_2}\right]+\text{c.c.}\\
   \mathcal{I}_{\text{L},>}^{p_1p_2}(r_1,r_2)
   =&~ \mathcal{C}_{\wt\nu}^{p_1p_2}\Big(\FR{r_1}{r_2}\Big)^{\ii\wt\nu} 
    \mb F_{\wt\nu}^{p_1}(r_1) \mb F_{-\wt\nu}^{p_2}(r_2) + \text{c.c.},\\
   \mathcal{I}_{\text{NL}}^{p_1p_2}(r_1,r_2)
   =&~ \mathcal{C}_{\wt\nu}^{p_1p_2} \Big(\FR{r_1r_2}4\Big)^{\ii\wt\nu} 
    \mb F_{\wt\nu}^{p_1}(r_1) \mb F_{\wt\nu}^{p_2}(r_2) + \text{c.c.}.
\end{align}
\end{keyeqn} 
Here $\wt\nu=\sqrt{m^2/H^2-9/4}$ is the mass parameter of the intermediate massive scalar, which we always assume real. $(z)_n$ is the Pochhammer symbol, defined in (\ref{eq_pochhammer}), $\mathcal{F}$ is the dressed hypergeometric function, defined in (\ref{eq_HyperGeoDressed}). The function $\mb{F}_{\wt\nu}^{p}$ is defined in (\ref{eq_Fbold}) and the coefficient $\mathcal{C}_{\wt\nu}^{p_1p_2}$ is defined in (\ref{eq_Ccoeff}).

\section{Helical Spin-1 Exchange}
\label{sec_vec} 
 
In this section we compute the vector seed integral (\ref{eq_vecIab}) using the partial MB representation, which leads to the main result of this paper, namely an exact and analytical expression for the helical 4-point correlators. This section will be rather technical. Readers uninterested in the computation details are invited to go directly to Sec.\ \ref{sec_vec_summary} for a summary of the result. 

The vector seed integral (\ref{eq_vecIab}) is defined for each intermediate polarization, labeled by $\lam=\pm,L$. As one can see from (\ref{eq_BmodePM}) and (\ref{eq_BmodeL}), the two transverse polarizations with $\lam=\pm$ are quite different from the longitudinal polarization with $\lam=L$. Therefore we shall treat these two cases separately, with the transverse integral $\mathcal{I}_{\aa\bb}^{(\pm)p_1p_2}$ computed in Sec.\ \ref{sec_vec_trans} and the longitudinal integral $\mathcal{I}_{\aa\bb}^{(L)p_1p_2}$ in Sec.\ \ref{sec_vec_long}.

\subsection{Transverse integral}
\label{sec_vec_trans}

We now use the partial Mellin transformation to carry out the transverse integral $\mathcal{I}_{\aa\bb}^{({\pm})p_1p_2}$. As one can see from (\ref{eq_Bpm}), for the two transverse components, the helicity basis is in one-to-one relationship with the polarization basis, and therefore we are free to use the helicity label $(h)$ to replace the polarization label $(\lam)$ in various expressions. 

To perform the partial Mellin transformation on the transverse mode function (\ref{eq_BmodePM}), we could use the following Mellin-Barnes representation for the Whittaker function, which we call the completely resolved representation \cite{nist:dlmf}:
\begin{align}
\label{eq_WhitFullMB}
  \Wh_{\ka,\nu}(az)= \int_{-\ii\infty}^{+\ii\infty}\FR{\di s}{2\pi\ii}\, \mathcal{F} \left[\bgm s-\nu,s+\nu \\ s-\ka+\fr12 \edm \middle|\fr{1}{2}\right](az)^{-s+1/2},
\end{align}
where $\mathcal{F}$ is the dressed hypergeometric function defined in (\ref{eq_HyperGeoDressed}). The advantage of this representation is that all $z$-dependences in the Whittaker function have been completely resolved into power functions, which trivializes the time integral. The drawback, however, is that the integrand involves a hypergeometric function. Although this posts no essential difficulty for us to perform the contour integral, the resulting expression has a somewhat more complicated form. We will present the result derived from this completely resolved representation in App.\ \ref{app_AlterMB}. Here, instead, we make use of the fact that the Whittaker function, when multiplied by an appropriate exponential factor, has a far simpler Mellin-Barnes representation. There are two ways to add this exponential factor, which lead to the following two alternative results, which we call partially resolved representations \cite{nist:dlmf}:
\begin{align}
\label{eq_WhitExpPlusMB}
  \Wh_{\ka,\nu}(az)=&~e^{az/2}\int_{-\ii\infty}^{+\ii\infty}\FR{\di s}{2\pi\ii}\,\Gamma\left[\bgm s-\nu,s+\nu \\ s-\ka+\fr12 \edm\right](az)^{-s+1/2}.\\
\label{eq_WhitExpMinusMB}
  \Wh_{\ka,\nu}(az)=&~e^{-az/2}\int_{-\ii\infty}^{+\ii\infty}\FR{\di s}{2\pi\ii}\,\Gamma\left[\bgm s-\nu,s+\nu,-s-\ka+\fr12 \\ \fr{1}{2}-\ka-\nu,\fr{1}{2}-\ka+\nu \edm\right](az)^{-s+1/2}.
\end{align}
We have the freedom to use either of them when computing the seed integral so long as all parameters stay in the validity ranges of the above two representations. (See \cite{nist:dlmf} for detailed specification of validity ranges of all MB representations quoted here.)

\paragraph{Opposite-sign integrals.} First consider the opposite-sign integrals $\mathcal{I}_{\pm\mp}^{(h)p_1p_2}$: 
\begin{equation} 
\label{eq_vecIpm}
     \mathcal{I}^{(h)p_1p_2}_{\pm\mp}(r_1,r_2) = k_s^{3+p_1+p_2} \int_{-\infty}^0 \di\tau_1\di\tau_2\,
     (-\tau_1)^{p_1}(-\tau_2)^{p_2}e^{\pm\ii k_{12}\tau_1\mp\ii k_{34}\tau_2}D_{\pm\mp}^{(h)}(k_s;\tau_1,\tau_2).
\end{equation} 

The two time integrals in these expressions are already factorized and so the integration is straightforward. For a reason that will be clear soon, we choose (\ref{eq_WhitExpMinusMB}) for the both Whittaker functions in $D_{\pm\mp}^{(h)}$:
\begin{align}
D_{\pm\mp}^{(h)}(k;\tau_1,\tau_2)=&~
\FR{e^{-h\pi\wt\mu}}{2\pi^2}(\cosh2\pi\wt\mu+\cosh2\pi\wt\nu)e^{\pm\ii k(\tau_1-\tau_2)}\n\\
&\times\int_{-\ii\infty}^{\ii\infty} \FR{\di s_1}{2\pi\ii}\FR{\di s_2}{2\pi\ii}\,
e^{\mp\ii\pi(s_1-s_2)/2}(2k_s)^{-s_{12}}(-\tau_1)^{-s_1+1/2}(-\tau_2)^{-s_2+1/2}\n\\
&\times \Gamma\Big[-s_1+\fr12\pm \ii h\wt\mu,-s_2+\fr12\mp \ii h\wt\mu,s_1-\ii\wt\nu,s_1+\ii\wt\nu,s_2-\ii\wt\nu,s_2+\ii\wt\nu\Big].
\end{align}
Inserting the MB representation into \eqref{eq_vecIpm} and completing the time integrals, we obtain: 
\begin{align}
\label{eq_IpmmpInter1}
    {\mathcal I}_{\pm\mp}^{(h)p_1p_2}=&~\FR{e^{-h\pi\wt\mu}}{2\pi^2}(\cosh2\pi\wt\mu+\cosh2\pi\wt\nu)
    \big(\FR{u_1}2\big)^{3/2+p_1}\big(\FR{u_2}2\big)^{3/2+p_2}e^{\mp\ii(p_1-p_2)\pi/2}\n\\
    &\times\int_{-\ii\infty}^{\ii\infty}\FR{\di s_1}{2\pi\ii}\FR{\di s_2}{2\pi\ii}\,
    u_1^{-s_1}u_2^{-s_2}\Gamma\big[p_1+\fr32-s_1,p_2+\fr32-s_2\big]\n\\ 
    &\times \Gamma\big[-s_1+\fr12\pm\ii h\wt\mu,-s_2+\fr12\mp\ii h\wt\mu,s_1-\ii\wt\nu,s_1+\ii\wt\nu,s_2-\ii\wt\nu,s_2+\ii\wt\nu\big],
\end{align} 
where we have used the previously defined momentum ratios $u_1=2r_1/(1+r_1)$ and $u_2=2r_2/(1+r_2)$.

At this point we can explain why we choose (\ref{eq_WhitExpMinusMB}) to represent the propagator: Had we chosen the other representation (\ref{eq_WhitExpPlusMB}), the integral (\ref{eq_IpmmpInter1}), and also all the subsequent results, would be expressed in terms of $w_1=2r_{1}/(1-r_1)$ and $w_2=2r_{2}/(1-r_2)$ instead of $u_{1,2}$. As we shall see below, the result of the vector seed integral involves the hypergeometric function ${}_2\mathrm{F}_1$ of the momentum ratio. We know that ${}_2\mathrm{F}_1(a,b;c;z)$ for generic orders $(a,b,c)$ is singular at $z=1$. Thus, had we used (\ref{eq_WhitExpPlusMB}) which gives rise to results in terms of $w_{1,2}$, we would encounter spurious divergences at $w_{1,2}=1$, namely $r_{1,2}=1/3$, a configuration of no special physical significance. Of course this superficial divergence must get canceled in the final result, but its presence would make the final expression rather ugly, and also make the numerical implementation hard. Therefore, here and below, we will always make choices from (\ref{eq_WhitExpPlusMB}) and (\ref{eq_WhitExpMinusMB}) such that we always get $u_{1,2}$ instead of $w_{1,2}$ in the final expressions.\footnote{Of course, our choice to represent the result in terms of $u_{1,2}$ does not remove the singularity of the hypergeometric function at $u_{1,2}=1$. However, $u_{1,2}=1$, namely $r_{1,2}=1$, correspond to the folded limit of the momentum configuration, which does have a physical significance. Requiring no divergences at the folded limit corresponds to choosing the Bunch-Davies vacuum as the initial states for all the quantum fields.}

Given $0<u_{1,2}<1$, we should close the contours of MB integrals in (\ref{eq_IpmmpInter1}) from left with a large semicircle. Applying the residue theorem, we pick up the following left poles of the integrand:
\begin{align}
\label{eq_VecPole1}
    s_1 = -n_1 \mp \ii\wt\nu,\quad s_2 = -n_2 \pm \ii\wt\nu,\\
\label{eq_VecPole2}
    s_1 = -n_1 \mp \ii\wt\nu,\quad s_2 = -n_2 \mp \ii\wt\nu.
\end{align}
The result for the MB integral is thus obtained by summing over the residues of all the above left poles: 
\begin{align}
    & {\mathcal I}_{\pm\mp}^{(h)p_1p_2}
    =\FR{e^{-h\pi\wt\mu}}{2\pi^2}(\cosh2\pi\wt\mu+\cosh2\pi\wt\nu)
    \big(\FR{u_1}2\big)^{3/2+p_1}\big(\FR{u_2}2\big)^{3/2+p_2}e^{\mp\ii(p_1-p_2)\pi/2}\n\\ 
    &\times \sum_{n_1=0}^\infty\bigg\{\FR{(-1)^{n_1}}{n_1!} u_1^{\ii\wt\nu+n_1}\Gamma\big[n_1+p_1+\fr32+\ii\wt\nu,n_1+\ii\wt\nu+\fr12\pm\ii h\wt\mu, -n_1-2\ii\wt\nu\big]+(\wt\nu\to-\wt\nu)\bigg\}\n\\
    &\times\sum_{n_2=0}^\infty\bigg\{\FR{(-1)^{n_2}}{n_2!}u_2^{\ii\wt\nu+n_2}\Gamma\big[n_2+p_2+\fr32+\ii\wt\nu,n_2+\ii\wt\nu+\fr12\mp\ii h\wt\mu, -n_2-2\ii\wt\nu\big]+(\wt\nu\to-\wt\nu)\bigg\}.
\end{align} 
The summation can be done in closed form, and the result can be expressed as:
\begin{align}
    {\mathcal I}_{\pm\mp}^{(h)p_1p_2} 
    =&~\FR{e^{-h\pi\wt\mu}}{2\pi^2}(\cosh2\pi\wt\mu+\cosh2\pi\wt\nu)e^{\mp\ii(p_1-p_2)\pi/2}\n\\ 
    &\times\Big[\mb{G}^{p_1}_{\pm h\wt\mu,\wt\nu}(u_1)u_1^{\ii\wt\nu}+\mb{G}^{p_1}_{\pm h\wt\mu,-\wt\nu}(u_1)u_1^{-\ii\wt\nu} \Big]
    \Big[\mb{G}^{p_2}_{\mp h\wt\mu,\wt\nu}(u_2)u_2^{\ii\wt\nu}+\mb{G}^{p_2}_{\mp h\wt\mu,-\wt\nu}(u_2)u_2^{-\ii\wt\nu} \Big],
\end{align}
where we have defined 
\begin{align}
\label{eq_Gbold}
  \mb{G}^p_{h\wt\mu,\wt\nu}(u)\equiv \ii\pi\text{csch}(2\pi\wt\nu)\Big(\FR{u}2\Big)^{3/2+p}\mathcal{F}\left[\bgm \fr{3}{2}+p+\ii\wt\nu,\fr{1}{2}+\ii h \wt\mu+\ii\wt\nu\\ 1+2\ii\wt\nu\edm\middle|u\right].
\end{align}

\paragraph{Same-sign integrals: factorized part.} 

We focus on the double-plus integral $\mathcal{I}^{(h)p_1p_2}_{++}$, and still considering the case of $k_{12}>k_{34}$. The result for $k_{12}<k_{34}$ can be obtained by switching the variables $(u_1,p_1)\leftrightarrow (u_2,p_2)$ as before, and the result for $\mathcal{I}^{(h)p_1p_2}_{--}$ can be obtained by taking complex conjugation. 

Similar to the case of the scalar integral discussed in the previous section, the same-sign integral can be separated into a factorized part and a time-ordered part:
\bge
  \mathcal{I}_{\pm\pm}^{(h)p_1p_2}(r_1,r_2)=\mathcal{I}_{\pm\pm,\text{F},>}^{(h)p_1p_2}(r_1,r_2)+\mathcal{I}_{\pm\pm,\text{TO},>}^{(h)p_1p_2}(r_1,r_2).~~~~(r_1<r_2)
\ede

The factorized integral is given by: 
\begin{equation} 
\label{eq_vecIppF}
     \mathcal{I}^{(h)p_1p_2}_{++,\text{F},>}(r_1,r_2) = -k_{s}^{3+p_1+p_2} \int_{-\infty}^0 \di\tau_1\di\tau_2\,
     (-\tau_1)^{p_1}(-\tau_2)^{p_2}e^{+\ii k_{12}\tau_1+\ii k_{34}\tau_2}D_{>}^{(h)}(k_s;\tau_1,\tau_2),
\end{equation} 
The computation of this integral is much similar to the opposite-sign integral, except that we should now use a different MB representation for the Wightman function $D_>$, since the exponential factor in \eqref{eq_vecIppF} is $e^{+\ii k_{12} \tau_1 +\ii k_{34}\tau_2}$:
\begin{align}
D_{>}^{(h)}(k;\tau_1,\tau_2)=&~
\FR{e^{-h\pi\wt\mu}}{\pi\Gamma[\fr12+\ii\wt\nu+\ii h\wt\mu,\fr12-\ii\wt\nu+\ii h\wt\mu]}e^{+\ii k(\tau_1+\tau_2)}\n\\
&\times\int_{-\ii\infty}^{\ii\infty} \FR{\di s_1}{2\pi\ii}\FR{\di s_2}{2\pi\ii}\,
e^{\ii\pi(s_1-s_2)/2}\cos\pi(s_1-\ii h\wt\mu)(2k_s)^{-s_{12}}(-\tau_1)^{-s_1+1/2}(-\tau_2)^{-s_2+1/2}\n\\
&\times \Gamma\Big[-s_1+\FR12+ \ii h\wt\mu,-s_2+\FR12+ \ii h\wt\mu,s_1-\ii\wt\nu,s_1+\ii\wt\nu,s_2-\ii\wt\nu,s_2+\ii\wt\nu\Big].
\end{align}
Substituting this representation into (\ref{eq_vecIppF}) and carrying out the time integral, we get: 
\begin{align}
    {\mathcal I}^{(h)p_1p_2}_{++,\text{F},>}=&-\FR{e^{-h\pi\wt\mu}}{\pi\Gamma\big[\fr12+\ii\wt\nu+\ii h\wt\mu,\fr12-\ii\wt\nu+\ii h\wt\mu\big]}
    \big(\FR{u_1}2\big)^{3/2+p_1}\big(\FR{u_2}2\big)^{3/2+p_2}    e^{-\ii\pi(p_1+p_2)/2}\n\\
    &\times
    \int_{-\ii\infty}^{\ii\infty}\FR{\di s_1}{2\pi\ii}\FR{\di s_2}{2\pi\ii}\, e^{+\ii\pi (s_1+1/2)}\cos\pi(s_1-\ii h\wt\mu)u_1^{-s_1}u_2^{-s_2}
    \Gamma\big[p_1+\fr32-s_1,p_2+\fr32-s_2\big]\n\\
    &\times \Gamma\big[-s_1+\fr12+\ii h\wt\mu,-s_2+\fr12+\ii h\wt\mu,s_1-\ii\wt\nu,s_1+\ii\wt\nu,s_2-\ii\wt\nu,s_2+\ii\wt\nu\big].
\end{align} 
Then, following the same procedure as before, we close the contour from the left, pick up all enclosed (left) poles, namely \eqref{eq_VecPole1} and \eqref{eq_VecPole2}, and sum over the residues at these poles, to get the final result:
\begin{align}
    {\mathcal I}^{(h)p_1p_2}_{++,\text{F},>}
    =&~\FR{-\ii e^{-h\pi\wt\mu}}{\pi\Gamma\big[\fr12+\ii\wt\nu+\ii h\wt\mu,\fr12-\ii\wt\nu+\ii h\wt\mu\big]}
   e^{-\ii\pi(p_1+p_2)/2}\n\\
    &\times\Big[ e^{+\pi\wt\nu}\cosh[\pi(\nu+ h\wt\mu)] \mb{G}^{p_1}_{h\wt\mu,\wt\nu}(u_1)u_1^{\ii\wt\nu}+e^{-\pi\wt\nu}\cosh[\pi(-\nu+ h\wt\mu)] \mb{G}^{p_1}_{h\wt\mu,-\wt\nu}(u_1)u_1^{-\ii\wt\nu}\Big]\n\\ 
    &\times\Big[\mb{G}^{p_2}_{h\wt\mu,\wt\nu}(u_2)u_2^{\ii\wt\nu}+\mb{G}^{p_2}_{h\wt\mu,-\wt\nu}(u_2)u_2^{-\ii\wt\nu}\Big],
\end{align}
where $\mb{G}_{h\wt\mu,\wt\nu}^{p}(u)$ is the same function as is defined in (\ref{eq_Gbold}).

\paragraph{Local and nonlocal signals.}
It is not surprising that the factorized part of the same-sign integral has very similar structure as the opposite-sign integrals. In particular, both results are nonanalytic in $u_1$ and $u_2$ as $u_{1,2}\to 0$. Similar to the case of scalar integral, these two parts contribute to the local and nonlocal signals: 
\begin{align}
  \mathcal{I}_{++,\text{F},>}^{(h)p_1p_2}+\mathcal{I}_{--,\text{F},>}^{(h)p_1p_2}+\mathcal{I}_{+-}^{(h)p_1p_2}+\mathcal{I}_{-+}^{(h)p_1p_2}=\mathcal{I}_{\text{L},>}^{(h)p_1p_2} +\mathcal{I}_\text{NL}^{(h)p_1p_2} ,
\end{align}
with the local signal $\mathcal{I}_{\text{L},>}^{(h)p_1p_2}$ and the nonlocal signal $\mathcal{I}_\text{NL}^{(h)p_1p_2}$ given by:
\begin{align}
   \mathcal{I}_{\text{L},>}^{(h)p_1p_2}(u_1,u_2) 
    =&~\mathcal G^{(h)p_1p_2}_{\text{L}}(u_1,u_2)\Big(\FR{u_1}{u_2}\Big)^{\ii\wt\nu}+\text{c.c.}\n\\
   \mathcal{I}_{\text{NL}}^{(h)p_1p_2}(u_1,u_2) 
    =&~\mathcal G^{(h)p_1p_2}_{\text{NL}}(u_1,u_2)(u_1u_2)^{\ii\wt\nu}+\text{c.c.}
\end{align}
where we have defined:
\begin{align}
    \mathcal{G}_{\text{L}}^{(h)p_1p_2}(u_1,u_2)
  =&~\FR{e^{-\pi h\wt\mu}}{2\pi^2}
  \bigg\{\big(\cosh2\pi\wt\mu+\cosh2\pi\wt\nu\big) 
  \Big[e^{-\ii(p_1-p_2)\pi/2}\mb{G}^{p_1}_{h\wt\mu,\wt\nu}(u_1)\mb{G}^{p_2}_{-h\wt\mu,-\wt\nu}(u_2)\n\\
  &+ e^{+\ii(p_1-p_2)\pi/2}\mb{G}^{p_1}_{-h\wt\mu,\wt\nu}(u_1)\mb{G}^{p_2}_{h\wt\mu,-\wt\nu}(u_2) \Big] \n\\ 
  &~+\FR{-2\ii\pi e^{\pi\wt\nu}\cosh[\pi(h\wt\mu+\wt\nu)]}{\Gamma[\fr{1}{2}+\ii h\wt\mu+\ii\wt\nu,\fr{1}{2}+\ii h\wt\mu-\ii\wt\nu]}e^{-\ii(p_1+p_2)\pi/2}\mb{G}^{p_1}_{h\wt\mu,\wt\nu}(u_1)\mb{G}^{p_2}_{h\wt\mu,-\wt\nu}(u_2)\n\\
  &~+\FR{2\ii\pi e^{-\pi\wt\nu}\cosh[\pi(h\wt\mu-\wt\nu)]}{\Gamma[\fr{1}{2}-\ii h\wt\mu-\ii\wt\nu,\fr{1}{2}-\ii h\wt\mu+\ii\wt\nu]}e^{+\ii(p_1+p_2)\pi/2}\mb{G}^{p_1}_{-h\wt\mu,\wt\nu}(u_1)\mb{G}^{p_2}_{-h\wt\mu,-\wt\nu}(u_2) 
   \bigg\},
\end{align}
\begin{align}
  \mathcal{G}_\text{NL}^{(h)p_1p_2}(u_1,u_2)
  =&~\FR{e^{-\pi h\wt\mu}}{2\pi^2}  \bigg\{ 
  \big(\cosh2\pi\wt\mu+\cosh2\pi\wt\nu\big) 
  \Big[e^{-\ii(p_1-p_2)\pi/2}\mb{G}^{p_1}_{h\wt\mu,\wt\nu}(u_1)\mb{G}^{p_2}_{-h\wt\mu,\wt\nu}(u_2)\n\\
  &+ e^{+\ii(p_1-p_2)\pi/2}\mb{G}^{p_1}_{-h\wt\mu,\wt\nu}(u_1)\mb{G}^{p_2}_{h\wt\mu,\wt\nu}(u_2) \Big] \n\\
  &~+\FR{-2\ii\pi e^{\pi\wt\nu}\cosh[\pi(h\wt\mu+\wt\nu)]}{\Gamma[\fr{1}{2}+\ii h\wt\mu-\ii\wt\nu,\fr{1}{2}+\ii h\wt\mu+\ii\wt\nu]}e^{-\ii(p_1+p_2)\pi/2}\mb{G}^{p_1}_{h\wt\mu,\wt\nu}(u_1)\mb{G}^{p_2}_{h\wt\mu,\wt\nu}(u_2)\n\\
  &~+\FR{2\ii\pi e^{-\pi\wt\nu}\cosh[\pi(h\wt\mu-\wt\nu)]}{\Gamma[\fr{1}{2}-\ii h\wt\mu-\ii\wt\nu,\fr{1}{2}-\ii h\wt\mu+\ii\wt\nu]}e^{+\ii(p_1+p_2)\pi/2}\mb{G}^{p_1}_{-h\wt\mu,\wt\nu}(u_1)\mb{G}^{p_2}_{-h\wt\mu,\wt\nu}(u_2) 
   \bigg\}.
\end{align}
Note that the nonlocal signal is symmetric with respect to $(u_1,p_1)\leftrightarrow (u_2,p_2)$ , and thus the result applies well to the other side $k_{12}<k_{34}$. So, we do not put a subscript $>$ for the nonlocal signal.

\paragraph{Same-sign integrals: time-ordered part.} 

Now we come to the time-ordered part: 
\begin{align}
\label{eq_vecIppTO}
    \mathcal{I}_{++,\text{TO},>}^{p_1p_2}(r_1,r_2)
  \equiv & - k_{s}^{3+p_1+p_2}\int_{-\infty}^{0} \di\tau_2\int_{-\infty}^{\tau_2}\di\tau_1\,(-\tau_1)^{p_1}(-\tau_2)^{p_2} e^{\pm \ii(k_{12}\tau_1+k_{34}\tau_2)}\n\\
  &~\times\Big[D_<(k_s;\tau_1,\tau_2)-D_>(k_s;\tau_1,\tau_2)\Big].
\end{align} 
We use the following MB representations for the Wightman functions $D_>$ and $D_<$ to ensure that all results are expressed in terms of $u_{1,2}$:
\begin{align}
D_{>}^{(h)}(k;\tau_1,\tau_2)=&~
\FR{e^{-h\pi\wt\mu}}{\pi\Gamma[\fr12+\ii\wt\nu+\ii h\wt\mu,\fr12-\ii\wt\nu+\ii h\wt\mu]}e^{+\ii k(\tau_1+\tau_2)}\n\\
&\times\int_{-\ii\infty}^{\ii\infty} \FR{\di s_1}{2\pi\ii}\FR{\di s_2}{2\pi\ii}\,
e^{\ii\pi(s_1-s_2)/2}\cos\pi(s_1-\ii h\wt\mu)(2k)^{-s_{12}}(-\tau_1)^{-s_1+1/2}(-\tau_2)^{-s_2+1/2}\n\\
&\times \Gamma\Big[-s_1+\FR12+ \ii h\wt\mu,-s_2+\FR12+ \ii h\wt\mu,s_1-\ii\wt\nu,s_1+\ii\wt\nu,s_2-\ii\wt\nu,s_2+\ii\wt\nu\Big], 
\end{align} 
\begin{align}
D_{<}^{(h)}(k;\tau_1,\tau_2)=&~
\FR{e^{-h\pi\wt\mu}}{\pi\Gamma[\fr12+\ii\wt\nu+\ii h\wt\mu,\fr12-\ii\wt\nu+\ii h\wt\mu]}e^{+\ii k(\tau_1+\tau_2)}\n\\
&\times\int_{-\ii\infty}^{\ii\infty} \FR{\di s_1}{2\pi\ii}\FR{\di s_2}{2\pi\ii}\,
e^{-\ii\pi(s_1-s_2)/2}\cos\pi(s_2-\ii h\wt\mu)(2k)^{-s_{12}}(-\tau_1)^{-s_1+1/2}(-\tau_2)^{-s_2+1/2}\n\\
&\times \Gamma\Big[-s_1+\FR12+ \ii h\wt\mu,-s_2+\FR12+ \ii h\wt\mu,s_1-\ii\wt\nu,s_1+\ii\wt\nu,s_2-\ii\wt\nu,s_2+\ii\wt\nu\Big].
\end{align} 
Substituting these MB representations into the time-ordered integral (\ref{eq_vecIppTO}), we can finish the time integral to get the following result: 
\begin{align}
    {\mathcal I}^{(h)p_1p_2}_{++,\text{TO},>}
    =&~\FR{-\ii e^{-h\pi\wt\mu}}{\pi\Gamma[\fr12+\ii\wt\nu+\ii h\wt\mu,\fr12-\ii\wt\nu+\ii h\wt\mu]}
    \FR{e^{-\ii\pi(p_1+p_2)/2}}{2^{3+p_1+p_2}}
    \int_{-\ii\infty}^{\ii\infty} \FR{\di s_1}{2\pi\ii}\FR{\di s_2}{2\pi\ii} \n\\
    &\times  \Big[ e^{\ii\pi s_2}\cos\pi(s_2-\ii h\wt\mu)- e^{\ii\pi s_1} \cos\pi(s_1-\ii h\wt\mu)\Big]u_1^{-s_{12}+3+p_1+p_2}\n\\ 
    &\times \Gamma\Big[-s_1+\fr12+ \ii h\wt\mu,-s_2+\fr12+ \ii h\wt\mu,s_1-\ii\wt\nu,s_1+\ii\wt\nu,s_2-\ii\wt\nu,s_2+\ii\wt\nu\Big]\n\\ 
    &\times\mathcal F \left[\bgm p_2+\fr32-s_2,p_1+p_2+3-s_{12}\\ p_2+\fr52-s_2 \edm\middle|-\FR{u_1}{u_2}\right],
\end{align} 
where $\mathcal{F}$ is the dressed hypergeometric function defined in (\ref{eq_HyperGeoDressed}). The function $\mathcal{F}$ does possess poles in $s_{1}$ and $s_2$ but they are right poles, and thus do not contribute to the result, since we should close the contour from the left. (Notice that the contribution from the poles in \eqref{eq_VecPole2} vanishes.) The result of the time-ordered integral is fully analytic in $u_1$ and $u_2$ at $u_{1,2}=0$ (when $p_1+p_2$ are integers no smaller than $-3$), as opposed to the opposite-sign integral and the factorized same-sign integral. Thus the time-ordered integral exactly gives rise to the background piece of correlator. Let us write
\bge
  {\mathcal I}^{(h)p_1p_2}_{\text{BG},>}={\mathcal I}^{(h)p_1p_2}_{++,\text{TO},>}+{\mathcal I}^{(h)p_1p_2}_{--,\text{TO},>}={\mathcal I}^{(h)p_1p_2}_{++,\text{TO},>}+\text{c.c.}.
\ede
Then the result for the background integral can be directly obtained as:
\begin{align}
     {\mathcal I}^{(h)p_1p_2}_{\text{BG},>}  
  =&~
 \FR{e^{-\ii\pi(p_1+p_2)/2}}{2^{3+p_1+p_2}}\sum_{n_1,n_2=0}^\infty \FR{(-1)^{n_{12}}}{n_1!n_2!} u_1^{n_{12}+3+p_1+p_2}\n\\
 &\times
     \bigg\{
   \FR{\ii}{2\wt\nu} 
     \big(\fr{1}{2}+\ii\wt\nu+\ii h\wt\mu \big)_{n_1}\big(\fr{1}{2}-\ii\wt\nu+\ii h\wt\mu \big)_{n_2}\big(-2\ii\wt\nu\big)_{-n_1}\big(+2\ii\wt\nu\big)_{-n_2}\n\\
    &\times  {\mathcal{F}}\left[\bgm n_2+p_2+\fr32-\ii\wt\nu,n_{12}+p_1+p_2+3\\n_2+p_2+\fr52-\ii\wt\nu\edm\middle|\,-\FR{u_1}{u_2}\right] 
     +(\wt\nu\to-\wt\nu)\bigg\}+\text{c.c.}.
\end{align}
 
\subsection{Longitudinal integral}
\label{sec_vec_long}

Next we consider the longitudinal component ($\lam=L$) of the vector seed integral (\ref{eq_vecIab}). The longitudinal polarization is independent of the chemical potential, and the mode function is quite similar to that of a massive scalar field. The computation thus follows exactly the same procedure as in Sec.\ \ref{sec_PMB}, and is also relatively easier than the transverse integral. Therefore we will not spell out all the details, but only present main steps and the final results. 

The longitudinal vector seed integral reads:
\begin{equation}
    {\mathcal I}^{(L)p_1p_2}_{\mathsf{ab}}(r_1,r_2) = -\mathsf{ab} 
    k_{s}^{3+p_1+p_2} \int_{-\infty}^0 \di\tau_1\di\tau_2\,
    (-\tau_1)^{p_1}(-\tau_2)^{p_2}e^{\ii\mathsf a k_{12}\tau_1+\ii\mathsf b k_{34}\tau_2}D_{\mathsf{ab}}^{(L)}(k_s;\tau_1,\tau_2),
\end{equation}
Comparing (\ref{eq_VecPropT}), (\ref{eq_VecPropL}), and (\ref{eq_ScalarDGreater}), we see that the longitudinal propagator $D^{(L)}_>$ is related to the massive scalar propagator $D_{>}$ of the same mass parameter $\wt\nu$ (but not the same mass) via:
\begin{align}
\label{eq_DLtoD}
    D_{>}^{(L)}(k_s;\tau_1,\tau_2)
    = \FR{1}{\wt\nu^2+\fr{1}{4}}\Big(\pd_{\tau_1}-\FR{2}{\tau_1}\Big)\Big(\pd_{\tau_2}-\FR{2}{\tau_2}\Big)D_{>}(k_s;\tau_1,\tau_2),
\end{align}
where we have explicitly expressed the mass as $m^2=\wt\nu^2+1/4$. Therefore we can recycle the MB representation for the scalar propagator, which leads us to the following expression:
\begin{align}
    D_{\lessgtr}^{(L)}(k;\tau_1,\tau_2) =&~ \FR{1}{4\pi(\wt\nu^2+\fr{1}{4})}
    \int_{-\ii\infty}^{\ii\infty}
    \FR{\di s_1}{2\pi\ii}\FR{\di s_2}{2\pi\ii}\,
    e^{\mp\ii\pi(s_1-s_2)}\Big(\FR{k}2\Big)^{-2s_{12}}
    (-\tau_1)^{-2s_1+1/2}(-\tau_2)^{-2s_2+1/2}\n\\
    &\times \Big(2s_1+\FR12\Big)\Big(2s_2+\FR12\Big) \Gamma\Big[s_1-\FR{\ii\wt\nu}2,s_1+\FR{\ii\wt\nu}2,s_2-\FR{\ii\wt\nu}2,s_2+\FR{\ii\wt\nu}2\Big].
\end{align} 
Then, following the same procedure as before, we calculate each component of the longitudinal vector seed integral, and find the following results. Again, we only present the result for $r_1<r_2$, and the result for $r_1>r_2$ can be obtained by switching the variables $(r_1,p_1)\leftrightarrow (r_2,p_2)$.

First, the opposite-sign integral is:
\begin{align}
    \mathcal I^{(L)p_1p_2}_{\pm\mp}=&~
    \FR{e^{\mp\ii\pi(p_1-p_2)/2}}{4\pi(\wt\nu^2+\fr{1}{4})}
    \bigg[{\mb H}^{p_1}_{\wt\nu}(r_1)\Big(\FR{r_1}2\Big)^{\ii\wt\nu}+{\mb H}^{p_1}_{-\wt\nu}(r_1)\Big(\FR{r_1}2\Big)^{-\ii\wt\nu}\bigg]\n\\
    &\times \bigg[{\mb H}^{p_2}_{\wt\nu}(r_2)\Big(\FR{r_2}2\Big)^{\ii\wt\nu}+{\mb H}^{p_2}_{-\wt\nu}(r_2)\Big(\FR{r_2}2\Big)^{-\ii\wt\nu}\bigg],
\end{align}
where we have defined: 
\begin{align}
\label{eq_Hbold}
    \mb H^p_{\wt\nu}(r)\equiv&~ r^{3/2+p}  \bigg\{\Big(\FR12-\ii\wt\nu\Big) \Gamma\Big[-\ii\wt\nu,p+\FR32+\ii\wt\nu\Big]
    {}_2\mathrm{F}_1\left[\bgm \fr p2+\fr34+\fr{\ii\wt\nu}2,\fr p2+\fr54+\fr{\ii\wt\nu}2\\1+\ii\wt\nu\edm\middle|\,r^2\right]\n\\
    &+\FR{r^2}2 \Gamma\Big[-1-\ii\wt\nu,p+\FR72+\ii\wt\nu\Big]
    {}_2\mathrm{F}_1\left[\bgm \fr p2+\fr74+\fr{\ii\wt\nu}2,\fr p2+\fr94+\fr{\ii\wt\nu}2\\2+\ii\wt\nu\edm\middle|\,r^2\right]\bigg\}.
\end{align} 
Second, the factorized part of the same-sign integral is:
\begin{align}
    \mathcal I^{(L)p_1p_2}_{\pm\pm,\text{F},>}=&~
    \FR{-\ii e^{\mp\ii\pi(p_1+p_2)/2}}{4\pi(\wt\nu^2+\fr{1}{4})}
    \bigg[e^{\pi\wt\nu}{\mb H}^{p_1}_{\wt\nu}(r_1)\Big(\FR{r_1}2\Big)^{\ii\wt\nu}+e^{-\pi\wt\nu}{\mb H}^{p_1}_{-\wt\nu}(r_1)\Big(\FR{r_1}2\Big)^{-\ii\wt\nu}\bigg]\n\\
    &\times \bigg[{\mb H}^{p_2}_{\wt\nu}(r_2)\Big(\FR{r_2}2\Big)^{\ii\wt\nu}+{\mb H}^{p_2}_{-\wt\nu}(r_2)\Big(\FR{r_2}2\Big)^{-\ii\wt\nu}\bigg].
\end{align}
Finally, the time-ordered part of the same-sign integral is:
\begin{align}
    {\mathcal I}_{\pm\pm,\text{TO},>}^{(L)p_1p_2}(r_1,r_2)
    =&~ \FR{e^{\mp\ii\pi(p_1+p_2)/2}}{2(\wt\nu^2+\fr{1}{4})} r_1^{3+p_1+p_2} \sum_{n_1,n_2=0}^{\infty}
    \FR{(-1)^{n_{12}}}{n_1!n_2!}\Big(\FR {r_1}2 \Big)^{2n_{12}}\n\\
    &\times\FR{\ii}{\wt\nu}
    \Big(2n_1-\FR12+\ii\wt\nu\Big)\Big(2n_2-\FR12-\ii\wt\nu\Big)(-\ii\wt\nu)_{-n_1}(\ii\wt\nu)_{-n_2}\n\\ &\times{\mathcal{F}}\left[\bgm 2n_2+p_2+\fr32-\ii\wt\nu,2n_{12}+p_1+p_2+3\\2n_2+p_2+\fr52-\ii\wt\nu\edm\middle|\,-\FR{r_1}{r_2}\right]+(\wt\nu\to-\wt\nu).
\end{align}
Again, we can sum up all SK components, and separate the result into a background piece, a local-signal piece, and a nonlocal-signal piece:
\begin{align}
  \sum_{\aa,\bb=\pm}\mathcal{I}_{\aa\bb}^{(L)p_1p_2}(r_1,r_2)=\mathcal{I}_{\text{BG},>}^{(L)p_1p_2}(r_1,r_2)+\mathcal{I}_{\text{L},>}^{(L)p_1p_2}(r_1,r_2)+\mathcal{I}_\text{NL}^{(L)p_1p_2}(r_1,r_2).
\end{align}
More explicitly, the local signal is given by:
\begin{equation}
     {\mathcal I}^{(L)p_1p_2}_{\text{L},>} (r_1,r_2)
    =\mathcal D^{p_1p_2}_{\wt\nu}
    \Big(\FR{r_1}{r_2}\Big)^{\ii\wt\nu}
    \mb H^{p_1}_{\wt\nu}(r_1) \mb H^{p_2}_{-\wt\nu}(r_2) + \text{c.c.},
\end{equation}
where
\bge
  \mathcal{D}_{\wt\nu}^{p_1p_2}\equiv\FR{1}{2\pi(\wt\nu^2+\fr{1}{4})}\Big[\cos\big(\fr{\pi}{2}(p_1-p_2)\big)-\sin\big(\fr{\pi}{2}(p_1+p_2+2\ii\wt\nu)\big)\Big].
\ede
The nonlocal signal is given by:
\begin{equation}
     {\mathcal I}^{(L)p_1p_2}_{\text{NL}} (r_1,r_2)
    =\mathcal D^{p_1p_2}_{\wt\nu}
    \Big(\FR{r_1r_2}{4}\Big)^{\ii\wt\nu}
    \mb H^{p_1}_{\wt\nu}(r_1) \mb H^{p_2}_{\wt\nu}(r_2) + \text{c.c.},
\end{equation}
Finally, the background is given by:
\begin{align}
    {\mathcal I}_{\text{BG},>}^{(L)p_1p_2}(r_1,r_2)
    =&~ \FR{1}{\wt\nu^2+\fr{1}{4}}\cos\big[\fr{\pi}{2}(p_1+p_2)\big] r_1^{3+p_1+p_2} \sum_{n_1,n_2=0}^{\infty}
    \FR{(-1)^{n_{12}}}{n_1!n_2!}\Big(\FR {r_1}2 \Big)^{2n_{12}}\n\\
    &\times\FR{\ii}{\wt\nu}
    \Big(2n_1-\FR12+\ii\wt\nu\Big)\Big(2n_2-\FR12-\ii\wt\nu\Big)(-\ii\wt\nu)_{-n_1}(\ii\wt\nu)_{-n_2}\n\\ &\times{\mathcal{F}}\left[\bgm 2n_2+p_2+\fr32-\ii\wt\nu,2n_{12}+p_1+p_2+3\\2n_2+p_2+\fr52-\ii\wt\nu\edm\middle|\,-\FR{r_1}{r_2}\right]+\text{c.c.}.
\end{align}

\subsection{Summary}
\label{sec_vec_summary}

After the rather long and technical calculations, it would be useful to summarize and present the result we have obtained in a self-contained manner. In this section, we have computed the vector seed integral (\ref{eq_vecIab}) for all three polarizations $\lam=\pm,L$ and for arbitrary power indices $(p_1,p_2)$. For general  applications, it is more convenient to present these results with the SK indices summed. As we have shown before, the summed integral can be broken into three pieces, the background piece $\mathcal{I}_{\aa\bb}^{(\lam)p_1p_2}$, the local signal  
$\mathcal{I}_\text{L}^{(\lam)p_1p_2}$, and the nonlocal signal $\mathcal{I}_\text{NL}^{(\lam)p_1p_2}$, according to their analytic properties at $r_{1,2}=0$:
\begin{keyeqn}
\begin{align}
\label{eq_vecSeedIntResult}
    \sum_{\aa,\bb=\pm}\mathcal{I}_{\aa\bb}^{(\lam)p_1p_2}(r_1,r_2)
   =&~\mathcal{I}_\text{BG}^{(\lam)p_1p_2}(r_1,r_2)+\mathcal{I}_\text{L}^{(\lam)p_1p_2}(r_1,r_2)+\mathcal{I}_\text{NL}^{(\lam)p_1p_2}(r_1,r_2),\\
    \mathcal{I}_\text{BG}^{(\lam)p_1p_2}(r_1,r_2)
    =&~\mathcal{I}_{\text{BG},>}^{(\lam)p_1p_2}(r_1,r_2)\theta(r_2-r_1)+\mathcal{I}_{\text{BG},>}^{(\lam)p_2p_1}(r_2,r_1)\theta(r_1-r_2),\\
    \mathcal{I}_\text{L}^{(\lam)p_1p_2}(r_1,r_2)
    =&~\mathcal{I}_{\text{L},>}^{(\lam)p_1p_2}(r_1,r_2)\theta(r_2-r_1)+\mathcal{I}_{\text{L},>}^{(\lam)p_2p_1}(r_2,r_1)\theta(r_1-r_2),
\end{align}
\end{keyeqn}
Below we summarize the results for each of the three polarizations. 

\paragraph{Transverse integrals.} The two transverse integrals correspond to $\lam=h=\pm$. They are dependent on the chemical potential. For convenience, we introduced new variables $u_1=2r_1/(1+r_1)$ and $u_2=2r_2/(1+r_2)$. Then, the three integrals in (\ref{eq_vecSeedIntResult}) can be written as:
\begin{keyeqn}
\begin{align}
\label{eq_TransSeedBG}
     {\mathcal I}^{(h)p_1p_2}_{\text{BG},>}  (u_1,u_2)
  =&~
 \FR{e^{-\ii\pi(p_1+p_2)/2}}{2^{3+p_1+p_2}}\sum_{n_1,n_2=0}^\infty \FR{(-1)^{n_{12}}}{n_1!n_2!} u_1^{n_{12}+3+p_1+p_2}\n\\
 &\times
     \bigg\{
   \FR{\ii}{2\wt\nu} 
     \big(\fr{1}{2}+\ii\wt\nu+\ii h\wt\mu \big)_{n_1}\big(\fr{1}{2}-\ii\wt\nu+\ii h\wt\mu \big)_{n_2}\big(-2\ii\wt\nu\big)_{-n_1}\big(+2\ii\wt\nu\big)_{-n_2}\n\\
    &\times  {\mathcal{F}}\left[\bgm n_2+p_2+\fr32-\ii\wt\nu,n_{12}+p_1+p_2+3\\n_2+p_2+\fr52-\ii\wt\nu\edm\middle|\,-\FR{u_1}{u_2}\right] 
     +(\wt\nu\to-\wt\nu)\bigg\}+\text{c.c.},\\
\label{eq_TransSeedL}
   \mathcal{I}_{\text{L},>}^{(h)p_1p_2}(u_1,u_2) 
    =&~\mathcal G^{(h)p_1p_2}_{\text{L}}(u_1,u_2)\Big(\FR{u_1}{u_2}\Big)^{\ii\wt\nu}+\text{c.c.}, \\
\label{eq_TransSeedNL}
   \mathcal{I}_{\text{NL}}^{(h)p_1p_2}(u_1,u_2)
    =&~\mathcal G^{(h)p_1p_2}_{\text{NL}}(u_1,u_2)(u_1u_2)^{\ii\wt\nu}+\text{c.c.},
\end{align}
\end{keyeqn}
where we have defined:
\begin{align} 
\label{eq_calGL}
    \mathcal{G}_{\text{L}}^{(h)p_1p_2}(u_1,u_2)
  =&~\FR{e^{-\pi h\wt\mu}}{2\pi^2}
  \bigg\{\big(\cosh2\pi\wt\mu+\cosh2\pi\wt\nu\big) 
  \Big[e^{-\ii(p_1-p_2)\pi/2}\mb{G}^{p_1}_{h\wt\mu,\wt\nu}(u_1)\mb{G}^{p_2}_{-h\wt\mu,-\wt\nu}(u_2)\n\\
  &+ e^{+\ii(p_1-p_2)\pi/2}\mb{G}^{p_1}_{-h\wt\mu,\wt\nu}(u_1)\mb{G}^{p_2}_{h\wt\mu,-\wt\nu}(u_2) \Big] \n\\ 
  &~+\FR{-2\ii\pi e^{\pi\wt\nu}\cosh[\pi(h\wt\mu+\wt\nu)]}{\Gamma[\fr{1}{2}+\ii h\wt\mu+\ii\wt\nu,\fr{1}{2}+\ii h\wt\mu-\ii\wt\nu]}e^{-\ii(p_1+p_2)\pi/2}\mb{G}^{p_1}_{h\wt\mu,\wt\nu}(u_1)\mb{G}^{p_2}_{h\wt\mu,-\wt\nu}(u_2)\n\\
  &~+\FR{2\ii\pi e^{-\pi\wt\nu}\cosh[\pi(h\wt\mu-\wt\nu)]}{\Gamma[\fr{1}{2}-\ii h\wt\mu-\ii\wt\nu,\fr{1}{2}-\ii h\wt\mu+\ii\wt\nu]}e^{+\ii(p_1+p_2)\pi/2}\mb{G}^{p_1}_{-h\wt\mu,\wt\nu}(u_1)\mb{G}^{p_2}_{-h\wt\mu,-\wt\nu}(u_2) 
   \bigg\},
\end{align}
\begin{align}
\label{eq_calGNL}
  \mathcal{G}_\text{NL}^{(h)p_1p_2}(u_1,u_2)
  =&~\FR{e^{-\pi h\wt\mu}}{2\pi^2}  \bigg\{ 
  \big(\cosh2\pi\wt\mu+\cosh2\pi\wt\nu\big) 
  \Big[e^{-\ii(p_1-p_2)\pi/2}\mb{G}^{p_1}_{h\wt\mu,\wt\nu}(u_1)\mb{G}^{p_2}_{-h\wt\mu,\wt\nu}(u_2)\n\\
  &+ e^{+\ii(p_1-p_2)\pi/2}\mb{G}^{p_1}_{-h\wt\mu,\wt\nu}(u_1)\mb{G}^{p_2}_{h\wt\mu,\wt\nu}(u_2) \Big] \n\\
  &~+\FR{-2\ii\pi e^{\pi\wt\nu}\cosh[\pi(h\wt\mu+\wt\nu)]}{\Gamma[\fr{1}{2}+\ii h\wt\mu-\ii\wt\nu,\fr{1}{2}+\ii h\wt\mu+\ii\wt\nu]}e^{-\ii(p_1+p_2)\pi/2}\mb{G}^{p_1}_{h\wt\mu,\wt\nu}(u_1)\mb{G}^{p_2}_{h\wt\mu,\wt\nu}(u_2)\n\\
  &~+\FR{2\ii\pi e^{-\pi\wt\nu}\cosh[\pi(h\wt\mu-\wt\nu)]}{\Gamma[\fr{1}{2}-\ii h\wt\mu-\ii\wt\nu,\fr{1}{2}-\ii h\wt\mu+\ii\wt\nu]}e^{+\ii(p_1+p_2)\pi/2}\mb{G}^{p_1}_{-h\wt\mu,\wt\nu}(u_1)\mb{G}^{p_2}_{-h\wt\mu,\wt\nu}(u_2) 
   \bigg\}.
\end{align}
and the function $\mb{G}$ is defined by
\begin{align}
  \mb{G}^p_{h\wt\mu,\wt\nu}(u)\equiv \ii\pi\text{csch}(2\pi\wt\nu) \Big(\FR{u}2\Big)^{3/2+p} \mathcal{F}\left[\bgm \fr{3}{2}+p+\ii\wt\nu,\fr{1}{2}+\ii h \wt\mu+\ii\wt\nu\\ 1+2\ii\wt\nu\edm\middle|u\right].
\end{align}

\paragraph{Longitudinal integrals.} The longitudinal integral correspond to $\lam=L$. This part is independent of the chemical potential, and the results for the three integrals in (\ref{eq_vecSeedIntResult}) are:
\begin{keyeqn}
\begin{align}
\label{eq_LongSeedBG}
    {\mathcal I}_{\text{BG},>}^{(L)p_1p_2}(r_1,r_2)
    =&~ \FR{1}{ \wt\nu^2+\fr14}\cos\big[\fr{\pi}{2}(p_1+p_2)\big] r_1^{3+p_1+p_2} \sum_{n_1,n_2=0}^{\infty}
    \FR{(-1)^{n_{12}}}{n_1!n_2!}\Big(\FR {r_1}2 \Big)^{2n_{12}}\n\\
    &\times\FR{\ii}{\wt\nu}
    \Big(2n_1-\FR12+\ii\wt\nu\Big)\Big(2n_2-\FR12-\ii\wt\nu\Big)(-\ii\wt\nu)_{-n_1}(\ii\wt\nu)_{-n_2}\n\\ &\times{\mathcal{F}}\left[\bgm 2n_2+p_2+\fr32-\ii\wt\nu,2n_{12}+p_1+p_2+3\\2n_2+p_2+\fr52-\ii\wt\nu\edm\middle|\,-\FR{r_1}{r_2}\right]+\text{c.c.},
    \\
\label{eq_LongSeedL}
     {\mathcal I}^{(L)p_1p_2}_{\text{L},>} (r_1,r_2)
    =&~\mathcal G^{(L)p_1p_2}_{\text{L}}(r_1,r_2) 
    \Big(\FR{r_1}{r_2}\Big)^{\ii\wt\nu}
     + \text{c.c.},
    \\
\label{eq_LongSeedNL}
     {\mathcal I}^{(L)p_1p_2}_{\text{NL}} (r_1,r_2)
    =&~\mathcal G^{(L)p_1p_2}_{\text{NL}}(r_1,r_2) 
    \Big(\FR{r_1r_2}{4}\Big)^{\ii\wt\nu}
      + \text{c.c.},
\end{align}
\end{keyeqn}
where we have defined the following functions:
\begin{align}
\label{eq_calGLongL}
  \mathcal G^{(L)p_1p_2}_{\text{L}}(r_1,r_2)
  \equiv&~ \FR{1}{2\pi (\wt\nu^2+\fr14)}\Big[\cos\big(\fr{\pi}{2}(p_1-p_2)\big)-\sin\big(\fr{\pi}{2}(p_1+p_2+2\ii\wt\nu)\big)\Big]\mb H^{p_1}_{\wt\nu}(r_1) \mb H^{p_2}_{-\wt\nu}(r_2),\\ 
\label{eq_calGLongNL}
  \mathcal G^{(L)p_1p_2}_{\text{NL}}(r_1,r_2)
  \equiv&~ \FR{1}{2\pi (\wt\nu^2+\fr14)}\Big[\cos\big(\fr{\pi}{2}(p_1-p_2)\big)-\sin\big(\fr{\pi}{2}(p_1+p_2+2\ii\wt\nu)\big)\Big]\mb H^{p_1}_{\wt\nu}(r_1) \mb H^{p_2}_{\wt\nu}(r_2),
\end{align}
and the function $\mb{H}$ is defined by 
\begin{align}
    \mb H^p_{\wt\nu}(r)\equiv&~ r^{3/2+p} \bigg\{\Big(\FR12-\ii\wt\nu\Big) \Gamma\Big[-\ii\wt\nu,p+\FR32+\ii\wt\nu\Big]
    {}_2\mathrm{F}_1\left[\bgm \fr p2+\fr34+\fr{\ii\wt\nu}2,\fr p2+\fr54+\fr{\ii\wt\nu}2\\1+\ii\wt\nu\edm\middle|\,r^2\right]\n\\
    &+\FR{r^2}2 \Gamma\Big[-1-\ii\wt\nu,p+\FR72+\ii\wt\nu\Big]
    {}_2\mathrm{F}_1\left[\bgm \fr p2+\fr74+\fr{\ii\wt\nu}2,\fr p2+\fr94+\fr{\ii\wt\nu}2\\2+\ii\wt\nu\edm\middle|\,r^2\right]\bigg\}.
\end{align} 

\paragraph{Tree-level signal cutting rule.}
At the end of this section, we briefly comment on the relation between our calculation and a recently proposed cutting rule for computing CC signals at the tree level \cite{Tong:2021wai}. With our notations, it was proposed in \cite{Tong:2021wai} that the oscillating signals in inflation correlators are contributed by two sources: 1) The opposite-sign integrals $\mathcal{I}_{\pm\mp}$; 2) the factorized part of the same-sign integrals $\mathcal{I}_{\pm\pm,\text{F}}$. On the other hand, the time-ordered part of the same-sign integrals $\mathcal{I}_{\pm\pm,\text{TO}}$ do not contribute to the signal.  The upshot is that the computation of CC signals involves no genuine time-ordered integral, hence the name cutting rule. A cutting rule of this sort can be phenomenologically useful, since the computation of factorized integrals are almost always much simpler than that of the time-ordered integrals. The cutting rule proposed in \cite{Tong:2021wai} was based on physical arguments of particle production in the bulk, and was further checked with numerical evaluations. Our results in this paper agree with the physical observations made in \cite{Tong:2021wai}, and can be viewed as an analytical derivation for the cutting rule in the presence of the chemical potential. (The dS covariant case without chemical potential can be verified with previously known results  \cite{Arkani-Hamed:2018kmz}.) Therefore, the partial MB representation can be used to give a more rigorous and more satisfactory derivation of the cutting rule for the CC signals. We shall pursue such a proof in a future work.

\section{Bootstrapping Helical Correlators}
\label{sec_boot}

In this section we provide an alternative derivation for both the scalar seed integral (\ref{eq_ScalarSeedInt}) introduced in Sec.\ \ref{sec_PMB} and the vector seed integral (\ref{eq_vecIab}) introduced in Sec.\ \ref{sec_HCP}, using a ``bootstrapping'' method developed in recent years by several groups of authors \cite{Arkani-Hamed:2018kmz,Baumann:2019oyu,Baumann:2020dch,Pajer:2020wnj,Pajer:2020wxk,Cabass:2021fnw,Sleight:2019hfp,Pimentel:2022fsc,Jazayeri:2022kjy}. See also  \cite{Baumann:2022jpr} for a recent review.

At the conceptual level, the cosmological bootstrap program aims at building inflation correlators directly at the future boundary without doing bulk time integral at each interaction vertex. This is achieved by exploiting the symmetry properties of the inflation correlators, together with a set of constraints and boundary conditions from the bulk unitarity, locality, as well as the bulk initial condition. When boiled down to a practical problem of computing inflation correlators, the bootstrap program suggests that we can derive a set of differential equations satisfied by the correlators, so that we can find the correlators by solving the differential equations instead of computing bulk integrals. The boundary conditions that are required to uniquely pin down the solution can be found from physical considerations in the bulk. For simplicity, we will collectively call such differential equations satisfied by the correlators the \emph{bootstrap equations}, although its meaning is a bit shifted from the original conformal bootstrap program in conformal field theories. 

The idea of solving the inflation correlators from a set of differential equations has been put forward originally in \cite{Arkani-Hamed:2015bza}. There, the authors get the signal part of the whole correlator, which corresponds to a solution to the homogeneous equation. The background solution, corresponding to a particular solution to the inhomogeneous equation, was later obtained in \cite{Arkani-Hamed:2018kmz}. In \cite{Arkani-Hamed:2018kmz} it was further shown that the bootstrap equation for the 4-point tree diagram with massive exchange can be derived from the Ward identity of dS isometries. The bootstrap program was subsequently generalized to the case of boost-breaking theories for purely massless correlators \cite{Pajer:2020wnj,Pajer:2020wxk} and for correlators involving massive exchange \cite{Pimentel:2022fsc,Jazayeri:2022kjy}. 

Of course, when the dS boosts are broken by the background, the bootstrap equations are not directly from the Ward identities, although we suspect that they can be related to the Ward identity with additional symmetry breaking terms. Nonetheless, one can still derive the bootstrap equations for the correlators using the very observation originally made in \cite{Arkani-Hamed:2015bza}. The essential idea is the following. The tree-level exchange digram for the 4-point correlator can be expressed as SK integrals with the massive propagator in the integrand. As the Green function for the corresponding massive field, the massive propagator solves the field equation with either a $\de$-function source or no source. Therefore, if we insert the differential operator for the field equation in front of the massive propagator in the SK integral, the integral will be reduced to either 0 or a much simpler integral over a single time variable. Then, given that the factors multiplied to the massive propagators (including the bulk-to-boundary propagators and the coupling factors) are simple power functions or exponential functions, it is very often possible to commute the differential operator in front of the massive propagator with the integral itself, resulting in an differential equation for the whole integral, which is the desired bootstrap equation.

\begin{figure}
\centering
  \includegraphics[width=0.85\textwidth]{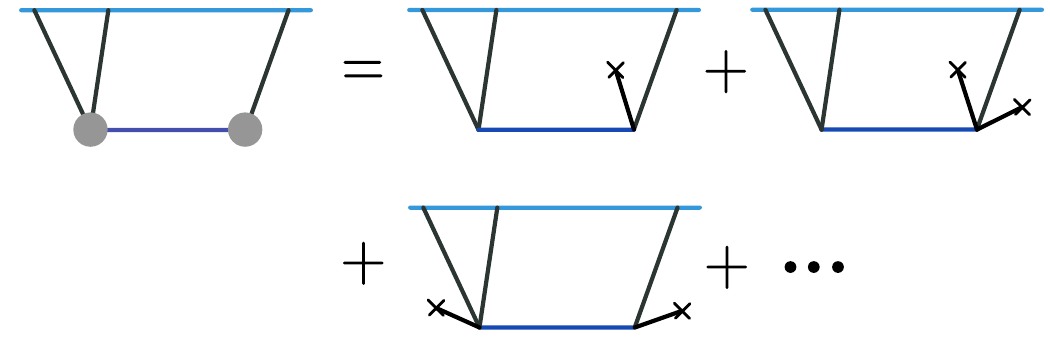} 
  \caption{An illustration of boostless bootstrap program developed in \cite{Pimentel:2022fsc,Jazayeri:2022kjy}, remade from a similar figure in \cite{Pimentel:2022fsc}. In this plot, the upper blue line represents the future boundary of dS and the space below represents the bulk. The blob represents the summed vertices with arbitrary number of background value $\dot\phi_0$ insertions, represented by crosses. Note that, while the number of crosses are arbitrary, they are all inserted at the two vertices. The propagators in these figures can have non-unit sound speed, but this can be implemented by simple change of the variable $k \to c_s k$. }
  \label{fig_bootless}
\end{figure}

\begin{figure}
 \centering
  \includegraphics[width=0.85\textwidth]{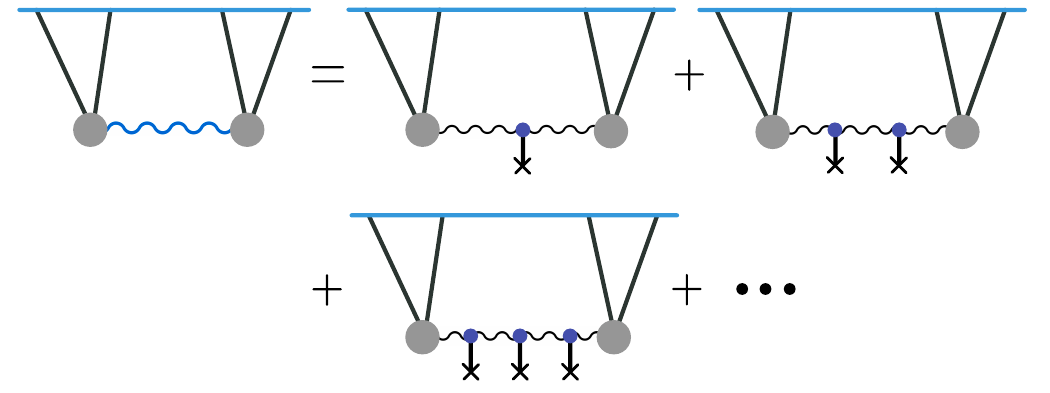} 
  \caption{An illustration of helical inflation correlators from resumming chemical-potential insertions. The blue wiggly line on the left hand side represents the massive spin-1 propagator dressed by the chemical potential, and the black wiggly lines on the right hand side represent the dS covariant spin-1 propagator without chemical potentials. While each gray blob already contains arbitrary number of inflaton background insertions, there are further boost-breaking insertions attached to the black propagator. A nonperturbative treatment of chemical potential effects requires summing over all such insertions. Technically, this resummation is done by directly solving the equation of motion. }
  \label{fig_vecboot}
\end{figure}

This bootstrap program was recently generalized to boost-breaking theories with massive exchanges \cite{Pimentel:2022fsc,Jazayeri:2022kjy}. These works allowed for non-unit sound speed of the massive particle and the external massless states, as well as more general boost-breaking couplings. However, their results are not directly applicable to cases with helical chemical potential, owing to the more complicated mode function for helical spinning fields, which require either a direct manipulation of Whittaker functions, or a resummation of infinite number of boost-breaking insertions in the propagator. We highlight the difference between the two cases in Figs.\ \ref{fig_bootless} and \ref{fig_vecboot}.

The presentation below will be separated into 3 subsections, corresponding to the case of scalar change, transverse vector change, and longitudinal vector change, respectively. The procedure will be identical: We first derive the bootstrap equation, and then try to solve it. The final result will be the sum of a particular solution to the inhomogeneous equation and a solution to the homogeneous equation with proper boundary conditions. We determine the coefficients of the homogeneous solutions by imposing boundary conditions at the squeezed limit $r_{1,2}\to 0$. We then check our results by looking at the folded limit $r_{1,2}\to 1$. Along the way, we will also get new analytical results for the 3-point and 2-point functions in completely closed analytical form.

\subsection{Scalar}

Again we begin with the known example of a massive scalar exchange. We do not pursue the most general results in this section. Instead, we will use a specific example to illustrate the idea of bootstrap equations. For this purpose, we derive a set of differential equations satisfied by the scalar seed integral (\ref{eq_ScalarSeedInt}) with $p_1=p_2=-2$. This corresponds to the four external states being directly coupled conformal scalars. The results for other values of $(p_1,p_2)$ can be obtained by either following the same procedure, or acting appropriate differential operators on the result with $p_1=p_2=-2$. 

Recall that the scalar seed integral with $p_1=p_2=-2$ is:
\begin{align}
\label{eq_ScalarSeed00}
  \mathcal{I}_{\aa\bb}^{-2,-2}(r_1,r_2)\equiv -\mathsf{ab}\, k_s \int_{-\infty}^{0} \FR{\di \tau_1}{\tau_1^2}\FR{\di \tau_2}{\tau_2^2}\,e^{\ii\mathsf{a}k_{12}\tau_1+\ii\mathsf{b}k_{34}\tau_2}D_{\mathsf{ab}}(k_s;\tau_1,\tau_2).
\end{align}
One useful observation here is that the following combination,
\bge
\label{eq_hattedScalarD}
  \wh{D}_{\aa\bb}(k_s\tau_1,k_s\tau_2)\equiv k_s^3 D_{\aa\bb}(k_s;\tau_1,\tau_2),
\ede
depends on the three variables $k_s$, $\tau_1$, and $\tau_2$ only through the combinations $k_s\tau_1$ and $k_s\tau_2$. Therefore, if we further define $z_{1}=k_{12}\tau_1$ and $z_2=k_{34}\tau_2$, then the seed integral (\ref{eq_ScalarSeed00}) can be rewritten in the following form:
\begin{align}
\label{eq_ScalarBootInt}
  r_1r_2 \mathcal{I}_{\aa\bb}^{-2,-2}(r_1,r_2)\equiv -\mathsf{ab}\,\int_{-\infty}^{0} \FR{\di z_1}{z_1^2}\FR{\di z_2}{z_2^2}\, e^{\ii\mathsf{a}z_1+\ii\mathsf{b}z_2}\wh{D}_{\mathsf{ab}}(r_1z_1,r_2z_2),
\end{align}
where again $r_1=k_s/k_{12}$ and $r_2=k_s/k_{34}$.

To derive an equation satisfied by the integral (\ref{eq_ScalarBootInt}), we act the Klein-Gordon operator to the massive scalar propagator $D_{\aa\bb}(k_s;\tau_1,\tau_2)$, which, by definition, annihilates the opposite-sign propagator $D_{\pm\mp}$ and reduces the same-sign propagator $D_{\pm\pm}$ to a $\de$ function:
\begin{align}
&(\tau_1^2 \partial_{\tau_1}^2 - 2\tau_1 \partial_{\tau_1} + k_s^2\tau_1^2 + m^2)D_{\pm\mp}(k_s;\tau_1,\tau_2)=0,\\
&(\tau_1^2 \partial_{\tau_1}^2 - 2\tau_1 \partial_{\tau_1} + k_s^2\tau_1^2 + m^2)D_{\pm\pm}(k_s;\tau_1,\tau_2)=\mp\ii \tau_1^2\tau_2^2\delta(\tau_1-\tau_2).
\end{align} 
These equations can be rewritten as equations satisfied by the ``hatted'' propagator $\wh{D}_{\aa\bb}(r_1z_1,r_2z_2)$:
\begin{align}
&(r_1^2\pd_{r_1}^2-2r_1\pd_{r_1}+r_1^2z_1^2+m^2)\wh D_{\pm\mp}(r_1z_1,r_2z_2)=0,\\
&(r_1^2\pd_{r_1}^2-2r_1\pd_{r_1}+r_1^2z_1^2+m^2)\wh D_{\pm\pm}(r_1z_1,r_2z_2)=\mp\ii (r_1z_1)^2(r_2z_2)^2\delta(r_1z_1-r_2z_2).
\end{align}
Now we insert the operator $r_1^2\pd_{r_1}^2-2r_1\pd_{r_1}+r_1^2z_1^2+m^2$ in front of $\wh D_{\aa\bb}$ in the integral (\ref{eq_ScalarBootInt}), and use the above two equations, which give rise to the following two equations:
\begin{align}
  &+\int_{-\infty}^{0} \FR{\di z_1}{z_1^2}\FR{\di z_2}{z_2^2}\, e^{\ii\mathsf{a}z_1+\ii\mathsf{b}z_2}\big(r_1^2\pd_{r_1}^2-2r_1\pd_{r_1}+r_1^2z_1^2+m^2\big)\wh{D}_{\pm\mp}(r_1z_1,r_2z_2)=0,\\
  &-\int_{-\infty}^{0} \FR{\di z_1}{z_1^2}\FR{\di z_2}{z_2^2}\, e^{\ii\mathsf{a}z_1+\ii\mathsf{b}z_2}\big(r_1^2\pd_{r_1}^2-2r_1\pd_{r_1}+r_1^2z_1^2+m^2\big)\wh{D}_{\pm\pm}(r_1z_1,r_2z_2)
  =\FR{(r_1r_2)^2}{r_1+r_2}.
\end{align}
At this point, if we can somehow pull the differential operator $r_1^2\pd_{r_1}^2-2r_1\pd_{r_1}+r_1^2z_1^2+m^2$ out of the integral, then we will get an differential equation for the whole integral. This can be done by using the following two relations:
\begin{align}
\label{eq_IntDiffCommut1}
\int_{-\infty}^0 \FR{\di z_1}{z_1^2}\, z_1 e^{\ii \mathsf a z_1}\wh D_{\mathsf{ab}}(r_1z_1,r_2z_2)
=&~\ii\mathsf a (r_1\pd_{r_1}-1)\int_{-\infty}^0 \FR{\di z_1}{z_1^2}\, e^{\ii \mathsf a z_1}\wh D_{\mathsf{ab}}(r_1z_1,r_2z_2),\\
\label{eq_IntDiffCommut2}
\int_{-\infty}^0 \FR{\di z_1}{z_1^2}\, z_1^2 e^{\ii \mathsf a z_1}\wh D_{\mathsf{ab}}(r_1z_1,r_2z_2)
=&-(r_1\pd_{r_1})(r_1\pd_{r_1}-1)\int_{-\infty}^0 \FR{\di z_1}{z_1^2} \,e^{\ii \mathsf a z_1}\wh D_{\mathsf{ab}}(r_1z_1,r_2z_2).
\end{align}
These relations are direct consequences of integration by parts. For instance, the first relation (\ref{eq_IntDiffCommut1}) can be derived as follows: 
\begin{align}
0=&\int_{-\infty}^0 \di z_1\,\pd_{z_1}\Big[ z_1^{-1}e^{\ii \mathsf a z_1} \wh D_{\mathsf{ab}}(r_1z_1,r_2z_2)\Big] 
=\int_{-\infty}^0 \di z_1\,  
\Big[-\FR{1}{z_1^2}+\FR{\ii \mathsf a }{z} +\FR{r_1\pd_{r_1}}{z_1^2}\Big]e^{\ii \mathsf a z_1}\wh D_{\mathsf{ab}}(r_1z_1,r_2z_2),
\end{align}
and the second relation (\ref{eq_IntDiffCommut2}) can be derived similarly. 

Then, using (\ref{eq_IntDiffCommut2}), we find that the integral $r_1r_2\mathcal{I}_{\aa\bb}^{-2,-2}(r_1,r_2)$ in (\ref{eq_ScalarBootInt}) satisfies the equations:
\begin{align}
&\bigg[(r_1^2-r_1^4)\pd_{r_1}^2-2r_1\pd_{r_1}+\Big(\wt\nu^2+\FR94\Big)\bigg]\Big( r_1r_2 \mathcal{I}_{\pm\mp}^{-2,-2}(r_1,r_2)\Big)=0,\\
&\bigg[(r_1^2-r_1^4)\pd_{r_1}^2-2r_1\pd_{r_1}+\Big(\wt\nu^2+\FR94\Big)\bigg]\Big( r_1r_2 \mathcal{I}_{\pm\pm}^{-2,-2}(r_1,r_2)\Big)=\FR{(r_1r_2)^2}{r_1+r_2}.
\end{align}
From these equations it is direct to get the following equations for the seed integral itself, which turns out a bit simpler than the above ones:
\begin{keyeqn}
\begin{align}
\label{eq_ScalarBootEqnHom}
&\bigg[(r_1^2-r_1^4)\pd_{r_1}^2-2r_1^3\pd_{r_1}+\Big(\wt\nu^2+\FR{1}4\Big)\bigg]  {\mathcal I}^{-2,-2}_{\pm\mp}(r_1,r_2)=0,\\
\label{eq_ScalarBootEqnInhom}
&\bigg[(r_1^2-r_1^4)\pd_{r_1}^2-2r_1^3\pd_{r_1}+\Big(\wt\nu^2+\FR{1}4\Big)\bigg]  {\mathcal I}^{-2,-2}_{\pm\pm}(r_1,r_2)=\FR{r_1r_2}{r_1+r_2}.
\end{align}
\end{keyeqn}
These are the bootstrap equations satisfied by the scalar seed integral, to which we shall  derive a proper solution satisfying all required boundary conditions.

\begin{figure}
 \centering
  \includegraphics[width=0.48\textwidth]{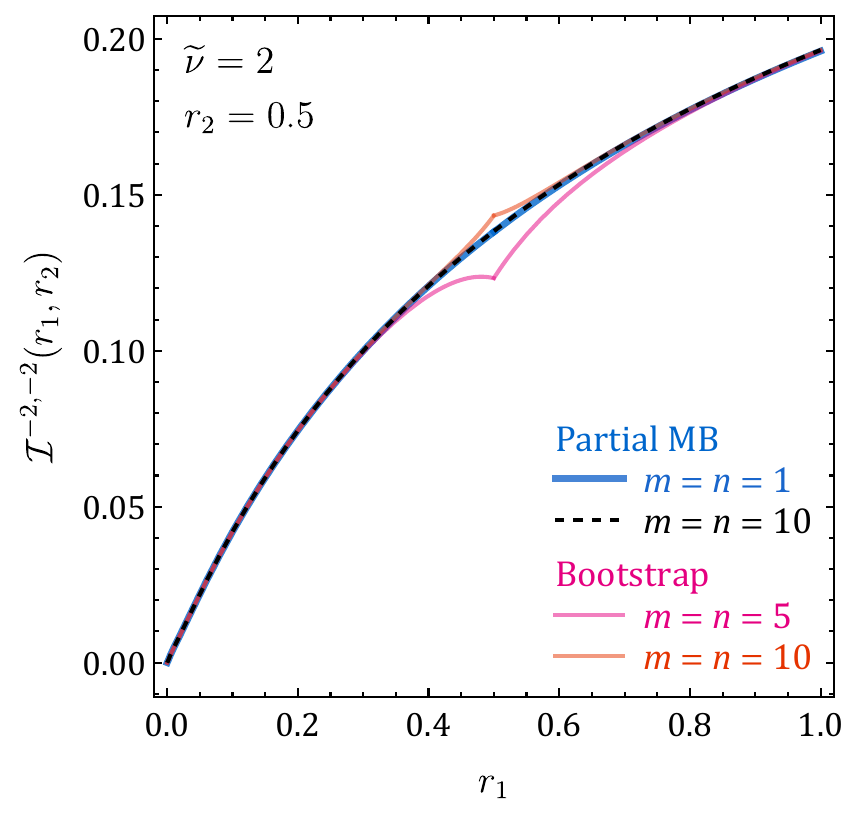}
  \includegraphics[width=0.48\textwidth]{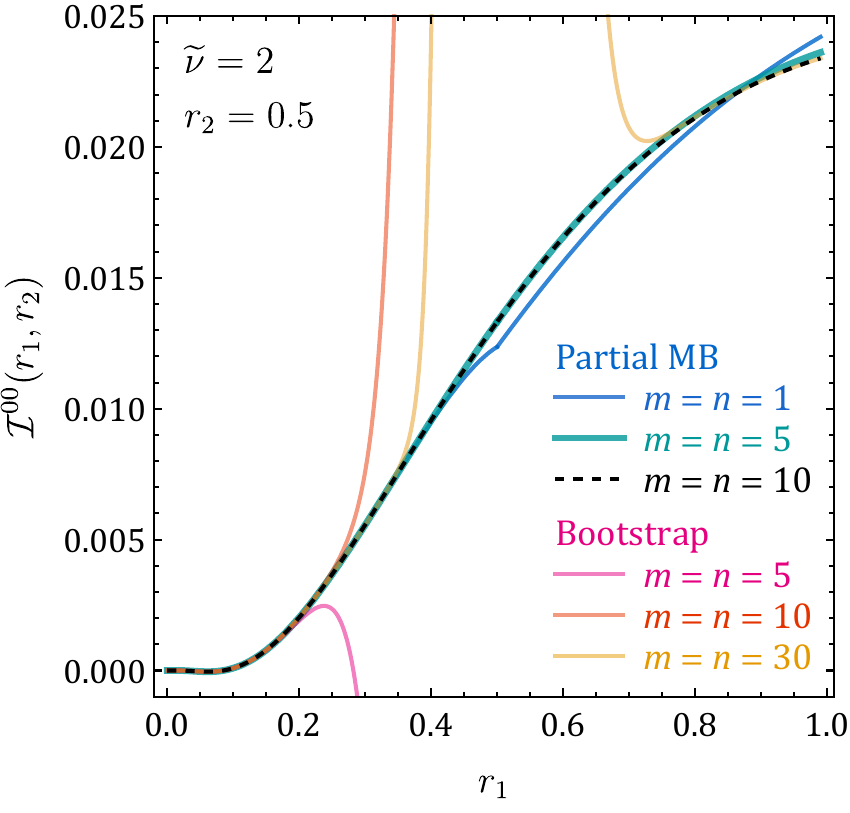}
  \caption{The summed scalar seed integrals $\sum_{\aa,\bb}\mathcal{I}_{\aa\bb}^{p_1p_2}(r_1,r_2)$ as functions of $r_1$ with fixed $r_2=0.5$ and $\wt\nu=2$. For the two cases $p_1=p_2=-2$ (left) and $p_1=p_2=0$ (right), we show the convergences of series from partial MB representation (\ref{eq_ScalarSeedBG}) and from the bootstrap method (\ref{eq_ScalarBootBG}), by including terms of $m$ and $n$ summations up to values indicated in the plots. }
  \label{fig_scalarBGConv}
\end{figure}

\paragraph{Particular solution to the inhomogeneous equation.}
We first try to find a particular solution to the inhomogeneous equation (\ref{eq_ScalarBootEqnInhom}). Following the method originally used in \cite{Arkani-Hamed:2018kmz}, we use double series expansion in both $r_1$ and $r_1/r_2$ as an ansatz and try to solve the series coefficients. To this end, we assume $r_1<r_2$, and Taylor expand the right hand side of the inhomogeneous equation (\ref{eq_ScalarBootEqnInhom}):
\begin{equation}
\label{eq_ScalarSourceTaylor}
   \FR{r_1r_2}{r_1+r_2} = r_1 \sum_{n=0}^\infty
    (-1)^n\Big(\FR{r_1}{r_2}\Big)^n.
\end{equation}
The form of this series motivates us to try the following ansatz for the inhomogeneous solution:
\begin{equation}
\label{eq_ScalarBootBGAnsatz}
    \mathcal Z(r_1,r_2) =  r_1\sum_{m,n=0}^\infty (-1)^n \mathcal{Z}_{m,n} r_1^{m} \Big(\FR{r_1}{r_2}\Big)^{n},
\end{equation}
Substituting the ansatz (\ref{eq_ScalarBootBGAnsatz}) and the expanded source term (\ref{eq_ScalarSourceTaylor}) back into the inhomogeneous equation (\ref{eq_ScalarBootEqnInhom}), and matching the powers of $r_1$ and $r_1/r_2$, we find the following set of recursion relations satisfied by the series coefficients $\mathcal{Z}_{m,n}$:
\begin{subequations}
\begin{align}
  \Big[\big(n+\fr{1}{2}\big)^2+\wt\nu^2\Big]\mathcal{Z}_{0,n}=&~1;\\
  \mathcal{Z}_{1,n}=&~0;\\
  \Big[\big(m+n+\fr{5}{2}\big)^2+\wt\nu^2\Big]\mathcal{Z}_{m+2,n}=&~(m+n+1)(m+n+2)\mathcal{Z}_{m,n}.
\end{align}
\end{subequations}
This recursion relation is straightforward to solve with the general term given by:
\begin{align}
    \mathcal{Z}_{2m,n} =&~ \FR{1}{(n+\fr12)^2+\wt\nu^2} \FR{(n+1)(n+2)}{(n+\fr52)^2+\wt\nu^2} \cdots  \FR{(n+2m-1)(n+2m)}{(n+2m+\fr12)^2+\wt\nu^2},\\
    \mathcal{Z}_{2m+1,n}=&~0,
\end{align}
or equivalently,
\begin{align}
  & \mathcal{Z}_{2m,n} =\FR{(n+1)_{2m}}{4^{1+m}(\fr{n}{2}+\fr{1}{4}+\fr{\ii\wt\nu}{2})_{m+1}(\fr{n}{2}+\fr{1}{4}-\fr{\ii\wt\nu}{2})_{m+1}},
  &&\mathcal{Z}_{2m+1,n}=0.
\end{align}
Note that the inhomogeneous solution we have found applies to both ${\mathcal{I}}^{-2,-2}_{++}$ and ${\mathcal{I}}^{-2,-2}_{--}$. Furthermore, this solution is analytic in both $r_1$ and $r_2$ as $r_{1,2}\to0$, and thus contributes only to the background piece of the 4-point correlator. As we shall see soon, the homogeneous solutions are all nonanalytic in $r_{1,2}$ at $r_{1,2}=0$, and therefore the above particular solution is the only contribution to the background. Therefore we conclude that the background piece of the scalar seed integral ${\mathcal{I}}^{-2,-2}_{\text{BG},>}(r_1,r_2)$ is twice of the above particular solution, namely:
\begin{keyeqn}
\begin{align}
\label{eq_ScalarBootBG}
  {\mathcal{I}}^{-2,-2}_{\text{BG},>}(r_1,r_2)=\sum_{m,n=0}^\infty 
  \FR{(-1)^n(n+1)_{2m}}{2^{2m+1}(\fr{n}{2}+\fr{1}{4}+\fr{\ii\wt\nu}{2})_{m+1}(\fr{n}{2}+\fr{1}{4}-\fr{\ii\wt\nu}{2})_{m+1}}
  r_1^{2m+1}\Big(\FR{r_1}{r_2}\Big)^{n}.
\end{align}
\end{keyeqn}
This recovers the same result originally found in \cite{Arkani-Hamed:2018kmz}. Following the same procedure, we can also derive the background part for other choices of $p_{1,2}$. In Fig.\ \ref{fig_scalarBGConv} we show the results for this series for both $p_1=p_2=-2$ and $p_1=p_2=0$, together with the results (\ref{eq_ScalarSeedBG}) from the partial MB representation, for fixed $r_2=0.5$. From this figure we can see that both methods yield convergent results for most of values of $r_1$, and they agree well with each other. We also observe that the results from partial MB representation have better convergence speed in the case of $p_1=p_2=0$, which corresponds to a realistic case with derivatively coupled external massless scalars. Therefore, we see that the partial MB result could be advantageous for numerical implementation at least in some cases.

\paragraph{Homogeneous solutions and boundary conditions.}

The full result for the scalar integral should be the sum of the above particular solution (\ref{eq_ScalarBootBG}) with a solution to the corresponding homogeneous equation (\ref{eq_ScalarBootEqnHom}). It is straightforward to solve (\ref{eq_ScalarBootEqnHom}) to get the following pair of independent solutions:
\begin{equation}
    \mathcal Y_\pm(r_1) = r_1^{1/2}\Big(\FR{r_1}2\Big)^{\pm\ii\wt\nu}\Gamma\Big[\mp\ii\wt\nu,\FR12\pm\ii\wt\nu\Big]
    {}_2\mathrm{F}_1\left[\bgm \fr14\pm\fr{\ii\wt\nu}2,\fr34\pm\fr{\ii\wt\nu}2\\1\pm\ii\wt\nu\edm\middle|\,r_1^2\right]=  \Big(\FR{r_1}2\Big)^{\pm\ii\wt\nu}\mb F_{\pm\wt\nu}^{-2}(r_1),
\end{equation}
where $\mb{F}_{\wt\nu}^p(r)$ is the function defined in (\ref{eq_Fbold}).
Here we have multiplied an $r_1$-independent coefficient for later convenience. 
Since $\mathcal Y_-(r_1)=\mathcal Y_+^*(r_1)$, the general symmetric and real homogeneous solution can be written as:
\begin{align}
\label{eq_hattedScalarSignal}
{\mathcal I}^{-2,-2}_{\text{S},>}(r_1,r_2)
  =\mathcal{Y}_+(r_1)\Big[\al_{++}\mathcal{Y}_+(r_2)+\al_{+-}\mathcal Y_-(r_2)\Big]+\text{c.c.},
  \end{align}
where $\al_{++}$ and $\al_{+-}$ are two arbitrary complex coefficients to be determined. This homogeneous solution is in fact the signal piece because of the non-analytical behaviors around $r_{1,2}\to0$ and thus we add a subscript S to ${\mathcal I}_{\text{S},>}^{-2,-2}$. The full correlator is then:
\bge
\label{eq_I00bootResult}
  {\mathcal I}^{-2,-2}_>(r_1,r_2)  = {\mathcal I}^{-2,-2}_{\text{BG},>}(r_1,r_2)+{\mathcal I}^{-2,-2}_{\text{S},>}(r_1,r_2).
\ede

To determine the coefficients $\al_{++}$ and $\al_{+-}$, we need to impose boundary conditions, which are usually found by considering some special limits of ${\mathcal I}^{-2,-2}_>(r_1,r_2)$ in $r_1$ and $r_2$. In previous works, the folded limits $r_{1,2}\to 1$ and the factorized limits $r_{1,2}\to -1$ were used to impose boundary conditions. Here let us adopt a different choice: We shall consider the squeezed limit $r_{1,2}\to0$, keeping $r_1\ll r_2$, and determine the coefficients by comparing the leading terms with the bulk integral.

In this squeezed limit, it is easy to check that the leading behaviors:
\begin{align}
{\mathcal I}^{-2,-2}_{\text{BG},>}(r_1,r_2)\to&~ \mathcal O(r_1),\\
{\mathcal I}^{-2,-2}_{\text{S},>}(r_1,r_2)\to&~
(r_1r_2)^{1/2}\bigg\{
\al_{++}\Gamma^2(-\ii\wt\nu)\Gamma^2\Big(\FR12+\ii\wt\nu\Big)\Big(\FR{r_1r_2}4\Big)^{\ii\wt\nu}\n\\
&~~~~~~~~~~+\al_{+-}|\Gamma(-\ii\wt\nu)|^2\Big|\Gamma\Big(\FR12+\ii\wt\nu\Big)\Big|^2\Big(\FR{r_1}{r_2}\Big)^{\ii\wt\nu}\bigg\}+\text{c.c.},
\end{align}
so the background contribution can be neglected.

On the other hand, we consider the same squeezed limit for the bulk integral, which is equivalent to taking the late-time limits for the propagators: $\tau_{1,2}\to 0$ with $\tau_1<\tau_2$. So the (anti-)time-ordered propagators become much simpler:
\begin{align}
D_{\pm\pm}(k_s;\tau_1,\tau_2)\to D_{\mp\pm}(k_s;\tau_1,\tau_2).
\end{align}
By comparing the result of the bulk integral (See Sec.\ \ref{sec_PMB}), we can finally determine the coefficients:
\bge
\al_{++}=\al_{+-}=\FR{1+\ii\sinh\pi\wt\nu}{2\pi}.
\ede
With these two coefficients determined, we reach the full solution (\ref{eq_I00bootResult}) to the bootstrap equation with appropriate boundary conditions. This completes the bootstrap for the scalar seed integral.

\paragraph{Folded limit and alternative version of scalar bootstrap equation.} Now let us check the folded limit $r_{1,2}\to 1^-$ of the above bootstrapped solution (\ref{eq_I00bootResult}). Our choice of Bunch-Davies initial condition for all fluctuations requires that the solution (\ref{eq_I00bootResult}) is regular in the folded limit. However,  both the signal and the background pieces are divergent in this limit. The cancellation of this superficial divergence is thus a useful consistency check of our result.

It is relatively complicated to analyze the behavior of the background piece in the folded limit with the expression \eqref{eq_ScalarBootBG}. Thus, inspired by the transverse vector integral derived in Sec.\ \ref{sec_vec}, we introduce two new variables $u_{1,2}=2r_{1,2}/(1+r_{1,2})$ as in (\ref{eq_Kratios}), and rewrite the (inhomogeneous) scalar bootstrap equation (\ref{eq_ScalarBootEqnInhom}) as:
\begin{equation}
  \label{eq_BgTaylorExpand}
    \bigg[(u_1^2-u_1^3)\pd_{u_1}^2-u_1^2\pd_{u_1}+(\wt\nu^2+\FR{1}{4}\Big)\bigg]{\mathcal I}^{-2,-2}_{\pm\pm}(u_1,u_2)=\FR12\FR{u_1u_2}{u_1+u_2-u_1u_2},
\end{equation}
 where we have rewritten ${\mathcal I}^{-2,-2}_{\pm\pm}$ as a function of $u_{1,2}$. 
Again, we Taylor expand the right hand side:
\begin{equation}
\label{eq_RHSTaylorExp}
    \FR12\FR{u_1u_2}{u_1+u_2-u_1u_2}=\FR{1}{2}\sum_{n=0}^\infty u_1^{n+1}\Big(1-\FR1{u_2}\Big)^n,
\end{equation}
and try the following ansatz which is analytical for both $u_1$ and $u_2$ (and thus both $r_1$ and $r_2$):
\begin{equation}
    \mathcal X_{\pm}(u_1,u_2)=\sum_{m,n=0}^\infty \mathcal X_{m,n}u_1^{m+n+1}\Big(1-\FR1{u_2}\Big)^n.
\end{equation}
Substituting the ansatz and the Taylor expansion back into the bootstrap equation, and matching the coefficients for each term, we obtain the following recursion relations:
\begin{subequations}
\begin{align}
  &\Big[\big(n+\fr12\big)^2+\wt\nu^2\Big]\mathcal X_{0,n}=\FR{1}{2};\\
    &\Big[\big(m+n+\fr32\big)^2+\wt\nu^2\Big]\mathcal X_{m+1,n}=(m+n+1)^2\mathcal X_{m,n},
\end{align}
\end{subequations}
and the solution can be expressed as: 
\bge
\mathcal X_{m,n}= \FR{(n+1)_{m}^2}{2(n+\fr12+\ii\wt\nu)_{m+1}(n+\fr12-\ii\wt\nu)_{m+1}},
\ede
and finally the background piece is the twice of our solution, but now in terms of $u_1$ and $u_2$:
\bge
\label{eq_ScalarBootBGu}
{\mathcal I}^{-2,-2}_{\text{BG},>}(u_1,u_2)=\sum_{m,n=0}^\infty \FR{(-1)^n(n+1)_{m}^2}{(n+\fr12+\ii\wt\nu)_{m+1}(n+\fr12-\ii\wt\nu)_{m+1}}
u_1^{m+1}(1-u_2)^n  \Big(\FR{u_1}{u_2}\Big)^n.
\ede
It is worth noting that the expression \eqref{eq_ScalarBootBGu} is equivalent to \eqref{eq_ScalarBootBG}, and we will use \eqref{eq_ScalarBootBGu} to analyze the background behavior in the folded limit.

Without loss of generality, we first consider the limit $r_2\to1^-$.
For the background piece, notice that $u_2\to1^-$ when $r_2\to1^-$ and thus only $n=0$ terms survive due to the factor $(1-u_2)^n$. Therefore, the background can be written in closed form in this limit:
\begin{align}
  \label{eq_ScaBGFolded}
\lim_{r_2\to 1^-} {\mathcal I}^{-2,-2}_{\text{BG},>}(u_1,u_2)=& \sum_{m=0}^\infty \FR{(1)_{m}^2}{(\fr12+\ii\wt\nu)_{m+1}(\fr12-\ii\wt\nu)_{m+1}}
u_1^{m+1}\n\\
=&~\pi\,\text{sech}(\pi\wt\nu) u_1\times  {}_3\wt{\mathrm{F}}_2\left[\bgm 1,1,1\\ \fr32-\ii\wt\nu,\fr32+\ii\wt\nu \edm\middle|\,u_1\right],
\end{align}
which converges for all $0\leq r_1<1$. This implies the divergences in the homogeneous solution (the signal part) must be canceled with each other. Indeed, the divergent parts for the homogeneous solutions are:
\bge
\lim_{r\to 1^-} \mathcal Y_\pm(r)  \sim \mp\ii \sqrt{\FR\pi2}\text{csch}(\pi\wt\nu)\log(1-r),
\ede
and thus these divergences get canceled in the combination:
\bge
\label{eq_ScaSignalFolded}
\lim_{r\to 1^-} \Big[\mathcal Y_+(r)+\mathcal Y_-(r)\Big] = \sqrt{2\pi^3}\,\text{sech}(\pi\wt\nu).
\ede
Therefore, the signal converges in the folded limit for $0\leq r_1 <1$:
\bge
\lim_{r_2\to1^-} {\mathcal I}^{-2,-2}_{\text{S},>}(r_1,r_2)=\sqrt{2\pi^3}\,\text{sech}(\pi\wt\nu)\al_{++} \mathcal Y_+(r_1)+\text{c.c.}.
\ede

Now we go on to take the second folded limit $r_1\to1^-$. This time, both the background and the signal diverge, but the divergences are canceled in the finally result:
\begin{align}
\lim_{r_1\to1^-} \mathcal I^{-2,-2}_{\text{BG},>}(u_1,1^-)
\sim&-\pi\,\text{sech}\log(1-u_1)\sim -\pi\,\text{sech}(\pi\wt\nu)\log(1-r_1),\\
\lim_{r_1\to1^-} {\mathcal I}^{-2,-2}_{\text{S},>}(r_1,1^-) \sim&~\pi\,\text{sech}(\pi\wt\nu)\log(1-r_1).
\end{align}
After the cancellation of divergences, we get a finite result in the double folded limit $r_{1,2}\to 1^-$:
\begin{keyeqn}
\begin{align}
\label{eq_Iscalar2pt}
 {\mathcal I}^{-2,-2}_>(1^{-},1^-)=&~2\pi\,\text{sech}(\pi\wt\nu)\text{Re}\,\Big[
\big(1+\ii\,\text{csch}(\pi\wt\nu)\big)\psi\big(\FR12-\ii\wt\nu\big)+ \gamma\Big]\n\\
&~+  \mathcal{F}\left[\bgm 1,1,\fr32-\ii\wt\nu,\fr32+\ii\wt\nu\\2,2,2 \edm\middle|\,1\right],
\end{align}
\end{keyeqn}
where $\psi(z)=\di\log\Gamma(z)/\di z$ is the digamma function. When supplemented with appropriate $\wt\nu$-independent prefactors (including the coupling and the momentum), this result is nothing but the 2-point correlator of massless scalar fields $\varphi$ corrected by an intermediate massive scalar $\si$ of mass parameter $\wt\nu$ with coupling $a^3\varphi'\si$. This result has been derived originally in \cite{Chen:2012ge}; See also \cite{Chen:2017ryl}. Although our result has a different look from the expressions in  \cite{Chen:2012ge,Chen:2017ryl}, we have checked numerically that the two results perfectly agree with each other.  

Thus we have shown that the full correlator is regular in both the single folded limit ($0\leq r_1<1$, $r_2\to 1^-$) and the double folded limit ($r_{1,2}\to1^-$), as expected.

\subsection{Transverse vector}

Next we consider the transverse components $(\lam=h=\pm)$ of the vector seed integral (\ref{eq_vecIab}), following the same procedure as before. Again we do not pursue the general case, but only focus on an example with $p_1=p_2=-1$. Generalizations to arbitrary $p_{1,2}$ should be straightforward. 

The goal here is to derive a differential equation for the following seed integral:
\begin{align}
\label{eq_vecBootInt}
  {\mathcal I}^{(h)-1,-1}_{\mathsf{ab}}(r_1,r_2) =& -\mathsf{ab} k_s \int_{-\infty}^0 \FR{\di\tau_1}{\tau_1}\FR{\di\tau_2}{\tau_2}\,
e^{\ii \mathsf a k_{12}\tau_1+\ii\mathsf b k_{34}\tau_2} D^{(h)}_{\mathsf{ab}}(k_s;\tau_1,\tau_2)\n\\
=&-\mathsf{ab} \int_{-\infty}^0 \FR{\di z_1}{z_1}\FR{\di z_2}{z_2}\, e^{\ii \mathsf a z_1+\ii\mathsf b z_2}\wh D^{(h)}_{\mathsf{ab}}(r_1z_1,r_2z_2),
\end{align}
where we have defined $z_1\equiv k_{12}\tau_1$, $z_2\equiv k_{34}\tau_2$. In the second line, we have introduced the hatted propagator for the transverse vector fields, which is again a function of $k_s\tau_1$ and $k_s\tau_2$:
\begin{equation}
\wh D^{(h)}_{\mathsf{ab}}(k_s\tau_1,k_s\tau_2)
\equiv k_s  D^{(h)}_{\mathsf{ab}}(k_s;\tau_1,\tau_2).
\end{equation}
Then, the equation of motion for the transverse spin-1 fields (\ref{eq_ModeEqTrans}) gives rise to the following equations satisfied by the propagators,
\begin{align}
&(\tau_1^2 \partial_{\tau_1}^2 +k_s^2\tau_1^2 -2h\wt\mu k_s\tau_1+ m^2)D^{(h)}_{\pm\mp}(k_s;\tau_1,\tau_2)=0,\\
&(\tau_1^2 \partial_{\tau_1}^2 +k_s^2\tau_1^2 -2h\wt\mu k_s\tau_1+ m^2)D^{(h)}_{\pm\pm}(k_s;\tau_1,\tau_2)=\mp\ii \tau_1\tau_2\delta(\tau_1-\tau_2),
\end{align}
which further imply the following equations for the hatted propagators:
\begin{align}
&(r_1^2\pd_{r_1}^2+r_1^2z_1^2-2h\wt\mu r_1z_1+m^2)\wh D_{\pm\mp}(r_1z_1,r_2z_2)=0,\\
&(r_1^2\pd_{r_1}^2+r_1^2z_1^2-2h\wt\mu r_1z_1+m^2)\wh D_{\pm\pm}(r_1z_1,r_2z_2)=\mp\ii (r_1z_1)(r_2z_2)\delta(r_1z_1-r_2z_2).
\end{align} 
We can now again insert the above differential operators in front of the hatted propagator in (\ref{eq_vecBootInt}), using the following
relations similar to (\ref{eq_IntDiffCommut1}) and (\ref{eq_IntDiffCommut2}) to commute  the differential operator with the integral:
\begin{align}
  \int_{-\infty}^0 \FR{\di z_1}{z_1}\, z_1 e^{\ii \mathsf a z_1}\wh D_{\mathsf{ab}}(r_1z_1,r_2z_2)
  =&~\ii\mathsf a (r_1\pd_{r_1})\int_{-\infty}^0 \FR{\di z_1}{z_1}\, e^{\ii \mathsf a z_1}\wh D_{\mathsf{ab}}(r_1z_1,r_2z_2),\\
  \int_{-\infty}^0 \FR{\di z_1}{z_1}\, z_1^2 e^{\ii \mathsf a z_1}\wh D_{\mathsf{ab}}(r_1z_1,r_2z_2)
  =&-(r_1\pd_{r_1}+1)(r_1\pd_{r_1})\int_{-\infty}^0 \FR{\di z_1}{z_1}\,e^{\ii \mathsf a z_1}\wh D_{\mathsf{ab}}(r_1z_1,r_2z_2),
  \end{align}
and we find the following equations satisfied by the integral: 
\begin{align}
&\bigg[(r_1^2-r_1^4)\pd_{r_1}^2-(\pm 2\ii h\wt\mu r_1^2 +2r_1^3)\pd_{r_1}+\Big(\wt\nu^2+\FR14\Big)\bigg]{\mathcal I}^{(h)-1,-1}_{\pm\mp}(r_1,r_2)=0,\\
&\bigg[(r_1^2-r_1^4)\pd_{r_1}^2-(\pm 2\ii h\wt\mu r_1^2 +2r_1^3)\pd_{r_1}+\Big(\wt\nu^2+\FR14\Big)\bigg]{\mathcal I}^{(h)-1,-1}_{\pm\pm}(r_1,r_2)=\FR{r_1r_2}{r_1+r_2}.
\end{align}
It is possible to solve this set of equation directly, with a set of homogeneous solutions and a particular solution to the inhomogeneous equation.  The resulting inhomogeneous solution has a double-series representation, with the coefficients satisfying a set of recursion relations that can be easily found. However, it turns out not trivial to find a general term formula from this set of recursion relations. Thus it is not easy to write down the explicit series solution in the same way as we did in (\ref{eq_ScalarBootBG}). Fortunately, the previous computation with partial MB representation in Sec.\ \ref{sec_vec} has hinted at a solution: We have learnt that the change of variables $u_{1,2}=2r_{1,2}/(1+r_{1,2})$ as in (\ref{eq_Kratios}) can further simplify the equations. With the new variables $u_{1,2}$, we finally reach the following set of vector bootstrap equations that are much easier to solve:
\begin{keyeqn}
\begin{align}
\label{eq_VecBootEqHomo}
&\bigg[(u_1^2-u_1^3)\pd_{u_1}^2-(1 \pm \ii h\wt\mu) u_1^2 \pd_{u_1}+\Big(\wt\nu^2+\FR14\Big)\bigg]{\mathcal I}_{\pm\mp}^{(h)-1,-1}(u_1,u_2)=0,\\
\label{eq_VecBootEqImhomo}
&\bigg[(u_1^2-u_1^3)\pd_{u_1}^2-(1\pm \ii h\wt\mu)  u_1^2 \pd_{u_1}+\Big(\wt\nu^2+\FR14\Big)\bigg]{\mathcal I}_{\pm\pm}^{(h)-1,-1}(u_1,u_2)=\FR12\FR{u_1u_2}{u_1+u_2-u_1u_2}.
\end{align}
\end{keyeqn}

\paragraph{Particular solution to the inhomogeneous equation.}

\begin{figure}
 \centering
  \includegraphics[width=0.48\textwidth]{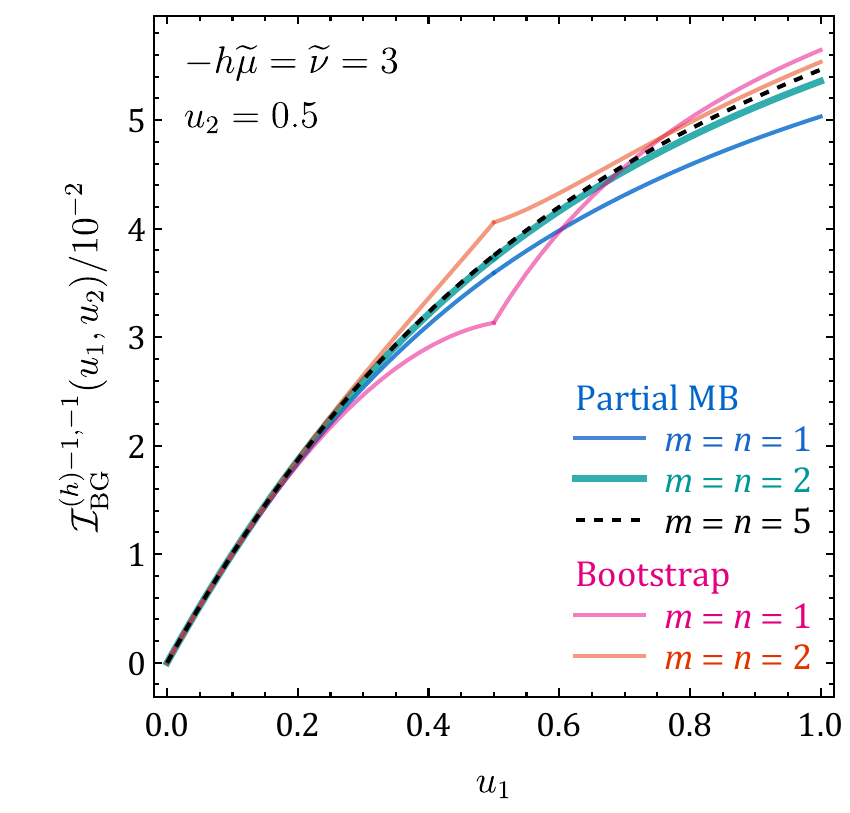}
  \includegraphics[width=0.48\textwidth]{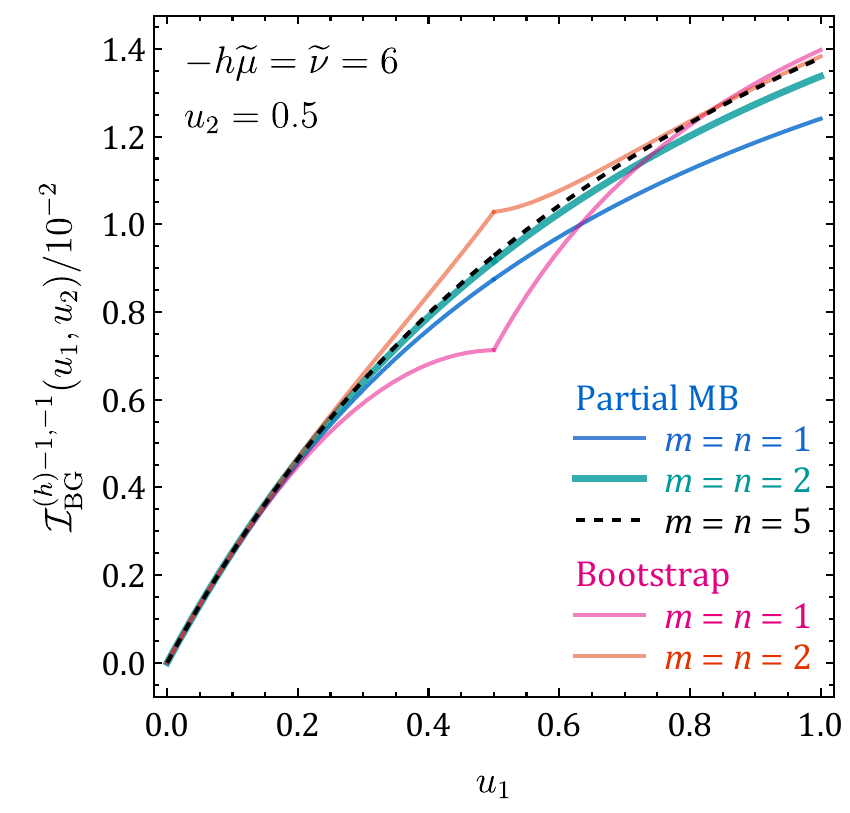}
    \caption{The background part of the vector seed integrals $\mathcal{I}_\text{BG}^{(h)-1,-1}(u_1,u_2)$ as functions of $u_1$ with fixed $u_2=0.5$. We show the convergences of series from partial MB representation (\ref{eq_TransSeedBG}) and from the bootstrap method (\ref{eq_TransBGBoot}), by including terms of $m$ and $n$ summations up to values indicated in the plots. }
  \label{fig_VecConv}
\end{figure}

Now we look for a particular solution to the inhomogeneous equation (\ref{eq_VecBootEqImhomo}). As before, we first Taylor expand the right hand side of (\ref{eq_VecBootEqImhomo}) as a series of $u_1$, already given in \eqref{eq_RHSTaylorExp}, 
and consider the following ansatz for the particular solution:  
\begin{equation}
    \mathcal V_\pm^{(h)}(u_1,u_2) = \sum_{m,n=0}^\infty \mathcal{V}_{\pm;m,n}^{(h)} u_1^{m+n+1} \Big(1-\FR{1}{u_2}\Big)^n,
\end{equation}
where the subscript in $\mathcal V_\pm^{(h)}(u_1,u_2)$ corresponds to the two choices of the SK indices in $ {\mathcal I}_{\pm\pm}^{(h)}$. (Unlike the scalar equation, the inhomogeneous equations for the vector seed integral are different for $ {\mathcal I}_{++}^{(h)}$ and $ {\mathcal I}_{--}^{(h)}$.)
Compare the coefficient for fixed $n$ and we find:
\begin{subequations}
\begin{align}
    & \Big[\big(n+\fr12\big)^2+\wt\nu^2\Big]\mathcal{V}_{\pm;0,n}^{(h)} = \FR{1}{2};\\
    & \Big[\big(m+n+\fr32\big)^2+\wt\nu^2\Big]\mathcal{V}_{\pm;m+1,n}^{(h)} = (m+n+1)(m+n+1\pm \ii h\wt\mu)\mathcal{V}_{\pm;m,n}^{(h)}.
\end{align}
\end{subequations}
This recursion relation is easy to solve, with the solution given by:
\begin{align}
    \mathcal{V}_{\pm;m,n}^{(h)} =& ~\FR{1}{2}\FR{1}{(n+\fr12)^2+\wt\nu^2} \FR{(n+1)(n+1\pm \ii h\wt\mu)}{(n+\fr32)^2+\wt\nu^2} \cdots  \FR{(n+m)(n+m\pm \ii h\wt\mu)}{(n+m+\fr12)^2+\wt\nu^2}\n\\
    =&~\FR{(n+1)_{m}(n+1\pm\ii h\wt\mu)_m}{2(n+\fr{1}{2}+\ii\wt\nu)_{m+1}(n+\fr{1}{2}-\ii\wt\nu)_{m+1}}.
\end{align}
As in the scalar solution considered before, this particular solution is the sole contribution to the background of the vector seed integral. Therefore, the full result for the background is obtained by adding up $\mathcal V_+^{(h)}$ and $\mathcal V_-^{(h)}$: 
\begin{keyeqn}
\begin{align}
\label{eq_TransBGBoot}
    {\mathcal{I}}_\text{BG}^{(h)-1,-1}(u_1,u_2)
    =&\sum_{m,n=0}^\infty\FR{(-1)^n(n+1)_{m}(n+1\pm\ii h\wt\mu)_m}{2(n+\fr{1}{2}+\ii\wt\nu)_{m+1}(n+\fr{1}{2}-\ii\wt\nu)_{m+1}}u_1^{m+1}(1-u_2)^n\Big(\FR{u_1}{u_2}\Big)^{n}+\text{c.c.}.
\end{align}
\end{keyeqn} 
In Fig.\ \ref{fig_VecConv}, we show the background series $\mathcal{I}_\text{BG}^{(h)-1,-1}$ in (\ref{eq_TransBGBoot}) together with the partial MB results (\ref{eq_TransSeedBG}). Again we see that both methods yield convergent results with good agreements.  Note that, for the parameters chosen in the figure, the signals are much greater than the background. Therefore we only show the background part of the vector seed integral in this figure, since the purpose here is to show the convergence of the background series.

\paragraph{Homogeneous solutions and limits.}
Next we consider the homogeneous solution. Note that, unlike the previous scalar case, the homogeneous and inhomogeneous equations are different for $\mathcal{I}_{+\pm}^{(h)-1,-1}$ and $\mathcal{I}_{-\mp}^{(h)-1,-1}$.  
The general solution to the upper-sign equation in (\ref{eq_VecBootEqHomo}) is the linear combination of $\mathcal{U}_{+}(u_1)$ and $\mathcal{U}_{-}(u_1)$, while the general solution to the lower-sign equation is the linear combination of $\mathcal{U}^*_{+}(u_1)$ and $\mathcal{U}^*_{-}(u_1)$, where $\mathcal{U}_{\pm}(u_1)$ is given by:
\begin{equation}
    \mathcal U_{\pm}(u_1) =u_1^{1/2\pm\ii\wt\nu}
    \mathcal F\left[\bgm \fr12\pm\ii\wt\nu,\fr12+\ii h\wt\mu\pm\ii\wt\nu\\1\pm2\ii\wt\nu\edm\middle|\,u_1\right]= \mp \ii \FR{\sqrt 2}{\pi}\sinh(2\pi\wt\nu)u_1^{\pm\ii\wt\nu} \mb{G}^{-1}_{h\wt\mu,\pm\wt\nu}(u_1).
\end{equation}
Here $\mb{G}_{h\wt\mu,\wt\nu}^{p}(u)$ is the function defined in (\ref{eq_Gbold}).
Note that we have expressed these solutions in terms of the dressed hypergeometric function (\ref{eq_HyperGeoDressed}) for later convenience. 
Furthermore, since amplitudes with different SK indices obey different equations, we can separately determine the homogeneous solutions for each ${\mathcal I}_{\mathsf{ab}}^{(h)-1,-1}$ by comparing the leading terms in the squeezed limit $r_{1,2}\to 0$ with $r_1\ll r_2$.

We first focus on the opposite-sign integrals $\mathcal I^{(h)-1,-1}_{\pm\mp}$. On one hand, $\mathcal I^{(h)-1,-1}_{+-}(u_1,u_2)$ satisfies the homogeneous equation \eqref{eq_VecBootEqHomo}
and thus must be a linear combination of $\mathcal U_+(u_1)$ and $\mathcal U_-(u_1)$. On the other hand, from (\ref{eq_vecBootInt}) we notice that $\mathcal I^{(h)-1,-1}_{+-}(u_1,u_2)=\mathcal I^{(h)-1,-1}_{-+}(u_2,u_1)$, which also satisfies \eqref{eq_VecBootEqHomo} but with the variable $u_2$.
This implies that $\mathcal I_{+-}^{(h)-1,-1}(u_1,u_2)$ should also be a linear combination of $\mathcal U_+^*(u_2)$ and $\mathcal U_-^*(u_2)$. Therefore, we can write $\mathcal I^{(h)-1,-1}_{\pm\mp}$ as:
\begin{align}
{\mathcal I}^{(h)-1,-1}_{+-}(u_1,u_2) =&~ \mathcal U_+(u_1)\Big[\beta_{++}\mathcal U^*_-(u_2)+\beta_{+-}\mathcal U^*_+(u_2)\Big]\n\\
&~+\mathcal U_-(u_1)\Big[\beta_{-+}\mathcal U^*_-(u_2)+\beta_{--}\mathcal U^*_+(u_2)\Big],\\
{\mathcal I}^{(h)-1,-1}_{-+}(u_1,u_2) =&~ {\mathcal I}^{(h)-1,-1,*}_{+-}(u_1,u_2),
\end{align}

We then consider the same-sign integrals $\mathcal I^{(h)-1,-1}_{\pm\pm}$. Similar to the scalar case, $\mathcal I^{(h)-1,-1}_{++}$ is the combination of an inhomogeneous solution (corresponding to the background piece) to \eqref{eq_VecBootEqImhomo}, and a factorized homogeneous solution denoted by ${\mathcal I}^{(h)-1,-1}_{\pm\pm,\text{F},>}$. Furthermore, from  (\ref{eq_vecBootInt}) we can see that ${\mathcal I}^{(h)-1,-1}_{++}(u_1,u_2)={\mathcal I}^{(h)-1,-1}_{++}(u_2,u_1)$, which further implies that ${\mathcal I}^{(h)-1,-1}_{++,\text{F},>}(u_1,u_2)$ can be written as a linear combination of $\mathcal U_\pm(u_1)$, as well as a linear combination of $\mathcal U_\pm(u_2)$. Therefore we can write:
\begin{align}
{\mathcal I}^{(h)-1,-1}_{++,\text{F},>}(u_1,u_2) =&~ \mathcal U_+(u_1)\Big[\gamma_{++}\mathcal U_+(u_2)+\gamma_{+-}\mathcal U_-(u_2)\Big]\n\\
&~+\mathcal U_-(u_1)\Big[\gamma_{-+}\mathcal U_+(u_2)+\gamma_{--}\mathcal U_-(u_2)\Big],\\
{\mathcal I}^{(h)-1,-1}_{--,\text{F},>}(u_1,u_2) =&~ {\mathcal I}^{(h)-1,-1,*}_{++,\text{F},>}(u_1,u_2).
\end{align}

To determine the coefficients, we again consider the squeezed limit $r_{1,2}\to0$ with $r_1\ll r_2$ and compare the behaviors of homogeneous solutions with the late-time approximation of the bulk integrals. Again, in the squeezed limit, the background piece
${\mathcal I}^{(h)-1,-1}_{\text{BG},>}(u_1,u_2)\to~ \mathcal O(u_1)$ and thus can be neglected. Comparing homogeneous solutions to the bulk integrals (See Sec.\ \ref{sec_vec}), we find:
\begin{align}
\beta_{++}=-\beta_{+-}=-\beta_{-+}=\beta_{--}=&-\FR{e^{-h\pi\wt\mu}}{4}\text{csch}^2(2\pi\wt\nu)(\cosh2\pi\wt\mu+\cosh2\pi\wt\nu),\\
\gamma_{++}=-\gamma_{+-}=&~ \FR{-\ii \pi e^{-h\pi\wt\mu}\text{csch}^2(2\pi\wt\nu)}{2 \Gamma[\fr12+\ii\wt\nu+\ii h\wt\mu,\fr12-\ii\wt\nu+\ii h\wt\mu]} e^{\pi\wt\nu}\cosh[\pi(\wt\nu+h\wt\mu)],\\
\gamma_{--}=-\gamma_{-+}=&~ \FR{-\ii \pi e^{-h\pi\wt\mu}\text{csch}^2(2\pi\wt\nu)}{2 \Gamma[\fr12+\ii\wt\nu+\ii h\wt\mu,\fr12-\ii\wt\nu+\ii h\wt\mu]} e^{-\pi\wt\nu}\cosh[\pi(-\wt\nu+h\wt\mu)],
\end{align}
and the full amplitude is given by:
\begin{align}
  {\mathcal I}_{>}^{(h)-1,-1}(u_1,u_2) =&~ \Big[{\mathcal I}_{+-}^{(h)-1,-1}(u_1,u_2)+{\mathcal I}^{(h)-1,-1}_{++,\text{F},>}(u_1,u_2)+\mathcal V_{+}(u_1,u_2)\Big]+\text{c.c.}\n\\
  =&~\bigg\{\Big[\big(\gamma_{++}\mathcal U_+(u_1)-\gamma_{--}\mathcal U_-(u_1)\big)-\beta_{++}\big(\mathcal U^*_+(u_1)-\mathcal U^*_-(u_1)\big)\Big]\Big(\mathcal U_+(u_2)-\mathcal U_-(r_2)\Big)\n\\
  &+\mathcal V_+(u_1,u_2)\bigg\}+\text{c.c.}.
\end{align}
This completes the bootstrap of the transverse vector seed integral. 

\paragraph{Folded limit.} Now we consider the folded limit $u_{1,2}\to 1^-$. First we take $u_2\to1^-$. The background piece is regular in this limit:
\begin{align}
\label{eq_VecTrans3ptBGhalf}
  \lim_{u_2\to 1^-} \mathcal V_{+}(u_1,u_2)=&~ \sum_{m=0}^\infty \mathcal V_{+;m,0}^{(h)}u_1^{m+1} 
  =\FR\pi 2\,\text{sech}(\pi\wt\nu) u_1\times  {}_3\wt{\mathrm{F}}_2\left[\bgm 1,1,1+\ii h\wt\mu\\ \fr32-\ii\wt\nu,\fr32+\ii\wt\nu \edm\middle|\,u_1\right].
\end{align}
So, any divergences in the signals must cancel among themselves. The divergent parts for the homogeneous solutions are:
\begin{align}
  \lim_{u\to1^-} \mathcal U_\pm(u) \sim
 - \FR{\ii \pi \text{csch}(h\pi\wt\mu)}{\Gamma(1-\ii h\wt\mu)}(1-u)^{-\ii h\wt\mu},
\end{align}
we find that this divergence is canceled in the following combination:
\begin{align}
  \lim_{u_2\to1^-}\Big[\mathcal U_+(r_2)-\mathcal U_-(r_2)\Big]=
  \FR{\ii\pi^2\text{csch}(h\pi\wt\mu)\big(\text{sech}[\pi(\wt\nu+h\wt\mu)]-\text{sech}[\pi(\wt\nu-h\wt\mu)]\big)}
  {\Gamma(1+\ii h\wt\mu)\Gamma(\fr12-\ii h\wt\mu-\ii\wt\nu)\Gamma(\fr12-\ii h\wt\mu+\ii\wt\nu)}.
\end{align}
On the other hand, the signal part in the folded limit $u_2\to 1^-$ can be written as
\begin{align}
  &\lim_{u_2\to 1^-} {\mathcal I}^{(h)-1,-1}_{\text{S},>}(u_1,u_2)\n\\
  =&~\Big[\big(\gamma_{++}\mathcal U_+(u_1)-\gamma_{--}\mathcal U_-(u_1)\big)-\beta_{++}\big(\mathcal U^*_+(u_1)-\mathcal U^*_-(u_1)\big)\Big]\lim_{u_2\to1^-}\Big[\mathcal U_+(u_2)-\mathcal U_-(r_2)\Big]
  +\text{c.c.},
\end{align}
which is therefore convergent, as expected. For later use, we also give an explicit expression for the signal in $u_2\to 1^-$ limit:
\begin{align}
\label{eq_VecTrans3ptSignal}
&~{\mathcal I}^{(h)-1,-1}_{\text{S},>}(u_1,1^-)\n\\
=&~ \FR{\sinh(h\pi\wt\mu)\sinh(\pi\wt\nu)\text{csch}(2\pi\wt\nu)}{\sqrt 2\pi}\n\\
&\times
\Big[
(e^{-2\pi\wt\nu}+e^{-2h\pi\wt\mu})\FR{\Gamma(\ii h\wt\mu)}{\Gamma(\fr12+\ii h\wt\mu+\ii\wt\nu)}
-\ii e^{\pi\wt\nu}(1+e^{-2\pi(\wt\nu+h\wt\mu)})\FR{\Gamma(-\ii h\wt\mu)}{\Gamma(\fr12-\ii h\wt\mu+\ii\wt\nu)}
\Big]\n\\
&\times u_1^{\ii\wt\nu} \bigg\{
\Gamma(\fr12+\ii h\wt\mu+\ii\wt\nu)\mb G^{-1}_{-h\wt\mu,\wt\nu}(u_1)
+ \ii e^{\pi\wt\nu}
\Gamma(\fr12-\ii h\wt\mu+\ii\wt\nu)\mb G^{-1}_{+h\wt\mu,\wt\nu}(u_1)
\bigg\}+\text{c.c.}.
\end{align} 

We then consider the double folded limit, by further taking $u_1\to 1^-$. Like the case of scalar correlator considered before, both the background and the signal diverge in this limit. The background diverges as
\begin{align}
  \lim_{r_1\to 1^-}{\mathcal I}^{(h)-1,-1}_{\text{BG},>}(u_1,1^-)
  \sim -\FR{\ii \pi\text{sech}(\pi\wt\nu)}{2 h\wt\mu}(1-u_1)^{-\ii h\wt\mu}+\text{c.c.},
\end{align}
while the signal diverges as
\begin{align}
  \lim_{u_1\to 1^-}{\mathcal I}^{(h)-1,-1}_{\text{S},>}(u_1,1^-)
  \sim &~(\gamma_{++}-\gamma_{-+})\FR{-\ii\pi \text{csch}(h\pi\wt\mu)}{\Gamma(1-\ii h\wt\mu)}(1-u_1)^{-\ii h\wt\mu} \times \lim_{r_2\to 1^-}\Big[\mathcal U_+(u_2)-\mathcal U_-(u_2)\Big]+\text{c.c.}\n\\
  =&~\FR{\ii \pi\text{sech}(\pi\wt\nu)}{2 h\wt\mu}(1-u_1)^{-\ii h\wt\mu}+\text{c.c.},
\end{align}
so the full correlator is regular in the double folded limit as well. After the cancellation, the finite result in the double folded limit is:
\begin{keyeqn}
\begin{align}
  &{\mathcal I}^{(h)-1,-1}_{>}(1^-,1^-)\n\\
=&~\FR{\pi \text{sech}^2(\pi\wt\nu)}{2h\wt\mu}\bigg\{\FR12(1-e^{-2h\pi\wt\mu})+\FR{\ii\cosh\pi\wt\nu}{\Gamma[\ii h\wt\mu,\fr12-\ii h\wt\mu-\ii\wt\nu,\fr12-\ii h\wt\mu+\ii\wt\nu]}\n\\
&\times \bigg(\pi e^{-h\pi\wt\mu}\Gamma(-\ii h\wt\mu)\text{sech}(\pi\wt\nu)
-\mathcal{F}\left[\bgm -\ii h\wt\mu,\fr12-\ii h\wt\mu-\ii\wt\nu,\fr12-\ii h\wt\mu+\ii\wt\nu\\1-\ii h\wt\mu,1-\ii h\wt\mu \edm\middle|\,1\right]
\bigg)
\bigg\}+\text{c.c.}.
\end{align}
\end{keyeqn}
The readers might have found it a bit unphysical to take the folded limit of the transverse seed integral. After all, the transverse component of the massive spin-1 exchange as shown in Fig.\ \ref{fd_vector} does not survive the folded limit. (For instance, the $r_2\to 1$ limit means that either $\mb k_s$ is parallel to $\mb k_4$ or $\mb k_4\to 0$. In either case, the contraction $\mb k_4\cdot \mb e_{\mb k_s}^{(\pm)^*}$ in (\ref{eq_T2}) vanishes.) However, there do exist cases where the folded limit of transverse integrals does not vanish, if we consider tensor external states. We will show such an example in Sec.\ \ref{sec_var}, where the above folded limit can be directly used.

\subsection{Longitudinal vector}

Finally we consider the longitudinal component $(\lam=L)$ of the vector seed integral (\ref{eq_vecIab}) with $p_1=p_2=-1$: 
\begin{equation}
  {\mathcal I}^{(L)-1,-1}_{\mathsf{ab}}(r_1,r_2) = -\mathsf{ab}\,k_s \int_{-\infty}^0 \FR{\di\tau_1}{\tau_1}\FR{\di\tau_2}{\tau_2}\,
e^{\ii \mathsf a k_{12}\tau_1+\ii\mathsf b k_{34}\tau_2} D^{(L)}_{\mathsf{ab}}(k_s;\tau_1,\tau_2).
\end{equation}
Similar to our computation with partial MB representation in Sec.\ \ref{sec_vec}, here we can recycle the result for the scalar bootstrap equation. Note again that the longitudinal propagator $D^{(L)}_>$ is related to the massive scalar propagator $D_{>}$ by (\ref{eq_DLtoD}). For convenience, we define the operator:
\begin{equation}
    \mathcal D_i \equiv \pd_{\tau_i} - \FR{2}{\tau_i}.\qquad (i=1,2)
\end{equation}
Then, (\ref{eq_DLtoD}) implies the following relations for the longitudinal propagators $D_{\aa\bb}^{(L)}$ and the scalar propagator $D_{\aa\bb}$ of the same mass parameter $\wt\nu$ (but not the same mass):
\begin{align}
\label{eq_DpmmpLtoD}
     D_{\pm\mp}^{(L)}(k_s;\tau_1,\tau_2) = &~\FR{1}{m^2}\mathcal D_1\mathcal D_2 D_{\pm\mp}(k_s;\tau_1,\tau_2),\\
     \label{eq_DpmpmLtoD}
     D_{\pm\pm}^{(L)}(k_s;\tau_1,\tau_2) =&~\FR{1}{m^2}
    \Big[\theta(\tau_1-\tau_2)\mathcal D_1\mathcal D_2 D_\gtrless(k_s;\tau_1,\tau_2)+\theta(\tau_2-\tau_1)\mathcal D_1\mathcal D_2 D_\lessgtr(k_s;\tau_1,\tau_2)\Big]\n\\
    =&~\FR{1}{m^2}\Big[\mathcal D_1\mathcal D_2 D_{\pm\pm}(k_s;\tau_1,\tau_2)\mp\ii  \tau_1\tau_2 \de(\tau_1-\tau_2)\Big].
\end{align}
Note in particular the appearance of a contact term $\de(\tau_1-\tau_2)$ in the last line of (\ref{eq_DpmpmLtoD}), which is a consequence of commuting the operator $\mathcal{
D}_{1,2}$ with the Heaviside $\theta$-functions in $D_{\pm\pm}$. See (in particular, App.\ B of) \cite{Chen:2017ryl} for more discussions about this contact term. 

Using the relations \eqref{eq_DpmmpLtoD} and \eqref{eq_DpmpmLtoD}, we can directly obtain the longitudinal vector seed integral by acting differential operators on the scalar seed ${\mathcal I}_{\aa\bb}^{-2,-2}$ in (\ref{eq_ScalarBootInt}). More explicitly, 
\begin{align}
\label{eq_ILfromIscalar1}
{\mathcal I}^{(L)-1,-1}_{\pm\mp}(r_1,r_2) 
=&~\FR{(r_1r_2)^{-1}}{m^2}(r_1\pd_{r_1}-2)(r_2\pd_{r_2}-2)\Big(r_1r_2{\mathcal I}_{\pm\mp}^{-2,-2}(r_1,r_2)\Big),\\
\label{eq_ILfromIscalar2}
{\mathcal I}^{(L)-1,-1}_{\pm\pm}(r_1,r_2)
=&~\FR{(r_1r_2)^{-1}}{m^2}(r_1\pd_{r_1}-2)(r_2\pd_{r_2}-2) \Big(r_1r_2{\mathcal I}_{\pm\pm}^{-2,-2}(r_1,r_2)\Big)+\FR{1}{m^2}\FR{{r_1r_2}}{r_1+r_2}.
\end{align} 
We can rewrite \eqref{eq_ILfromIscalar1} and \eqref{eq_ILfromIscalar2} as: 
\begin{align}
\label{eq_ILfromIscalar3}
{\mathcal I}^{(L)-1,-1}_{\pm\mp}(r_1,r_2) 
=&~\FR{1}{\wt\nu^2+1/4}\wt{\mathcal D}_1\wt{\mathcal D}_2{\mathcal I}_{\pm\mp}^{-2,-2}(r_1,r_2),\\
\label{eq_ILfromIscalar4}
{\mathcal I}^{(L)-1,-1}_{\pm\pm}(r_1,r_2)
=&~\FR{1}{\wt\nu^2+1/4}\bigg[\wt{\mathcal D}_1\wt{\mathcal D}_2{\mathcal I}_{\pm\pm}^{-2,-2}(r_1,r_2)+\FR{r_1r_2}{r_1+r_2}\bigg].
\end{align}
In these equations we have used the relation $m^2=\wt\nu^2+1/4$, and defined:
\begin{equation}
\label{eq_calDtilde}
    \wt{\mathcal D}_i \equiv r_i\pd_{r_i}-1.~~~~~(i=1,2)
\end{equation}
Summing over all SK indices, we can collectively write ${\mathcal{I}}_>^{(L)-1,-1}\equiv \sum_{\aa,\bb}{\mathcal{I}}_{\aa\bb}^{(L)-1,-1}$ in terms of the scalar hatted integral ${\mathcal{I}}^{-2,-2}_> \equiv \sum_{\aa,\bb}{\mathcal{I}}^{-2,-2}_{\aa\bb}$, as:
\bge
{\mathcal I}^{(L)-1,-1}_{>}(r_1,r_2)
=\FR{1}{\wt\nu^2+1/4}\bigg[\wt{\mathcal D}_1\wt{\mathcal D}_2{\mathcal I}^{-2,-2}_{>}(r_1,r_2)+\FR{2r_1r_2}{r_1+r_2}\bigg].
\ede
\paragraph{Intermediate steps.} Here we provide the intermediate steps leading to (\ref{eq_ILfromIscalar1}) and (\ref{eq_ILfromIscalar2}). Uninterested readers can skip this part. The strategy is to compute the following integral:
\begin{align}
& -\mathsf{ab}\times \FR{ k_s }{m^2}\times \int_{-\infty}^0 \FR{\di\tau_1}{\tau_1}\FR{\di\tau_2}{\tau_2}\,
e^{\ii \mathsf a k_{12}\tau_1+\ii\mathsf b k_{34}\tau_2} \Big(\pd_{\tau_1}-\FR{2}{\tau_1}\Big)
    \Big(\pd_{\tau_2}-\FR{2}{\tau_2}\Big) D_{\mathsf{ab}}(k_s;\tau_1,\tau_2)\n\\
=&-\mathsf{ab}\times \FR{ (r_1r_2)^{-1} }{m^2}\times \int_{-\infty}^0 \FR{\di z_1}{z_1}\FR{\di z_2}{z_2}\, e^{\ii \mathsf a z_1+\ii\mathsf b z_2}
\FR{r_1\pd_{r_1}-2}{z_1}
\FR{r_2\pd_{r_2}-2}{z_2}
\wh D_{\mathsf{ab}}(r_1z_1,r_2z_2)\n\\
=&~\FR{1}{m^2}\times  ( r_1r_2)^{-1} 
\times (r_1\pd_{r_1}-2)(r_2\pd_{r_2}-2)\times (-\mathsf{ab})
\int_{-\infty}^0 \FR{\di z_1}{z_1^2}\FR{\di z_2}{z_2^2}\, e^{\ii \mathsf a z_1+\ii\mathsf b z_2}
\wh D_{\mathsf{ab}}(r_1z_1,r_2z_2),
\end{align}
where $D_{\mathsf{ab}}(k_s;\tau_1,\tau_2)$ in the first line denotes the scalar propagator, and the hatted propagator $\wh D_{\aa\bb}$ is defined in (\ref{eq_hattedScalarD}). Similar to previous cases, we use the variables $z_1=k_{12}\tau_1$, $z_2=k_{34}\tau_2$. In the second line we substitute $z_i\pd_{z_i}$ with $r_i\pd_{r_i}$ acting on $\wh D_{\mathsf{ab}}(r_1z_1,r_2z_2)$.
Finally, the contact term for $\mathcal I^{(L)-1,-1}_{\pm\pm}$ arising from the delta-function is:
\bge
-\FR{ k_s }{m^2}\int_{-\infty}^0 \FR{\di\tau_1}{\tau_1}\FR{\di\tau_2}{\tau_2}\,\Big[\mp\ii\tau_1\tau_2\delta(\tau_1-\tau_2)\Big]e^{\pm\ii k_{12}\tau_1\pm\ii k_{34}\tau_2}=
-\FR{1}{m^2}\FR{ r_1r_2 }{r_1+r_2}.
\ede
Combining all results above, together with the definition of seed integrals, we can directly get (\ref{eq_ILfromIscalar1}) and (\ref{eq_ILfromIscalar2}).
\paragraph{Solutions.}
The above derivations show that the longitudinal correlator can be obtained by acting the operator (\ref{eq_calDtilde}) on the scalar correlator given in (\ref{eq_I00bootResult}), together with an extra contact term which is regular when $r_{1,2}\to 0$ and thus contributes to the background piece. Again, the longitudinal correlator can be separated into a homogeneous part (the signals) and a inhomogeneous particular solution (the background). We work them out in turn. Acting (\ref{eq_calDtilde}) on the homogeneous solution solutions to the scalar integral, we find a pair of solutions:
\bge
\mathcal W_\pm (r_1) = \wt{\mathcal D}_1 \mathcal Y_\pm(r_1)= -\Big(\FR{r_1}{2}\Big)^{\pm\ii\wt\nu} \mb H^{-1}_{\pm\wt\nu}(r_1),
\ede
where $\mb H^{p}_{\wt\nu}$ is defined in (\ref{eq_Hbold}). The homogeneous solution to the longitudinal bootstrap equation is thus a linear combination of these two solutions with appropriate boundary conditions: 
\bge
\mathcal I^{(L)-1,-1}_{\text{S},>} = \mathcal W_+(r_1)\Big[\delta_{++}\mathcal W_+(r_2)+\delta_{+-}\mathcal W_-(r_2)\Big]+\text{c.c.}.
\ede
Similar to the scalar bootstrap, we compare the solution in the squeezed limit $r_{1,2}\to 0$ with the late-time limit of the bulk integral. This determines the coefficients to be:
\bge
\de_{++}=\de_{+-} = \FR{1+\ii\sinh\pi\wt\nu}{2\pi(\wt\nu^2+1/4)}.
\ede
One can also obtain the same result by simply acting the operator $\wt{\mathcal D}_1 \wt{\mathcal D}_2/(\wt\nu^2+1/4)$ on \eqref{eq_hattedScalarSignal}. 
For the inhomogeneous solution (the background), we apply the differential operator to \eqref{eq_ScalarBootBG}, and add the contact term:
\begin{keyeqn}
\begin{align}
\label{eq_BGStoBGL}
&{\mathcal I}^{(L)-1,-1}_{\text{BG},>}(r_1,r_2) =  \FR{1}{\wt\nu^2+1/4}\bigg[\wt{\mathcal D}_1\wt{\mathcal D}_2 {\mathcal I}^{-2,-2}_{\text{BG},>}(r_1,r_2)+ \FR{2r_1r_2}{r_1+r_2}\bigg]\n\\
=&~\FR{1}{\wt\nu^2+1/4}\bigg[\sum_{m,n=0}^\infty 
  \FR{(-1)^{n+1}(n+1)_{2m}(2m+n)(n+1)}{2^{2m+1}(\fr{n}{2}+\fr{1}{4}+\fr{\ii\wt\nu}{2})_{m+1}(\fr{n}{2}+\fr{1}{4}-\fr{\ii\wt\nu}{2})_{m+1}}
  r_1^{2m+1}\Big(\FR{r_1}{r_2}\Big)^{n}+\FR{2r_1r_2}{r_1+r_2}\bigg].
\end{align}
\end{keyeqn}
Note that the last contact term is crucial for us to obtain the correct result for the background. This completes the derivation of the longitudinal correlator from the bootstrap method.

\paragraph{Folded limit.}
Finally let us check the folded limit of the longitudinal correlator. 
For simplicity, we again express the background piece in terms of $u_{1,2}$, namely inserting \eqref{eq_ScalarBootBGu} into the first line of \eqref{eq_BGStoBGL}:\begin{align}
&~{\mathcal I}^{(L)-1,-1}_{\text{BG},>}(u_1,u_2)\n\\
=&~\FR{1}{\wt\nu^2+1/4}\bigg[\wt{\mathcal D_1}\wt{\mathcal D}_2\sum_{m,n=0}^\infty \FR{[(n+1)_{m}]^2}{(n+\fr12+\ii\wt\nu)_{m+1}(n+\fr12-\ii\wt\nu)_{m+1}}
u_1^{m+n+1}\Big(1-\FR{1}{u_2}\Big)^n  + \FR{2r_1r_2}{r_1+r_2}\bigg]\n\\
=&~\FR{1}{\wt\nu^2+1/4}\bigg[\sum_{m,n=0}^\infty \FR{-(n+1)_{m}^2}{(n+\fr12+\ii\wt\nu)_{m+1}(n+\fr12-\ii\wt\nu)_{m+1}}
 \FR{(m+n-r_1)}{(1+r_1)}  \FR{n+1-r_2}{1-r_2}
\n\\
&\times u_1^{m+n+1}\Big(1-\FR{1}{u_2}\Big)^n + \FR{2r_1r_2}{r_1+r_2}\bigg].
\end{align}
We first take $r_2\to1^-$, then only terms with $n=0$ and $n=1$ survive, giving:
\begin{align}
\lim_{r_2\to1^-}{\mathcal I}^{(L)-1,-1}_{\text{BG},>}(u_1,u_2)
=&~\FR{u_1}{4\wt\nu^2+1} \bigg\{
4+u_1-\Big(\wt\nu^2-\FR74\Big)\pi\text{sech}(\pi\wt\nu)u_1\n\\
&\times
\bigg(\mathcal{F}\left[\bgm 1,1,1\\ \fr32-\ii\wt\nu,\fr32+\ii\wt\nu \edm\middle|\,u_1\right]
-(2-u_1)\mathcal{F}\left[\bgm 2,2,2\\ \fr52-\ii\wt\nu,\fr52+\ii\wt\nu \edm\middle|\,u_1\right]
\bigg)\bigg\},
\end{align}
where $\mathcal{F}$ is the dressed (generalized) hypergeometric function defined in (\ref{eq_HyperGeoDressed}).
We find the background piece converge for all $0\leq r_1<1$. On the other hand, the homogeneous solutions diverge:
\begin{align}
\lim_{r\to1^-} \mathcal W_{\pm}(r_1)\sim
\pm\ii \sqrt{\FR\pi2}\text{csch}(\pi\wt\nu)\Big[\FR{1}{1-r}-
\FR{4\wt\nu^2-7}8\log(1-r)\Big],
 \end{align}
but the combination:
\bge
\lim_{r_2\to 1^-}\Big[\mathcal W_+(r_2)+\mathcal W_-(r_2)\Big] = \FR{\pi^{3/2}}{2^{5/2}}(4\wt\nu^2-7)\text{sech}(\pi\wt\nu)
\ede
converges. Therefore,
\bge
\lim_{r_2\to1^-} {\mathcal I}^{(L)-1,-1}_{\text{S},>}(r_1,r_2)= \FR{\pi^{3/2}}{2^{5/2}}(4\wt\nu^2-7)\text{sech}(\pi\wt\nu)\delta_{++} \mathcal W_+(r_1)+\text{c.c.},
\ede
and thus the signal also converges for all $0\leq r_1<1$. 

We then take the double folded limit by also taking $r_1\to1^-$. Again, both the background and the signal diverge, but the divergences are canceled:
\begin{align}
\lim_{r_1\to1^-} {\mathcal I}^{(L)-1,-1}_{\text{BG},>}(u_1,1^-)
\sim& ~\FR{\pi(4\wt\nu^2-7)}{4(4\wt\nu^2+1)}\,\text{sech}(\pi\wt\nu)\Big[\FR{1}{1-u_1}-
\FR{4\wt\nu^2-7}4\log(1-u_1)\Big],\\
\sim& ~\FR{\pi(4\wt\nu^2-7)}{2(4\wt\nu^2+1)}\,\text{sech}(\pi\wt\nu)\Big[\FR{1}{1-r_1}-
\FR{4\wt\nu^2-7}8\log(1-r_1)\Big],\\
 {\mathcal I}^{(L)-1,-1}_{\text{S},>}(r_1,1^-) \sim&-\FR{\pi(4\wt\nu^2-7)}{2(4\wt\nu^2+1)}\,\text{sech}(\pi\wt\nu)\Big[\FR{1}{1-r_1}-
\FR{4\wt\nu^2-7}8\log(1-r_1)\Big],
\end{align}
which shows that the longitudinal correlator is regular in the folded limit, as expected.
More explicitly,
\begin{keyeqn}
\begin{align}
&{\mathcal I}^{(L)-1,-1}_>(1^-,1^-)\n\\
=&~\FR{5}{4\wt\nu^2+1}-\FR{3}{16}(4\wt\nu^2-7)\pi\,\text{sech}(\pi\wt\nu)\n\\
&+
\FR{(4\wt\nu^2-7)^2}{8(4\wt\nu^2+1)}\pi\,\text{sech}(\pi\wt\nu)\,
\text{Re}\,\Big[
\big(1+\ii\,\text{csch}(\pi\wt\nu)\big)\psi\big(\FR12-\ii\wt\nu\big)+\ga\Big]\n\\
& - \FR{4\wt\nu^2-7}{4(4\wt\nu^2+1)} \bigg\{
\mathcal{F}\left[\bgm 1,1,\fr32-\ii\wt\nu,\fr32+\ii\wt\nu\\2,2,2 \edm\middle|\,1\right]
-
\mathcal{F}\left[\bgm 1,1,\fr52-\ii\wt\nu,\fr52+\ii\wt\nu\\3,3,3 \edm\middle|\,1\right]\bigg\}.
\end{align}
\end{keyeqn}
Again, this result can be directly used to express the 2-point function mediated by a massive longitudinal spin-1 field. 

\section{Variations: Higher Spins and Bispectra}
\label{sec_var}

\begin{figure}
\centering
  \includegraphics[width=0.35\textwidth]{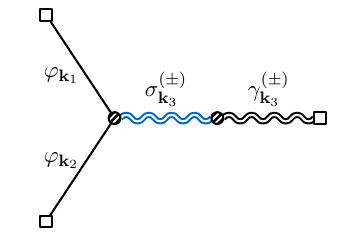}
  \caption{The helical 3-point correlator (\ref{eq_phiphigamma}). The straight lines represent external inflaton modes, the black doubly wiggly line represents the massless tensor mode, and the blue doubly wiggly line represents the massive spin-2 field with helical chemical potential.}
  \label{fd_mixed3pt}
\end{figure}

In this work, we have been mainly focusing on the 4-point correlators with spin-1 exchange. Nevertheless, it is straightforward to generalize our methods and results to other cases, such as the chemical potential for higher spin fields $(s\geq 2)$ and the 3-point function. As an example, in this section, we shall briefly consider the 3-point function mediated by a massive spin-2 field with a helicity-dependent chemical potential. Note that the 3-point functions with massive exchange such as in Fig.\ \ref{fig_fd_3ptScalar} and Fig.\ \ref{fd_mixed3pt} are formally obtained by taking $k_4\to 0$ in corresponding 4-point functions. In this limit, $k_s= k_3$, and thus $r_2=u_2=1$. Thus taking the 3-point limit in the seed integrals is formally equivalent to taking the folded limit $r_2\to 1$. The way we take folded limits of 4-point functions in Sec.\ \ref{sec_boot} allows us to find closed analytical results for the 3-point functions without any series expansion. Thus we also take this chance to present a similar closed-form analytical expression for the well-studied scale-exchange diagram (Fig.\ \ref{fig_fd_3ptScalar}), which seems new to us.

\paragraph{Helical chemical potential for higher spins.}

Before presenting the details, we first make two general remarks. 

The first remark is about arbitrary nonzero spin. In principle, the chemical potential reviewed in Sec.\ \ref{sec_HCP} has a natural generalization to massive fields of any nonzero spin. Let us consider integer spins for simplicity, and we can describe massive spin-$s$ states by a rank-$s$ symmetric tensor $\si_{\mu_1\cdots \mu_s}$. A helicity-dependent chemical potential for this field can thus be introduced from the following dimension-5 coupling between $\si_{\mu_1\cdots\mu_s}$ and a background scalar field $\phi$ \cite{Tong:2022cdz}:
\bge
\label{eq_CPspinS}
  \mathcal{O}^{(s)}=\FR{1}{2\Lambda}\ep^{\mu\nu\rho\si}\phi\nabla_\mu \si_{\nu\lam_{2}\cdots\lam_s}\nabla_\rho\si_{\si}{}^{\lam_2\cdots\lam_s}.
\ede
In particular, for a massive spin-2 field $\si_{\mu\nu}$, we have:
\bge
\label{eq_CPspin2}
  \mathcal{O}^{(2)}=\FR{1}{2\Lambda}\ep^{\mu\nu\rho\si}\phi\nabla_\mu \si_{\nu\lam}\nabla_\rho\si_{\si}{}^{\lam}.
\ede
Then, with the rolling background $\la\phi\ra=\dot\phi_0t+\text{const.}$, these operators generate terms proportional to the particle number operator weighted by the helicity. Unlike the chemical potential operator for spin-1 field in (\ref{eq_ActionVec}), the above operators break the gauge symmetry for the corresponding massless spin-$s$ fields. This does not place any obstacle, since the gauge symmetries are absent for massive fields anyway. In fact, for the spin-2 case, one might consider an alternative way of introducing the chemical potential via the operator $\phi\ep^{\mu\nu\rho\si}R_{\mu\nu\ka\lam}R_{\rho\si}{}^{\ka\lam}$ where $R_{\mu\nu\rho\si}$ is the Riemann tensor associated with $\si_{\mu\nu}$ (but not with the spacetime metric $g_{\mu\nu}$). This operator is indeed capable of generating a helicity-dependent chemical potential for the spin-2 field in the perturbative regime, as has been extensively studied in the literature. However, it is  known that this operator is pathologic in the UV \cite{Dyda:2012rj}, and our mechanism to boost the particle production (and thus the inflation correlators) works exactly at or beyond the UV cutoff scale of this operator. Therefore, we shall only consider (\ref{eq_CPspin2}). See \cite{Tong:2022cdz} for more discussions. 

The second remark is about the 3-point function. In a 3-point function with a massive spinning particle exchange at the tree level, it is necessary that the intermediate massive particle is attached to the external inflaton fluctuation through a two-point mixing. Then the angular momentum conservation at this two-point vertex implies that only the longitudinal component of the spinning particle can make nonzero contributions. On the other hand, the helicity-dependent chemical potential only boosts transverse components of the massive spinning particles, and does absolutely nothing to the longitudinal component. Therefore, we do not expect to see the chemical potential boosted helical correlators in tree-level 3-point functions. (It is possible to generate chemical potential boosted 3-point function at one-loop as has been studied in the literature \cite{Chen:2018xck,Wang:2019gbi,Hook:2019zxa,Hook:2019vcn,Wang:2020ioa}.) However, there is a unique chance where we can see the effect of the helical chemical potential in tree-level 3-point functions. That is, we can consider a mixed correlator $\la\varphi\varphi\ga\ra$ among two external inflaton fluctuations $\varphi$ and a tensor mode $\ga$. Then, we can form a two-point mixing between the massless tensor mode $\ga_{\mu\nu}$ and the massive spin-2 field $\si_{\mu\nu}$. Due to the spin-2 nature of the external tensor mode, this two-point mixing will pick out the helicity-2 component of $\si_{\mu\nu}$ rather than the longitudinal component. Therefore, there is a chance that this mixed operator is boosted by the chemical potential. This example has been considered in \cite{Tong:2022cdz}, in which the signal part of the mixed 3-point function has been calculated. With our results from previous sections, we can now easily derive the full analytical expression for this process. 

In this section, we shall focus only on the helicity-2 states of the massive spin-2 field, which is the dominant contribution to the inflation correlators when there is a nonzero  chemical potential. 
For the massive spin-2 field $\si_{\mu\nu}$ of mass $m$ and with a chemical potential term (\ref{eq_CPspin2}), the mode equation corresponding to the helicity-2 component $\Sigma^{(\pm2)}(k,\tau)$ reads:
\bge
  \Sigma^{(\pm2)}{}''-2a \Sigma^{(\pm2)}{}'+\Big[k^2\pm 2a\mu k  +a^2(m^2-2 )\Big]\Sigma^{(\pm2)}=0,
\ede 
where $\mu\equiv\dot\phi_0/\Lambda$ is the chemical potential. From this equation, we find the properly normalized mode function as (Our convention for normalizing the mode function and the polarization tensors is in accordance with \cite{Tong:2022cdz}.)
\bge
  \Sigma^{(\pm 2)}(k,\tau) = -\FR{e^{\mp \pi\wt\mu/2}}{2\sqrt{k}\tau}\mathrm{W}_{\pm \ii\wt\mu,\ii\wt\nu}(2\ii k\tau),
\ede 
where $\mathrm{W}_{\mu,\nu}(z)$ is again the Whittaker W function, and $\wt\mu\equiv \mu/H$ and $\wt\nu\equiv\sqrt{m^2/H^2-9/4}$. (Here we temporarily restore the Hubble parameter $H$. Note that the definition of the mass parameter $\wt\nu$ differs from the rest of the paper, due to the spin of the particle. In general, for a spin-$s$ state with $s\neq 0$, the mass parameter is defined by $\wt\nu\equiv\sqrt{m^2/H^2-(s-1/2)^2}$, and for $s=0$, it is defined by $\wt\nu\equiv\sqrt{m^2/H^2-9/4}$.)

With the above mode function, we can construct the transverse spin-2 propagator as in previous sections:
\begin{equation}
\label{eq_Dspin2}
    D^{(\pm 2)}_{>} (k;\tau_1,\tau_2)
    =\FR{1}{2\tau_1\tau_2} \FR{e^{\mp \pi\wt\mu}}{2k}\mathrm{W}_{\pm \ii\wt\mu,\ii\wt\nu}(2\ii k\tau_1)\mathrm{W}_{\mp \ii\wt\mu,-\ii\wt\nu}(-2\ii k\tau_2).
\end{equation}

\paragraph{Mixed bispectrum.}

To form a 3-point function at the tree level, we need a two-point mixing between the massive spin-2 field $\si_{\mu\nu}$ and the massless tensor mode $\ga_{ij}$, as well as 3-point vertex connecting $\si_{\mu\nu}$ with two massless inflaton fluctuations $\varphi$. In accordance with \cite{Tong:2022cdz}, we introduce the following couplings:
\begin{equation}
    \Delta \ld = \lam_2 a\gamma_{ij}'\si_{ij}-\FR{1}{2}\lam_3\si_{ij}\pd_i\varphi \pd_j\varphi.
\end{equation}

Then, the three-point function shown in Fig.\ \ref{fd_mixed3pt} can be written as
\begin{align}
\label{eq_phiphigamma}
  \la \varphi(\mb k_1)\varphi(\mb k_2)\gamma^{(\pm 2)}(\mb k_3)\ra'= \mathcal{K}^{(\pm 2)}(\theta_{13},\theta_{23})\mathcal{J}^{(\pm 2)}(k_1,k_2,k_3),
\end{align}
where $\mathcal{K}^{(\pm 2)}(\theta_{13},\theta_{23})$ is the kinematic factor arising from contracting the polarization tensor $e_{ij,\mb k_3}^{(\pm 2)}$ with the external momenta:
\bge
  \mathcal{K}^{(\pm 2)}(\theta_{13},\theta_{23})=e_{ij,\mb k_3}^{(\pm 2)}\wh{k}_{1i}\wh{k}_{2j}=-\sin\theta_{13}\sin\theta_{23},
\ede
with $\theta_{i3}$ $(i=1,2)$ being the angles between $\mb k_i$ and $\mb k_3$. So the kinematic factor depends only on the angles, and turns out to be independent of $h=\pm 2$. On the other hand, the factor $\mathcal J^{(\pm 2)}_{\mathsf{ab}}$ characterizes the shape dependence, and is our main focus. Using the diagrammatic rule, it can be written as the following SK integral:
\begin{align}
    \mathcal J^{(\pm 2)}_{\mathsf{ab}} 
    =&-\mathsf{ab} \lam_2\lam_3k_1k_2\int_{-\infty}^0 \di\tau_1 \FR{\di\tau_2}{(-\tau_2)}G_{\aa}(k_1,\tau_1)G_{\aa}(k_2,\tau_1)\pd_{\tau_2}T_\bb(k_3,\tau_2)
    D_{\mathsf{ab}}^{(\pm2)}(k_3;\tau_1,\tau_2).
\end{align}
Here $G_{\aa}(k,\tau)$ is the bulk-to-boundary propagator of the inflaton fluctuation, given in (\ref{eq_BtoBprop}), and $T_{\aa}(k,\tau)$ is the bulk-to-boundary propagator of the massless tensor mode, given by
\bge
  T_\aa(k,\tau)=\FR{1}{4k^3}(1-\ii\aa k\tau)e^{\ii\aa k\tau}.
\ede
Substituting the various propagators into above expression, and noting that the transverse spin-2 propagator (\ref{eq_Dspin2}) is related to the transverse spin-1 propagator (\ref{eq_VecPropPM}) via
\bge
  D_>^{(\pm 2)}(k;\tau_1,\tau_2)=\FR{1}{2\tau_1\tau_2} D_>^{(\pm)}(k;\tau_1,\tau_2),
\ede
it is straightforward to rewrite the above 3-point function in terms of the vector seed integral (\ref{eq_vecIab}) as:
 \begin{keyeqn}
\begin{align}
\label{eq_graviton3pt}
   \mathcal J^{(\pm 2)}_{\mathsf{ab}} = -\FR{\lam_2\lam_3}{32(k_1k_2k_3)^2}\Big(1+r\pd_{r}+\FR{k_1k_2}{k_{12}^2}(r^2\pd_r^2+2r\pd_r)\Big){\mathcal{I}}_{\aa\bb}^{(\pm)-1,-1}(r,1^-),
\end{align}
\end{keyeqn}
where we have defined $r\equiv k_3/k_{12}$.
 
\begin{figure}[t]
 \centering
  \includegraphics[height=0.38\textwidth]{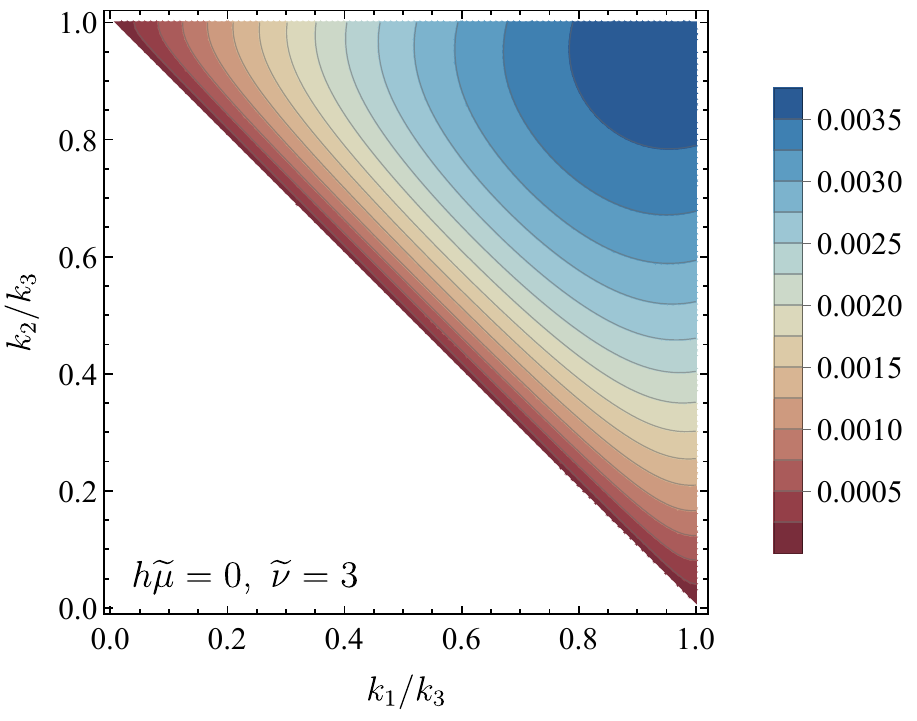} 
  \includegraphics[height=0.38\textwidth]{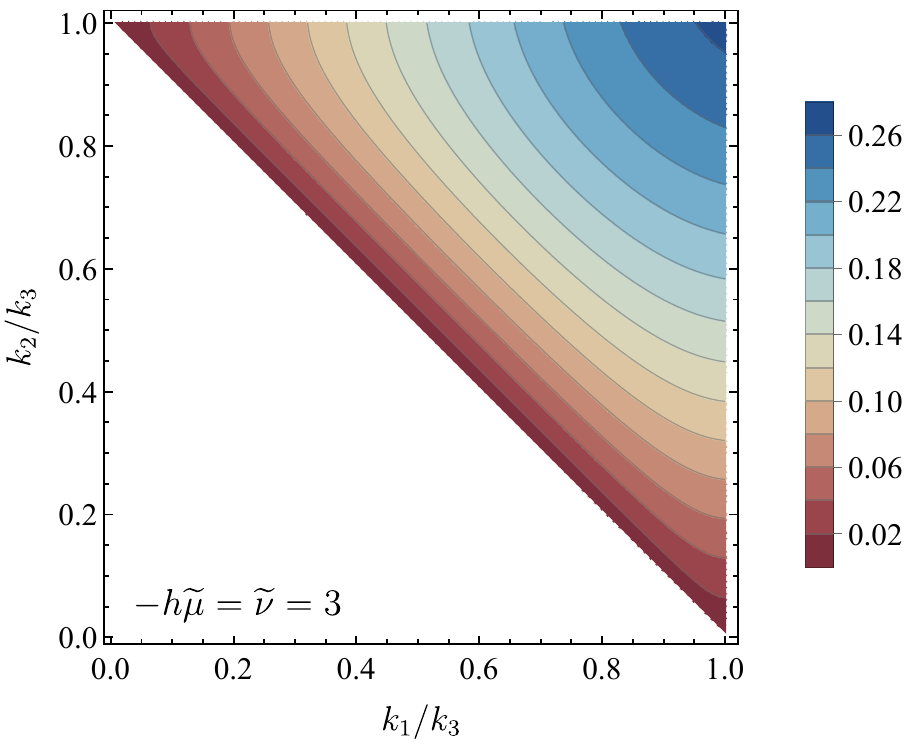}  

\caption{The mixed bispectrum $(k_1k_2k_3)^2\sum_{\aa,\bb}\mathcal{J}_{\aa\bb}^{(\pm 2)}$ mediated by the helicity $h=\pm2$ component of a massive spin-2 field with mass parameter $\wt\nu$ and chemical potential $\wt\mu$.}
\label{fig_3ptEquil} 
\end{figure}

\begin{figure}[t]
 \centering
  \includegraphics[width=0.49\textwidth]{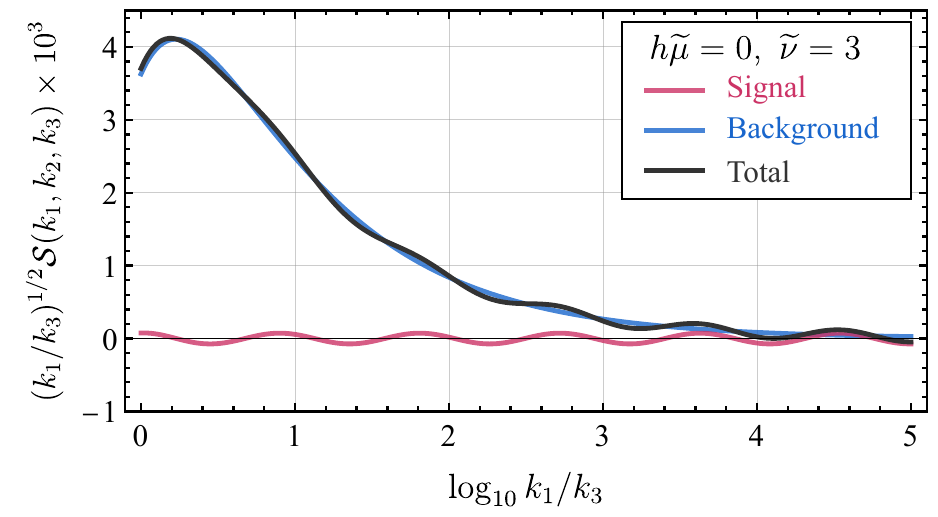} 
  \includegraphics[width=0.49\textwidth]{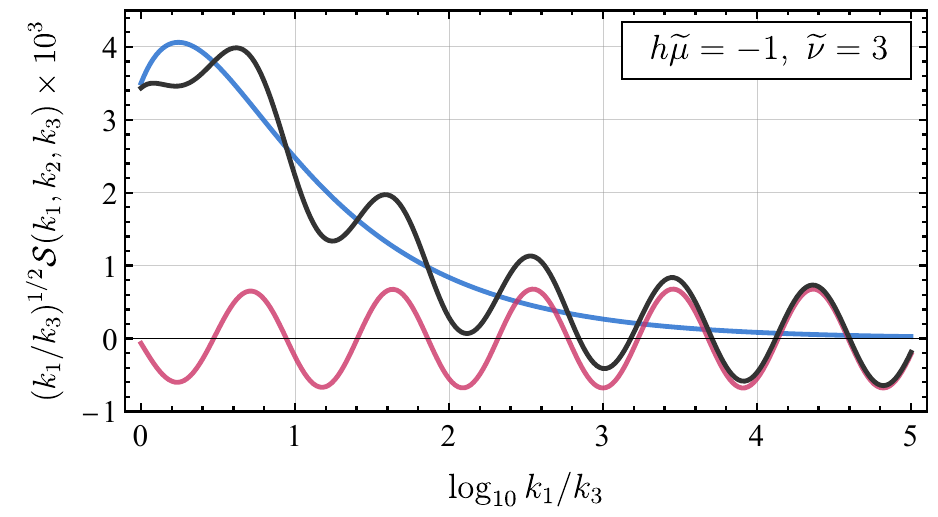} \\
  \includegraphics[width=0.49\textwidth]{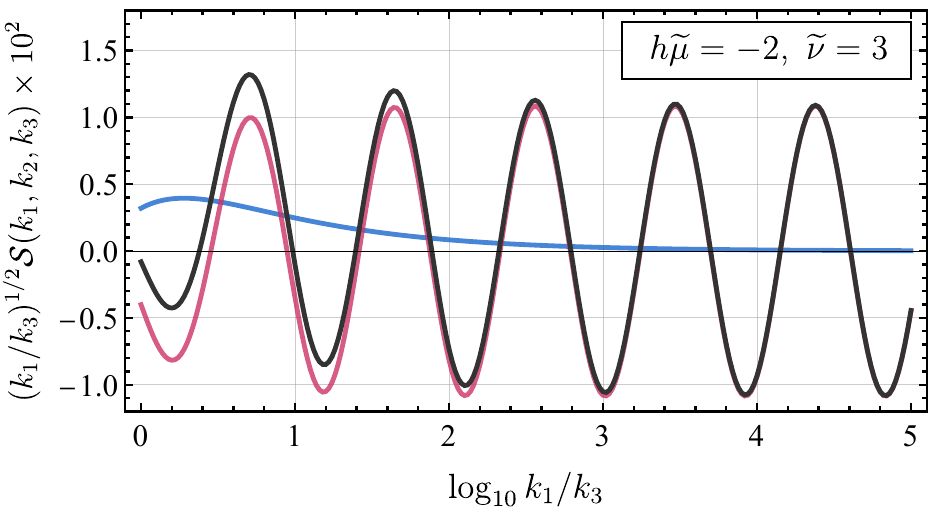} 
  \includegraphics[width=0.49\textwidth]{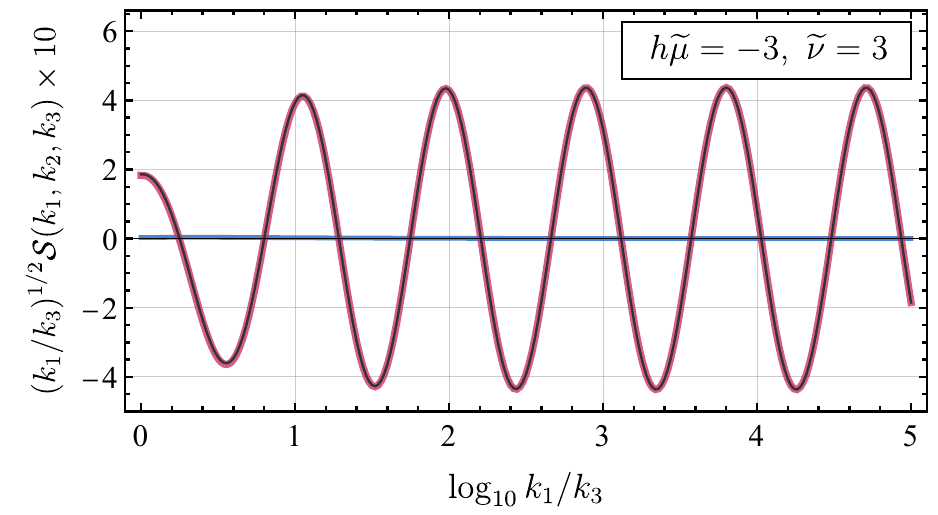} 

\caption{The mixed bispectrum $\mathcal{S}(k_1,k_2,k_3)\equiv(k_1k_2k_3)^2\sum_{\aa,\bb}\mathcal{J}_{\aa\bb}^{(\pm 2)}$ mediated by the helicity $h=\pm2$ component of a massive spin-2 field with mass parameter $\wt\nu$ and chemical potential $\wt\mu$. In this figure we choose isosceles and squeezed  configurations with $k_1=k_2>k_3$.}
\label{fig_3ptSqueezed} 
\end{figure}

Now we can make use of our folded limit results of the transverse vector seed integral derived in Sec.\ \ref{sec_boot}. Insert \eqref{eq_VecTrans3ptSignal} and \eqref{eq_VecTrans3ptBGhalf} into 
\eqref{eq_graviton3pt},\footnote{A gentle reminder that \eqref{eq_VecTrans3ptSignal} and \eqref{eq_VecTrans3ptBGhalf} are written as functions of $u$. We should rewrite all $u=2r/(1+r)$ when inserting them into \eqref{eq_graviton3pt}.} and also recall that ${\mathcal I}^{(h)-1,-1}_{\text{BG},>} = \mathcal V_+ + \text{c.c.}$, we will obtain the signal piece and the background piece of the 3-point function $\mathcal J^{(\pm)}$, respectively.
The final result is relatively long, thus we will not write it down explicitly. Instead, we plot the results for near equilateral configurations in Fig.\ \ref{fig_3ptEquil} and for squeezed configurations in Fig.\ \ref{fig_3ptSqueezed}. From Fig.\ \ref{fig_3ptEquil}, we see that the bispectrum is approximately in equilateral shape, and the shape has weak dependence on the chemical potential. On the contrary, we see from Fig.\ \ref{fig_3ptSqueezed} that the chemical potential has a huge impact on the oscillatory signal: The whole result is dominated by the background when $\wt\mu=0$ and dominated by the signal when $|\wt\mu|$ increases to $\wt\nu$.

\paragraph{Scalar bispectrum.} Finally, for completeness, we also write down an explicit expression for the scalar 3-point function mediated by a massive scalar as shown in Fig.\ \ref{fig_fd_3ptScalar}. This 3-point function has been written in terms of the scalar seed integral in (\ref{eq_BispecToI}), where we need to take the folded limit $r_2\to 1$.
In principle we can use our results for general scalar seed integrals, and set $r_2\to 1^-$ to obtain the correct bispectrum. However, the background will be expressed in the form of a series,
and the convergence speed becomes slower when $r_1\to1^-$.
On the other hand, in Sec.\ \ref{sec_boot} we have derived a closed form for the background piece in the folded limit, so it is more convenient to use the bootstrap results when we analyze the bispectrum.
We need only one more step to take $\mathcal{I}^{-2,-2}_>$ obtained in Sec.\ \ref{sec_boot} to $\mathcal{I}_>^{0,-2}$ as required by (\ref{eq_BispecToI}). This can be conveniently done by acting appropriate differential operators:
\bge
{\mathcal I}^{0,-2}_{>}(r_1,r_2) =- k_s^{2}\pd_{k_{12}}^2 {\mathcal I}^{-2,-2}_>\Big(\FR{k_s}{k_{12}},\FR{k_s}{k_{34}}\Big)
=-r_1^2 (r_1^2\pd_{r_1}^2+2r_1\pd_{r_1}){\mathcal I}^{-2,-2}_>(r_1,r_2).
\ede
Therefore, when we go to the folded limit $r_2\to 1^-$, we obtain:
\bge
{\mathcal I}^{0,-2}_{>}(r_1,1^-) =-r_1^2 (r_1^2\pd_{r_1}^2+2r_1\pd_{r_1}){\mathcal I}^{-2,-2}_>(r_1,1^-).
\ede
Now we insert \eqref{eq_ScaSignalFolded} and \eqref{eq_ScaBGFolded}, we obtain the bispectrum:
\begin{keyeqn}
\begin{align}
\label{eq_Scalar3ptFullResult}
\mathcal B_{\phi}
=& -\FR{\lam \lam_2}{8 k_1 k_2 k_3^4} \pi\text{sech}(\pi\wt\nu) \bigg\{\FR{1+\ii\sinh\pi\wt\nu}{\sqrt{2\pi}} \Big(\FR r2\Big)^{\ii\wt\nu}\mb F_{\wt\nu}^{0}(r)+
\FR{1-\ii\sinh\pi\wt\nu}{\sqrt{2\pi}}   \Big(\FR r2\Big)^{-\ii\wt\nu}\mb F_{-\wt\nu}^{0}(r)\n\\
&+\FR12 \Big(\FR{2r}{1+r}\Big)^3 
{}_3\wt{\mathrm{F}}_2\left[\bgm 1,1,1\\\fr32-\ii\wt\nu,\fr32+\ii\wt\nu\edm\middle|\,\FR{2r}{1+r}\right]
+\Big(\FR{2r}{1+r}\Big)^4 {}_3\wt{\mathrm{F}}_2\left[\bgm 2,2,2\\\fr52-\ii\wt\nu,\fr52+\ii\wt\nu\edm\middle|\,\FR{2r}{1+r}\right]\n\\
&+2\Big(\FR{2r}{1+r}\Big)^5 {}_3\wt{\mathrm{F}}_2\left[\bgm 3,3,3\\\fr72-\ii\wt\nu,\fr72+\ii\wt\nu\edm\middle|\,\FR{2r}{1+r}\right]
 \bigg\},
\end{align}
\end{keyeqn}
where $\mb{F}_{\wt\nu}^{p}(r)$ is defined in (\ref{eq_Fbold}), and $r\equiv k_3/k_{12}$.

\section{Phenomenology}
\label{sec_pheno}

A major motivation for the study of inflation correlators is their potential observability in future observations of primordial non-Gaussianities. In particular, the presence of the helicity-dependent chemical potential opens up the possibility of scale invariant oscillatory signals with large amplitudes and large oscillation frequency at the same time, unlike most of the non-Gaussian shapes considered before. More importantly, a measurement of these oscillatory shapes would provide invaluable information about heavy particles and their interactions at the inflation scale, which is the main idea of CC physics. 

The CC phenomenology of the helical inflation correlators has been studied in previous works in the context of 1-loop SM and BSM processes \cite{Chen:2018xck,Wang:2019gbi,Hook:2019zxa,Hook:2019vcn,Wang:2020ioa}, CP-violation \cite{Liu:2019fag}, higher spins \cite{Tong:2022cdz}, the signal phase \cite{Qin:2022lva}, etc. The exact and analytical results we obtained in this work enables us to fast and precisely determine the shapes and sizes of the signal and the background as functions various model parameters, which would be useful for future phenomenological studies. Below, we shall discuss implications of our results for the study of CC signals. Since there have been quite a few works studying the phenomenological aspects of the chemical potential, we shall not be comprehensive in this section, but will mostly focus on the aspects most directly related to the results obtained in previous sections.

\subsection{Cosmological collider signals in the squeezed limit}

The so-called CC signals consist of oscillatory shapes in the logarithms of various momentum ratios in certain soft configurations of the inflation correlators. In the case of 4-point function, the most relevant signals come from the squeezed limit $r_{1,2}\to 0$, or equivalently, $u_{1,2}\to 0$. Therefore, let us first look at the signal integrals, including the local signal (\ref{eq_TransSeedL}) and (\ref{eq_TransSeedNL}) for the transverse polarizations, and the corresponding signal integrals (\ref{eq_LongSeedL}) and (\ref{eq_LongSeedNL}) for the longitudinal polarization. 

\paragraph{Signals from transverse polarizations.} The transverse signals are of most interest, due to their exponential sensitivity to the chemical potential. To see this point, let us take the squeezed limit $u_{1,2}\to 0$ of (\ref{eq_TransSeedL}) and (\ref{eq_TransSeedNL}). For simplicity, we shall only consider the case of $p_1=p_2=0$.
Furthermore, notice that $\mathcal G^{(h)00}_{\text{L}/\text{NL}}(u_1,u_2)\propto (u_1u_2)^{3/2}$ in the squeezed limit. So, we remove the factor $(u_1u_2)^{3/2}$ by defining:
\bge
\label{eq_mathcalGtowhmathcalG}
\mathcal G^{(h)00}_{\text{L}}(u_1,u_2) = (u_1u_2)^{3/2}\wh{\mathcal G}^{(h)00}_{\text{L}}(u_1,u_2),\qquad
\mathcal G^{(h)00}_{\text{NL}}(u_1,u_2) = (u_1u_2)^{3/2}\wh{\mathcal G}^{(h)00}_{\text{NL}}(u_1,u_2).
\ede
We note that the factor $(u_1u_2)^{3/2}$ goes as $k_s^3$ in the squeezed limit $k_s\to 0$, which is canceled by the factor $1/k_s^3$ in the final result for the correlator such as $\mathcal{T}_2$ in (\ref{eq_T2fromDI}). So the signals do not decay away in the squeezed limit, where their sizes are controlled by the hatted quantities $\wh{\mathcal G}^{(h)00}_{\text{L}}$ and $\wh{\mathcal G}^{(h)00}_{\text{NL}}$.
Then the squeezed limit can be expressed as 
\begin{align}
  \lim_{u_1,u_2\to 0}\mathcal{I}_{\text{L},>}^{(h)00}(u_1,u_2)
  =&~(u_1u_2)^{3/2}\Big|\wh{\mathcal G}^{(h)00}_{\text{L}}(0,0)\Big|\cos\Big[\wt\nu\log\FR{u_1}{u_2}+\text{Arg}\,\wh{\mathcal G}^{(h)00}_{\text{L}}(0,0)\Big],\\
  \lim_{u_1,u_2\to 0}\mathcal{I}_{\text{NL},>}^{(h)00}(u_1,u_2)
  =&~(u_1u_2)^{3/2}\Big|\wh{\mathcal G}^{(h)00}_{\text{NL}}(0,0)\Big|\cos\Big[\wt\nu\log(u_1u_2)+\text{Arg}\,\wh{\mathcal G}^{(h)00}_{\text{NL}}(0,0)\Big].
\end{align}
Therefore, the local and nonlocal signals manifest as oscillations in $\log(u_1/u_2)$ and $\log(u_1u_2)$, respectively, with the frequency given by the mass parameter $\wt\nu$. Their sizes in the squeezed limit are respectively controlled by the coefficients $|\wh{\mathcal G}^{(h)00}_{\text{L}}(0,0)|$ and $|\wh{\mathcal G}^{(h)00}_{\text{NL}}(0,0)|$, while the complex phases $\text{Arg}\,\wh{\mathcal G}^{(h)00}_{\text{L}}(0,0)$ and $\text{Arg}\,\wh{\mathcal G}^{(h)00}_{\text{NL}}(0,0)$ control the phases of the signal. The signal phase of this process has been more carefully studied in \cite{Qin:2022lva}. Here we shall only focus on the signal size. 

From the explicit expressions (\ref{eq_calGL}) and (\ref{eq_calGNL}), together with the definition \eqref{eq_mathcalGtowhmathcalG}, we can easily read the squeezed limits $u_{1,2}\to 0$ of the functions $\wh{\mathcal{G}}_\text{L}^{(h)p_1p_2}(u_1,u_2)$ and $\wh{\mathcal{G}}_\text{NL}^{(h)p_1p_2}(u_1,u_2)$. The results are:
\begin{align}
  \wh{\mathcal{G}}_\text{L}^{(h)00}(0,0)
  =&~\FR{\pi(4\wt\nu^2+1)}{32\wt\nu(e^{4\pi\wt\nu}-1)}(e^{2\pi(\wt\nu-h\wt\mu)}+e^{2\pi\wt\nu}+\ii e^{\pi\wt\nu}-\ii e^{3\pi\wt\nu}), \\
  \wh{\mathcal{G}}_\text{NL}^{(h)00}(0,0)
  =&~\FR{\ii\pi^2 e^{\pi(3\wt\nu-h\wt\mu)}}{2(e^{2\pi\wt\nu}-1)^2(e^{\pi\wt\nu}-\ii)^2}\FR{\Gamma^2(\fr{3}{2}+\ii\wt\nu)}{\Gamma^2(1+2\ii\wt\nu)\Gamma(\fr{1}{2}+h\ii\wt\mu-\ii\wt\nu)\Gamma(\fr{1}{2}-h\ii\wt\mu-\ii\wt\nu)}.
\end{align}
In particular, we can find the absolute values of these functions, as
\begin{align}
\label{eq_calGLabs}
  \Big|\wh{\mathcal{G}}_\text{L}^{(h)00}(0,0)\Big|
  =&~\FR{\pi(4\wt\nu^2+1)}{32\wt\nu(e^{4\pi\wt\nu}-1)}\sqrt{e^{6\pi\wt\nu}+e^{4\pi\wt\nu}(e^{-4\pi h\wt\mu}+2e^{-2\pi h\wt\mu}-1)+e^{2\pi\wt\nu}}, \\
\label{eq_calGNLabs}
  \Big|\wh{\mathcal{G}}_\text{NL}^{(h)00}(0,0)\Big|
  =&~\FR{\pi e^{-\pi h\wt\mu}(4\wt\nu^2+1)}{64\wt\nu}\text{csch}(2\pi\wt\nu)\sqrt{2\cosh2\pi\wt\mu+2\cosh2\pi\wt\nu}.
\end{align}
Thus we see that the signal sizes are exponentially sensitive to both the mass parameter $\wt\nu$ and the chemical potential $\wt\mu$. The dependence on the chemical potential is of particular interest. We can see that, depending on the relative sizes of $\wt\mu$ and $
\wt\nu$, different terms in (\ref{eq_calGLabs}) and (\ref{eq_calGNLabs}) dominate the result. It is straightforward to get the following approximated expressions, which make clear the exponential dependences on $\wt\mu$ and $\wt\nu$ for different intervals of parameter space: 
\begin{keyeqn} 
\begin{align}
  \Big|\wh{\mathcal{G}}_\text{L}^{(h)00}(0,0)\Big|
  \simeq&~\FR{\pi(4\wt\nu^2+1)}{32\wt\nu} \times
  \begin{cases}
     e^{-2\pi(\wt\nu+h\wt\mu)},  &(h\wt\mu<-\fr{1}{2}\wt\nu)\\
     e^{-\pi\wt\nu}. &(h\wt\mu\geq -\fr{1}{2}\wt\nu)
  \end{cases}\\
  \Big|\wh{\mathcal{G}}_\text{NL}^{(h)00}(0,0)\Big| 
  \simeq&~\FR{\pi(4\wt\nu^2+1)}{32\wt\nu} \times
  \begin{cases}
     e^{-2\pi(\wt\nu+h\wt\mu)}, &(h\wt\mu<-\wt\nu)\\
     e^{-\pi(\wt\nu+h\wt\mu)},  &(-\wt\nu\leq h\wt\mu < \wt\nu)\\
     e^{-2\pi\wt\nu}.  &(h\wt\mu \geq \wt\nu)
  \end{cases} 
\end{align}
\end{keyeqn}

\begin{figure}
 \centering
  \includegraphics[width=0.45\textwidth]{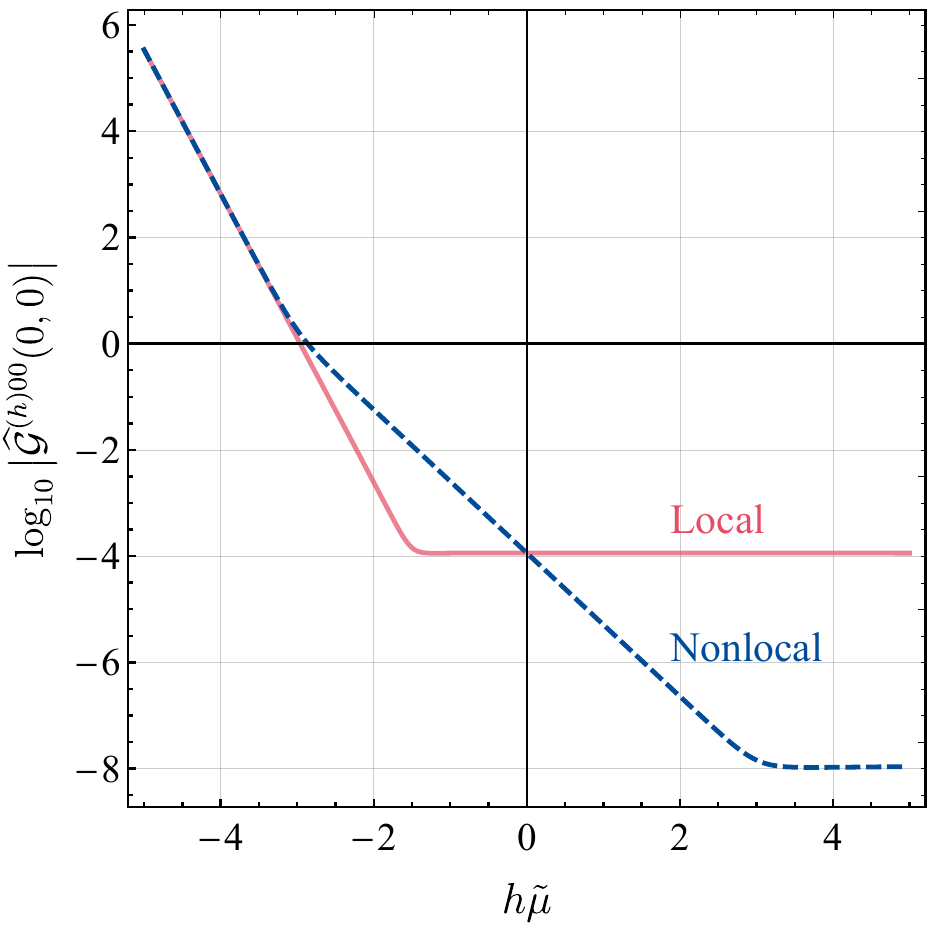} 
  \caption{The sizes of local and nonlocal signals from the helical 4-point function mediated by a massive spin-1 state with helicity $h=\pm 1$, mass parameter $\wt\nu=3$, and chemical potential $\wt\mu$.}
  \label{fig_SignalSize}
\end{figure}

In Fig.\ \ref{fig_SignalSize}, we plot the signal sizes (\ref{eq_calGLabs}) and (\ref{eq_calGNLabs}) for fixed mass parameter $\wt\nu=3$ as functions of helicity-weighted chemical potential $h\wt\mu$. The broken-exponential behavior is evident from the figure. As expected, the presence of a chemical potential will exponentially enhance the signal with $h\wt\mu<0$ and suppress the signal with $h\wt\mu>0$. On top of this general feature, there appears some finer structures as well. For instance, the degree of exponential enhancement is different for nonlocal signal with $h\wt\mu<0$ for $|\wt\mu|>\wt\nu$ and $|\wt\mu|<\wt\nu$. Also, the exponential suppressions for the nonlocal signal is saturated at $|\wt\mu|=\wt\nu$. On the other hand, the local signal never receives exponential suppression for $h\wt\mu>0$, but receives exponential enhancement for $h\wt\mu<-\wt\nu/2$. 

In particle model buildings, it often happens that the particles of both helicities $h=\pm 1$ are present in the spectrum, and therefore the overall signal size is the sum of the  contributions from both helicities. As a result, we can see from Fig.\ \ref{fig_SignalSize} that the nonlocal signal always dominates over the local signal when $|\wt\mu|<\wt\nu$, but is of comparable size with the local signal when $|\wt\mu|\geq \wt\nu$.\footnote{In \cite{Tong:2021wai} it was mentioned that the local signal is insensitive to the chemical potential. It seems to us that this conclusion was drawn by looking at the small chemical potential limit $\wt\mu\ll \wt\nu$. Our result shown in Fig.\ \ref{fig_SignalSize} confirms this observation around $h\wt\mu=0$ but shows otherwise for $|\wt\mu|>\wt\nu/2$. } 

\paragraph{Signals from the longitudinal polarization.} 
The signals from the longitudinal polarization is simpler, and is totally independent of the chemical potential. To find the signal sizes, we again expand the signal integrals (\ref{eq_LongSeedL}) and (\ref{eq_LongSeedNL}) in the squeezed limit $r_{1,2}\to 0$, and take $p_1=p_2=0$ for simplicity. Similar to the transverse case, the signal is governed by the functions $\mathcal{G}_\text{L}^{(L)00}$ and $\mathcal{G}_\text{NL}^{(L)00}$ at $r_1=r_2=0$. 
 Similarly, to remove the dilation factor in $\mathcal G^{(L)00}_{\text{L}}$ and  $\mathcal G^{(L)00}_{\text{NL}}$, we define:
\bge
\mathcal G^{(L)00}_{\text{L}}(r_1,r_2) = (r_1r_2)^{3/2}\wh{\mathcal G}^{(L)00}_{\text{L}}(r_1,r_2),\qquad
\mathcal G^{(L)00}_{\text{NL}}(r_1,r_2) = (r_1r_2)^{3/2}\wh{\mathcal G}^{(L)00}_{\text{NL}}(r_1,r_2).
\ede 
From the definition above and the expressions in (\ref{eq_calGLongL}) and (\ref{eq_calGLongNL}), we get 
\begin{align}
  \wh{\mathcal{G}}_\text{L}^{(L)00}(0,0) 
  =&~\FR{{\pi}(1+4\wt\nu^2)}{4\wt\nu}\big(1-\ii\sinh\pi\wt\nu\big)\,\text{csch}\, 2\pi\wt\nu,\\
   \wh{\mathcal{G}}_\text{NL}^{(L)00}(0,0) 
  =&~\FR{1+4\wt\nu^2}{8{\pi}}\Gamma^2(-\ii\wt\nu)\Gamma^2\big(\fr{1}{2}+\ii\wt\nu\big)\big(1-\ii\sinh\pi\wt\nu\big).
\end{align} 
In particular, the signal sizes are given by the absolute values of these functions, which turn out to be the same for both the local and the nonlocal signals: 
\begin{align}
   \Big|\wh{\mathcal{G}}_\text{L}^{(L)00}(0,0)\Big|
   =\Big|\wh{\mathcal{G}}_\text{NL}^{(L)00}(0,0)\Big|
  = \FR{{\pi}(1+4\wt\nu^2)}{8\wt\nu} \,\text{csch}\,  \pi\wt\nu .
\end{align}
As expected, the signal size is exponentially damped $\propto e^{-\pi\wt\nu}$ for large mass parameter $\wt\nu$. 

We note that in both the transverse and longitudinal signal expressions, there appears a divergence as the mass parameter $\wt\nu\to 0$, which corresponds to the borderline between the principal and complementary series. While some enhancement of the signal is expected for $\wt\nu\ll 1$, the divergence itself must be unphysical, since the perturbative expansion itself breaks down if we come too close to $\wt\nu=0$. Therefore we expect that the divergence around this point should be removed by resumming higher order diagrams. We leave this issue for future studies.

\subsection{Background} 

We next look at the background piece, given by (\ref{eq_TransSeedBG}) for the transverse polarizations and by (\ref{eq_LongSeedBG}) for the longitudinal polarization. Again we will only consider the case $p_1=p_2=0$. The background piece is not as easy to compute or even to estimate compared with the signal part. The late-time expansion of the massive propagator that is often used to estimate the signal size is inapplicable for background estimation. On the other hand, we know that the background piece corresponds to the local EFT contributions after we integrate out the intermediate heavy particle. Therefore, in the case of a dS covariant correlator mediated by a normal particle of mass $m$ (without dS boost-breaking chemical potentials), we know that the background contribution should be proportional to $1/m^2$ in the large $m$ limit. However, it is less clear if the chemical potential would affect this estimate, especially when the chemical potential becomes large. 

With the full analytical result at hand, we can give this question a definite answer. In short, the background piece is dependent on the chemical potential, but in a rather weak way: There is no exponential enhancement of the background even when the chemical potential is large. Therefore, with our exact results, we confirm a previous expectation that the CC signal could be parametrically larger than the background for scenarios with chemical potential enhancement.  

Since the longitudinal correlator is in any case independent of the chemical potential, here we will only consider the transverse background piece (\ref{eq_TransSeedBG}).

\begin{figure}[t]
 \centering
  \includegraphics[height=0.42\textwidth]{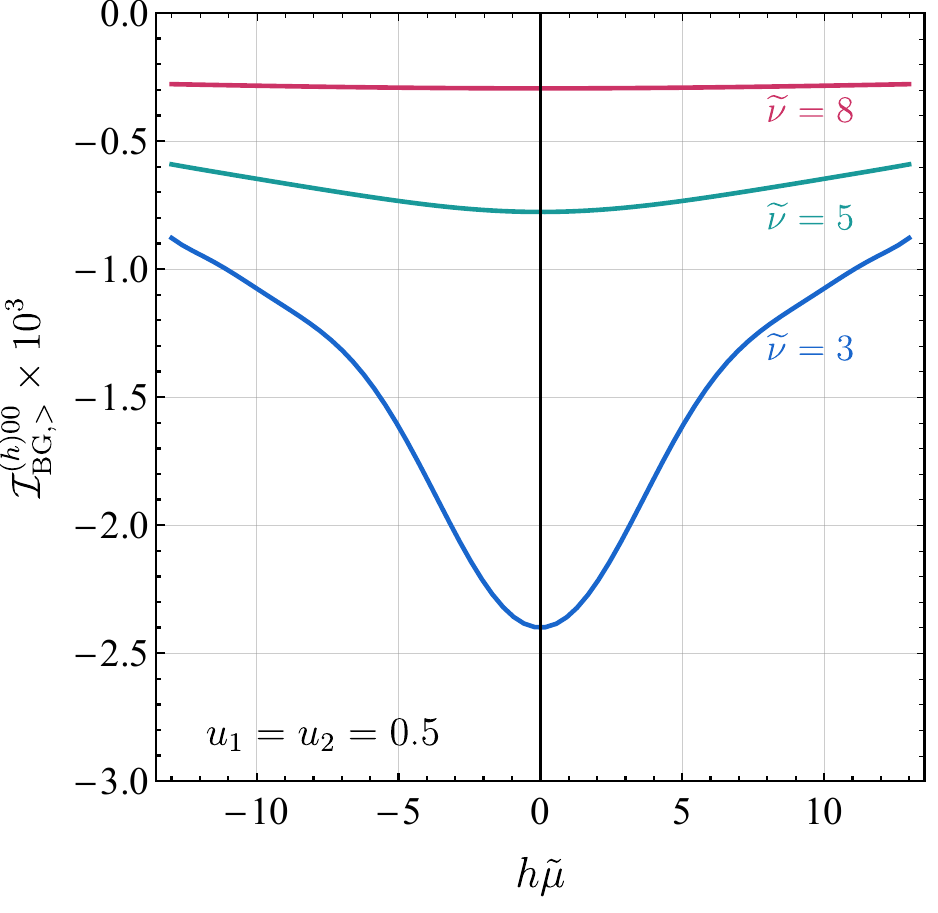} 
  \includegraphics[height=0.42\textwidth]{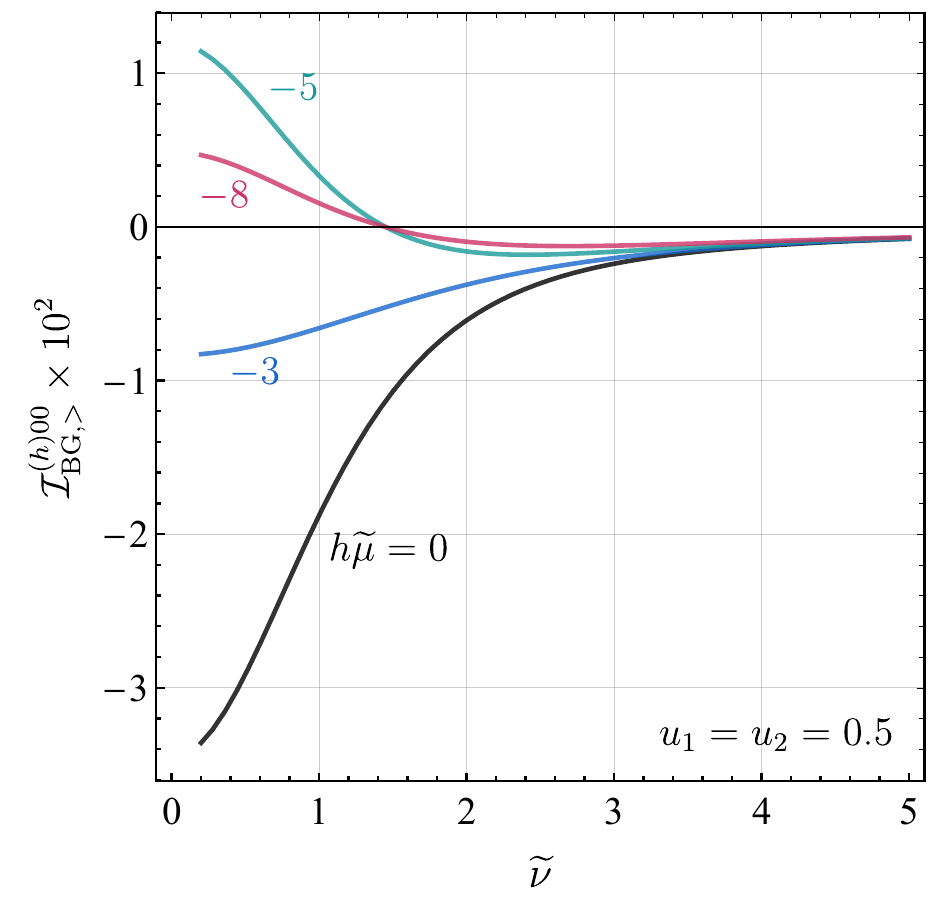} 
  \caption{The transverse component of the background integral $\mathcal{I}_{\text{BG},>}^{(h)00}$ in (\ref{eq_TransSeedBG}) at $u_1=u_2=0.5$. LEFT: $\mathcal{I}_{\text{BG},>}^{(h)00}$ as functions of helicity-weighted chemical potential $h\wt\mu$ for several choices of mass parameters $\wt\nu$. RIGHT: $\mathcal{I}_{\text{BG},>}^{(h)00}$ as functions of $\wt\nu$ for several values of $h\wt\mu$.}
  \label{fig_BGvsmu}
\end{figure}

A first hint that the background piece is insensitive to the chemical potential appears when we look at the squeezed limit $r_1<r_2\ll 1$, where only the $n_1=n_2=0$ term contribute to the series in (\ref{eq_TransSeedBG}): 
\begin{align}
  \lim_{r_1\ll 1} \mathcal{I}_{\text{BG},>}^{(h)00} 
  =\FR{\ii u_1^3}{8\wt\nu}\mathcal{F}\bigg[\bgm \fr{3}{2}-\ii\wt\nu,3\\ \fr{5}{2}-\ii\wt\nu\edm\bigg|-\FR{u_1}{u_2}\bigg]+\text{c.c.}.
\end{align} 
In particular, when we take a ``hierarchical'' squeezed limit $r_1\ll r_2$ (which implies $r_1\ll 1$ since $r_2<1$ but $r_2$ is not required small), the above expression gets further simplified to 
\begin{align}
  \lim_{r_1\ll r_2}\mathcal{I}_{\text{BG},>}^{(h)00} =-\FR{u_1^3}{9+4\wt\nu^2} .
\end{align} 
Therefore we see that the background piece is completely independent of the chemical potential in the squeezed limit. If the background piece were exponentially sensitive to chemical potential for nonsqueezed configurations, and in particular, were exponentially enhanced when the chemical potential is large, then it would follow that the background piece has an exponential dependence on the various momentum ratios, a behavior that is unlikely to be generated by local EFT operators. Therefore we expect that the chemical potential dependence in the background piece, if present at all, should be a weak one. 

\begin{figure}[t]
 \centering
  \includegraphics[width=0.48\textwidth]{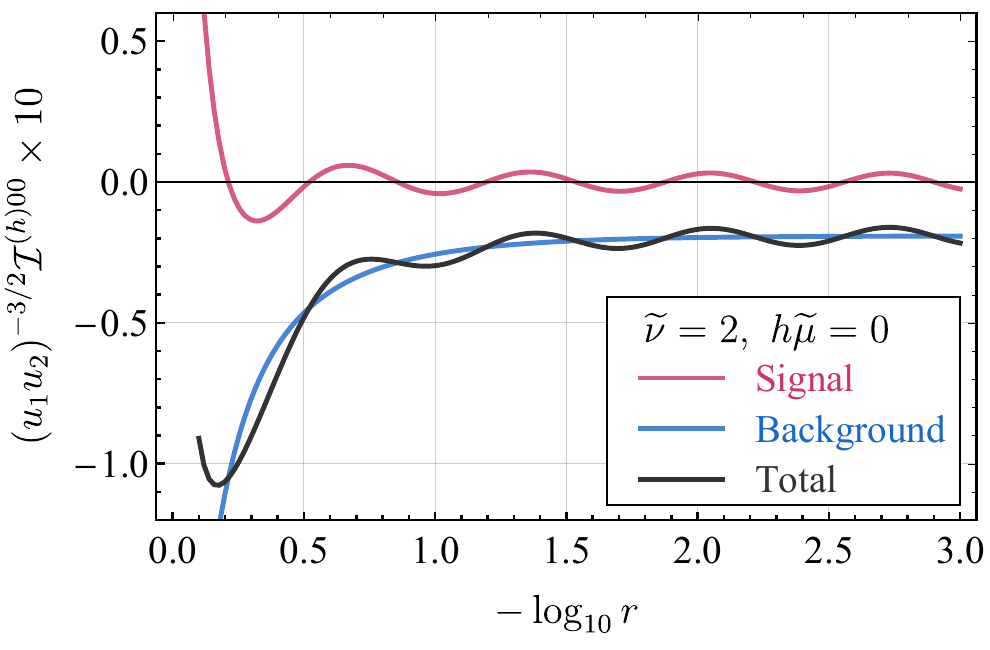} 
  \includegraphics[width=0.48\textwidth]{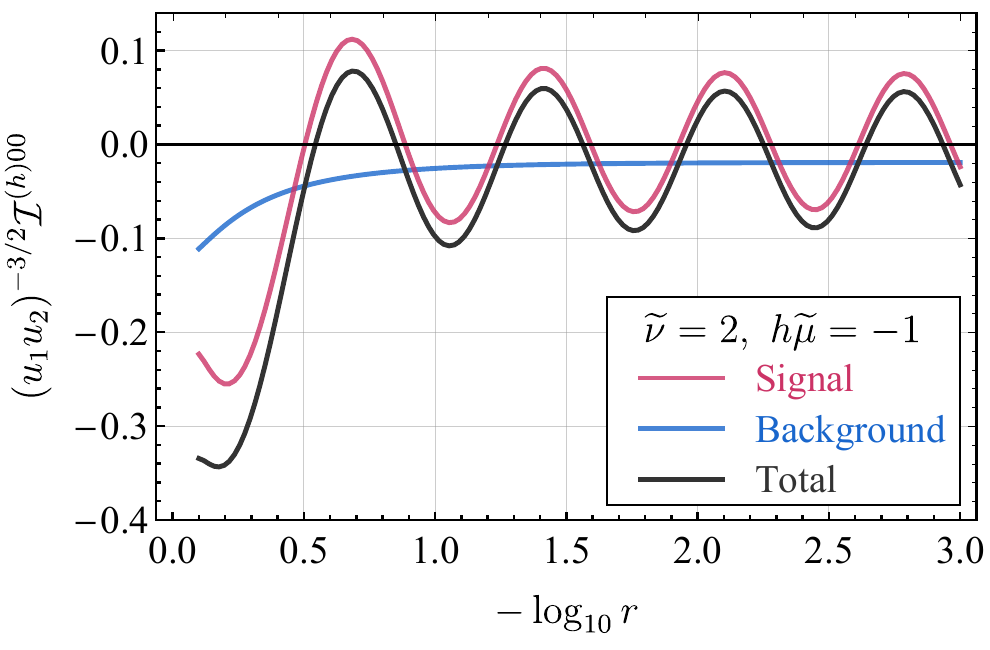}\\ 

\caption{The transverse vector seed integrals for $r_1=r_2\equiv r$ (namely $k_{12}=k_{34}$) as functions of $r$. The magenta curve corresponds to the signal integral, namely $\mathcal{I}_{\text{L},>}^{(h)00}+\mathcal{I}_{\text{NL},>}^{(h)00}$ in (\ref{eq_TransSeedL}) and (\ref{eq_TransSeedNL}), the blue curve shows the background integral $\mathcal{I}_{\text{BG},>}^{(h)00}$ in (\ref{eq_TransSeedBG}), and the black curves show the sum of the signal and the background. In both panels we choose $\wt\nu=2$, and we choose $h\wt\mu=0$ for the left panel and $h\wt\mu=-1$ for the right panel. }
\label{fig_NLsignal} 
\end{figure}

This expectation is confirmed by explicitly plotting the chemical potential dependence for nonsqueezed configurations. In Fig.\ \ref{fig_BGvsmu}, we show the dependence of the background integral (\ref{eq_TransSeedBG}) of the transverse component on the helicity-weighted chemical potential $h\wt\mu$ and also on the mass parameter $\wt\nu$. In this figure we choose a nonsqueezed configuration $u_1=u_2=0.5$. It is clear from the figure that the chemical potential affects the size of the background piece (for a nonsqueezed configuration) only in a rather mild way (i.e., no order of magnitude changes), even when the chemical potential $\wt\mu$ becomes greater than the mass parameter $\wt\nu$.

From the above discussion of the signal strength and the background size, we find a particularly interesting feature of the CC signals from chemical potential mechanism: When we increase the chemical potential from zero to a value comparable to or greater than the mass parameter, the size of the oscillatory signal will be exponentially enhanced, while the smooth background piece is largely insensitive to the change of the chemical potential. Therefore, the signal can be parametrically larger than the background. We illustrate this feature in Fig.\ \ref{fig_NLsignal} in which we show both the signal and the background for configurations with $r_1=r_2$. For the parameter $\wt\nu=2$ used in the figure, it is evident that the background dominates the total result when $h\wt\mu=0$ and the signal already dominates over the total result when we choose $h\wt\mu=-1$. Had we chosen $|h\wt\mu|\geq 2$, the background would be negligibly small compared to the signal and the difference between the signal and the total result would be invisible.

\begin{figure}
 \centering
  \includegraphics[height=0.4\textwidth]{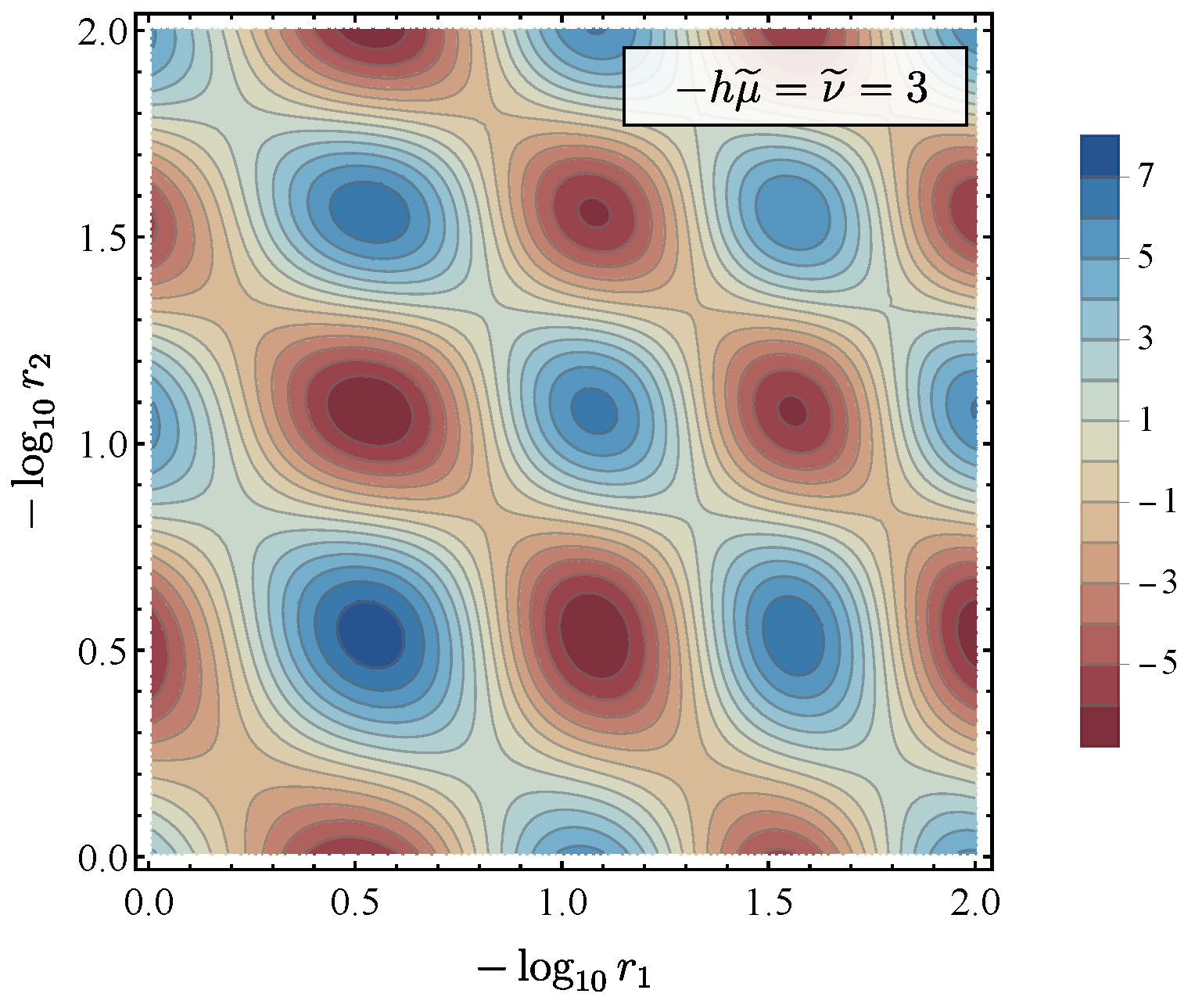} 
  \includegraphics[height=0.4\textwidth]{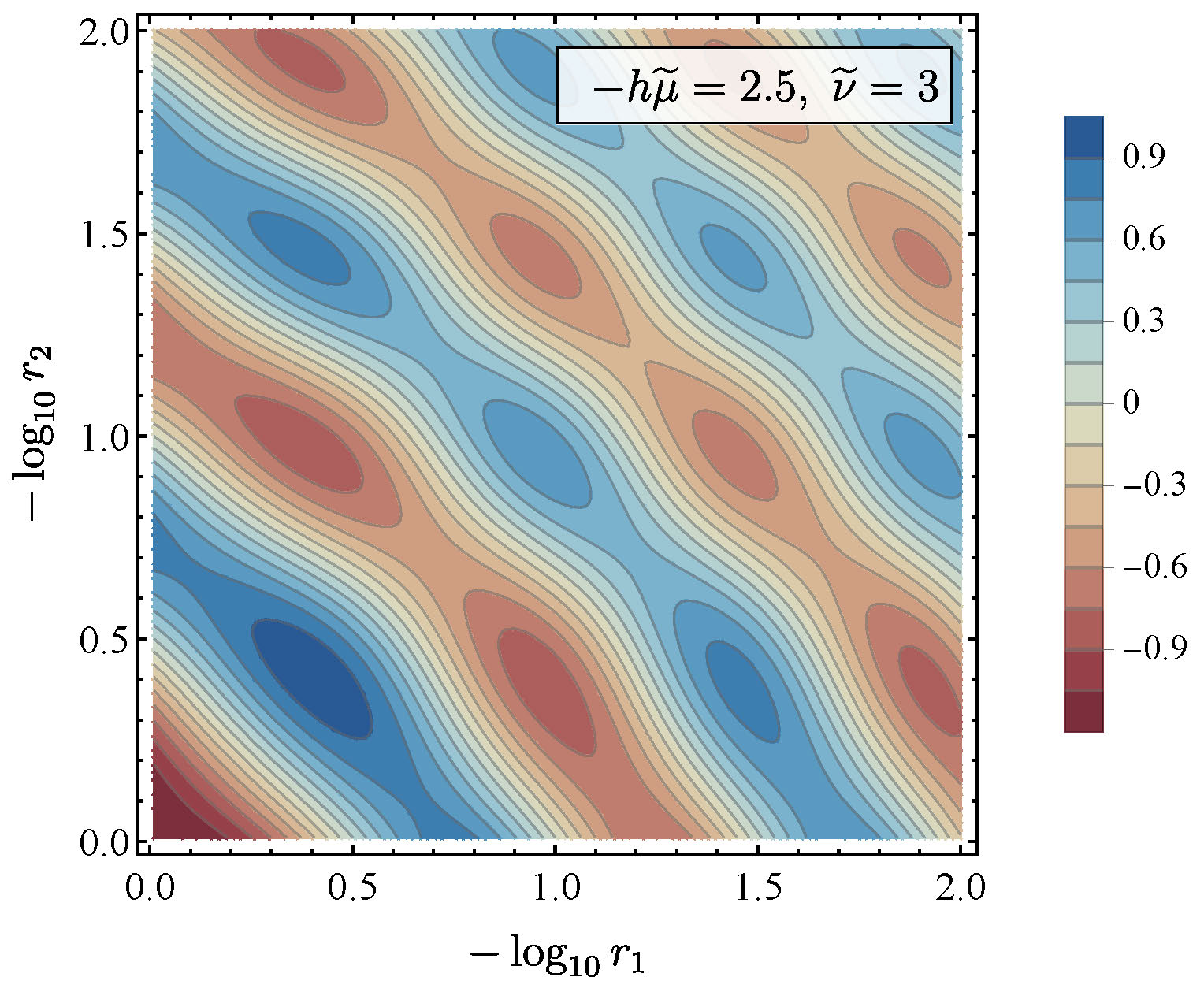}\\
  \includegraphics[height=0.4\textwidth]{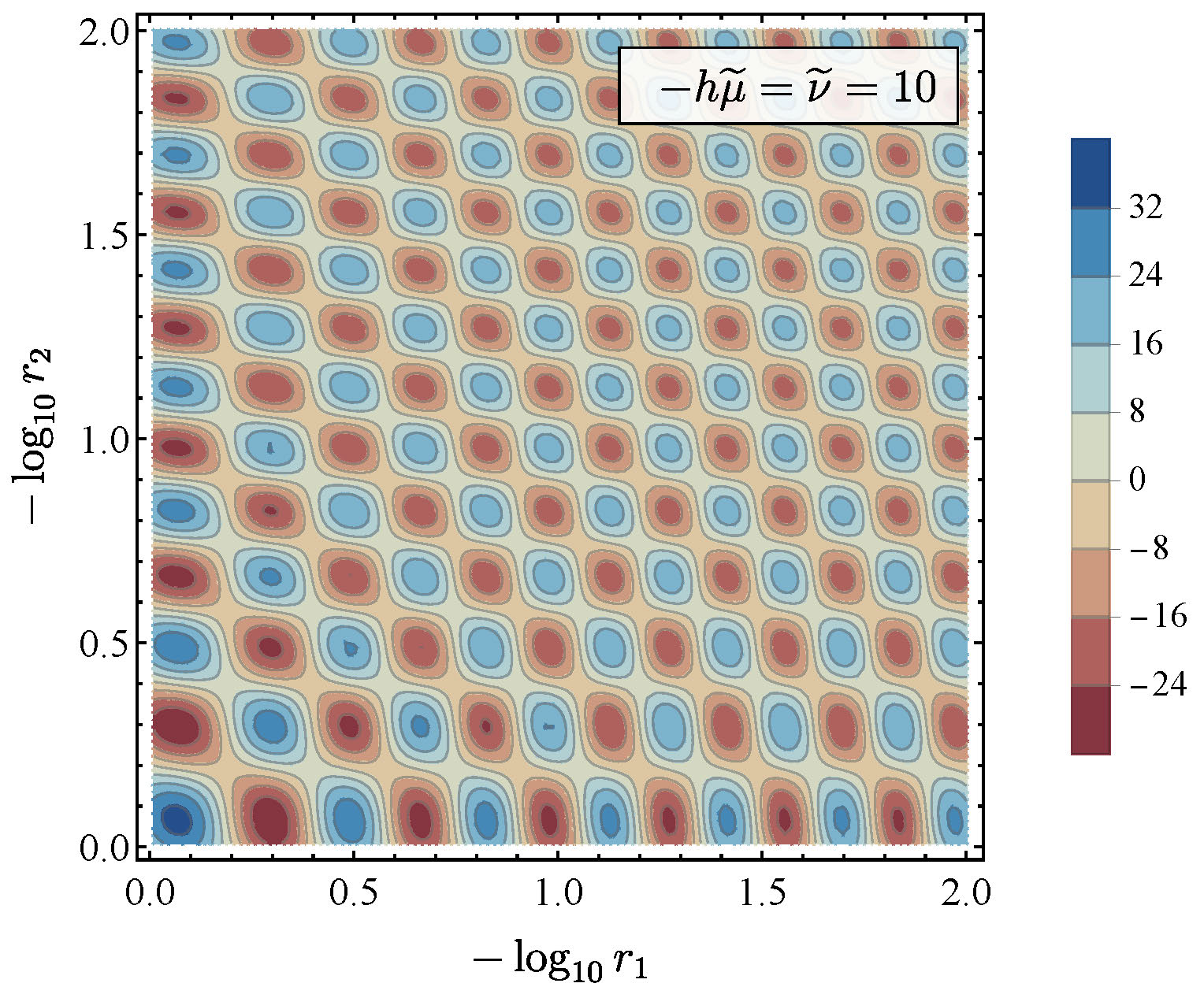} 
  \includegraphics[height=0.4\textwidth]{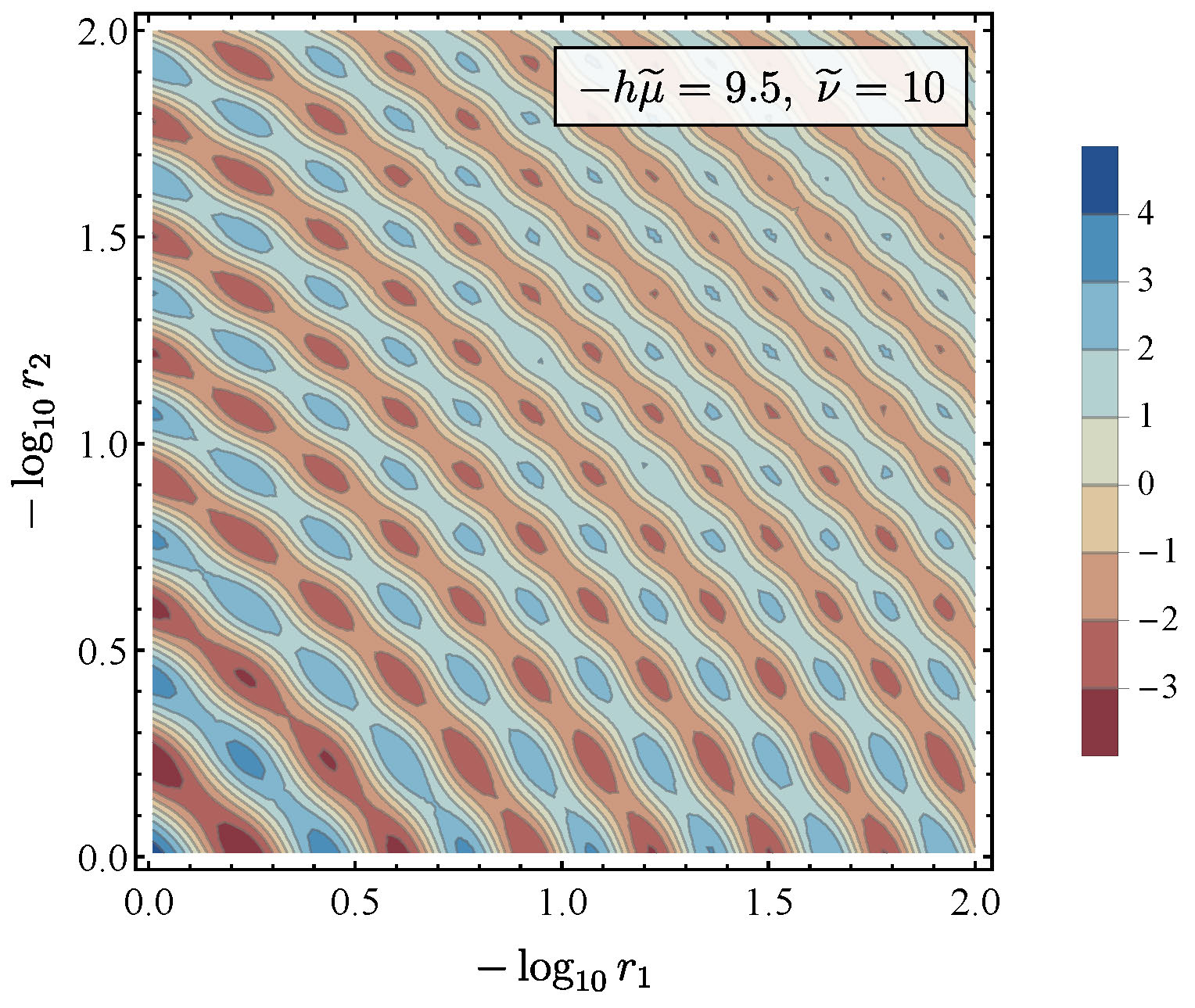} 

\caption{The transverse vector seed integral (\ref{eq_vecSeedIntResult}) as functions of the two momentum ratios $r_1=k_s/k_{12}$ and $r_2=k_s/k_{34}$. Roughly, the oscillations in the direction from lower-left to upper-right correspond to the nonlocal signal, while the oscillations in the direction from upper-left to lower-right correspond to the local signal.  }
\label{fig_signal} 
  
\end{figure}

The analytical expressions we obtained in this work enable us to easily generate plots for the shapes of the 4-point correlators. We show several examples in Fig.\ \ref{fig_signal}. In all these examples, we have chosen the chemical potential $\wt\mu$ to be equal or close to the mass parameter $\wt\nu$. In such cases, the signal always dominates over the background, as is manifested in the figure. By choosing two different mass parameters $\wt\nu=3$ (upper panels) and $\wt\nu=10$ (lower panels), we show how the frequency of the signal is changing with the mass. By choosing two different chemical potentials $\wt\mu$ for each mass parameter (left and right panels), we show the sensitive dependence of the signals on $\wt\mu$.

\section{Conclusions and Outlooks}
\label{sec_concl}

The helical inflation correlators are a special class of inflation correlators mediated by massive spinning particles and boosted by helicity-dependent chemical potentials. They arise naturally from a wide range of particle models for cosmological collider physics with distinct phenomenology and promising observation prospects. On the other hand, these correlators break the dS boosts and the space parity, which renders their analytical calculation more difficult than dS covariant correlators.

In this work, we have attacked the problem of analytical computation of helical inflation correlators with two independent methods. In the first method, we have used the partial Mellin-Barnes representation to resolve the complicated intermediate massive propagators into products of simple power functions. The originally difficult SK time integrals were thus transformed into more tractable contour integrals on the complex planes of Mellin variables.  In the second method, we have derived the equations, namely the bootstrap equations, satisfied by ordinary and helical inflation correlators, and showed that these correlators can be bootstrapped by solving the bootstrap equations with appropriate boundary conditions. In this process, we have reduced much of the computation of the correlators to that of a scalar seed integral and a vector seed integral. We further demonstrated how to use these building blocks to compute more general correlators, by working out a special example of mixed 3-point function with two scalar modes and one tensor mode. Finally, we have also discussed the phenomenology associated with the helical inflation correlators. 

Our work represents one step forward in the application of the partial MB representation in the computation of cosmic correlators. There are much left to be done. One obvious extension is to include both the non-unit sound speed and the nonzero chemical potential, so that one can build up the most general boost-breaking correlators mediated by spinning particles. Also, one could try out the method with more wide range of correlators, with different topologies (tree diagrams and loop diagrams), different couplings. As can be seen, the partial MB representation has few requirements on the symmetry of the problem, and thus can be particularly suitable to deal with problems with less symmetries. One can also imagine to use this method to compute correlators in non-inflation background, as long as the relevant mode functions have known MB representations.

Also, as mentioned before, our result has explicitly verified a recently proposed tree-level cutting for CC signals for helical inflation correlators. From our calculation one can see that the partial MB representation can neatly separate terms with different analytical properties in the squeezed limit, which makes it a possible tool for mathematically proving the tree-level signal cutting rule. We leave this proof for a future work.

When bootstrapping the helical correlators, we have imposed the boundary conditions from the squeezed limit, which corresponds to taking the late-time expansion of the intermediate propagator in a bulk calculation. This is an opposite approach compared with existing bootstrap works in the literature, where the boundary conditions were normally imposed from the factorized limit and the folded limit. While we have checked in this work that all folded limits of our results are regular, we have not considered the unphysical limits such as the zero-total-energy limit or the zero-partial-energy limit (namely the factorized limit). It would be interesting to look into these limits more carefully for helical correlators. 

In this work, we have only touched the spin-1 and spin-2 examples. It is possible to generalize the chemical potential to even higher spin fields, and also to half-integer spin fields. It would thus be interesting to understand more systematically the quantization and the dynamics of higher spin fields in the presence of such chemical potentials, and to use the existing analytical methods to compute the corresponding correlators with higher-spin or half-integer-spin mediations. 

Finally, there is a particularly interesting parameter range where the oscillating signal could be much greater than the background. Therefore, current search for non-Gaussianities in the trispectrum could be insensitive to these shapes unless one includes those scale-invariant oscillations into the templates. Since we have provided the full results for the 4-point function mediated by massive spinning particles, including the signal and the background, it should be straightforward to fast generate template banks for these shapes and use them to search for potential signals in the trispectrum or to put more stringent constraints. We leave this more phenomenological study for a future work as well.

\paragraph{Acknowledgments.} We thank Xi Tong, Dong-Gang Wang, Yi Wang, Yi-Ming Zhong, Yuhang Zhu for useful discussions. This work is supported by the National Key R\&D Program of China (2021YFC2203100), an Open Research Fund of the Key Laboratory of Particle Astrophysics and Cosmology, Ministry of Education of China, and a Tsinghua University Initiative Scientific Research Program.

\begin{appendix}

\section{Useful Formulae}
\label{app_formulae}

\paragraph{Useful functions.}
Below we list some of the frequently used functions in this paper. First, we use the following shorthand notation for the products and factions of Euler $\Gamma$ function:
\begin{align}
  \Gamma\left[ z_1,\cdots,z_m \right]
  \equiv&~ \Gamma(z_1)\cdots \Gamma(z_m) ,\\
  \Gamma\left[\bgm z_1,\cdots,z_m \\w_1,\cdots, w_n\edm\right]
  \equiv&~\FR{\Gamma(z_1)\cdots \Gamma(z_m)}{\Gamma(w_1)\cdots \Gamma(w_n)}.
\end{align}
We use the Pochhammer symbol $(z)_n$ to simplify some expressions, which is defined as
\begin{align}
\label{eq_pochhammer}
  (z)_n\equiv\Gamma\left[\bgm z+n \\ z\edm\right].
\end{align}
The (generalized) hypergeometric functions ${}_p\mathrm{F}_q$ is formally defined via the following  series.
\begin{align}
  {}_p\mathrm{F}_q\left[\bgm a_1,\cdots,a_p \\ b_1,\cdots,b_q \edm  \middle| z \right]=\sum_{n=0}^\infty\FR{(a_1)_n\cdots (a_p)_n}{(b_1)_n\cdots (b_q)_n}\FR{z^n}{n!}.
\end{align}
 In this work we will only encounter the case with $p=q+1$. In this case, the above series converges within the disk $|z|<1$. Outside this range, the function is defined via analytical continuation.

We employ the following \emph{regularized} hypergeometric function when we want to highlight the analytical structure of some MB integrand:
\bge
\label{eq_HyperGeoReg}
  {}_p\wt{\mathrm{F}}_q\left[\bgm a_1,\cdots,a_p \\ b_1,\cdots,b_q \edm \middle| z \right]=\FR{1}{\Gamma[b_1,\cdots,b_q]}\,{}_p\mathrm{F}_q\left[\bgm a_1,\cdots,a_p \\ b_1,\cdots,b_q \edm \middle| z \right],
\ede
which has the nice property that its principal branch is an entire function of all orders $a_1,\cdots,a_p$ and $b_1,\cdots,b_q$. 

We also use the following \emph{dressed} hypergeometric functions to simplify some of expressions. We shall suppress the subscripts $p$ and $q$ without causing confusions. 
\bge
\label{eq_HyperGeoDressed}
  {\mathcal{F}}\left[\bgm a_1,\cdots,a_p \\ b_1,\cdots,b_q \edm \middle| z \right]=\Gamma\left[\bgm a_1,\cdots,a_p \\ b_1,\cdots,b_q\edm\right] {}_p\mathrm{F}_q\left[\bgm a_1,\cdots,a_p \\ b_1,\cdots,b_q \edm \middle| z \right].
\ede

\paragraph{Useful integrals. }

\begin{align}
    &\int_{-\infty}^0 \di\tau\, e^{\pm \ii k\tau} (-\tau)^{p-1} = e^{\mp\ii p\pi/2} k^{-p} \Gamma(p),\\
    &\int_{-\infty}^0 \di\tau_2\int_{\tau_2}^{0}\di\tau_1\,
    e^{\pm (\ii k_{12} \tau_1 + \ii k_{34}\tau_2)} (-\tau_1)^{p-1}(-\tau_2)^{q-1}\n\\
    =&~e^{\mp\ii(p+q)\pi/2}k_{12}^{-p}k_{34}^{-q}\Gamma\Big[p,q\Big]-e^{\mp \ii (p+q)\pi/2} k_{12}^{-p-q} 
     \mathcal{F}\left[\bgm q,p+q\\1+q\edm\middle|\,-\FR{k_{34}}{k_{12}}\right],\\
    \label{eq_TOintFormula3}
    &\int_{-\infty}^0 \di\tau_2\int_{-\infty}^{\tau_2}\di\tau_1\,
    e^{\pm (\ii k_{12} \tau_1 + \ii k_{34}\tau_2)} (-\tau_1)^{p-1}(-\tau_2)^{q-1}\n\\
    =&~e^{\mp \ii (p+q)\pi/2} k_{12}^{-p-q} 
     \mathcal{F} \left[\bgm q,p+q\\1+q\edm\middle|\,-\FR{k_{34}}{k_{12}}\right].
\end{align}

\section{General Trilinear Couplings}
\label{app_general}

In this appendix we present the full expression for the 4-point correlator generated by the general coupling of the form given in (\ref{eq_CoupGeneral}). Specifically, we consider two such couplings, $\mathcal{O}_{\mathbb{P}_1}$ and $\mathcal{O}_{\mathbb{P}_2}$, with $\mathbb{P}_i=(J_i,K_i,M_i,N_i,R_i)$ $(i=1,2)$. Then, following the diagrammatic rule, we can write down the expression corresponding to  the $s$-channel exchange of a massive spin-1 field as:
\begin{align}
    \mathcal{T}_{\text{2}}^{(h)}
    =&\lam_{\mathbb{P}_1}\lam_{\mathbb{P}_2}(-\mb k_1\cdot\mb k_2)^{M_1}(-\mb k_3\cdot\mb k_4)^{M_2}(-k_1^2)^{N_1}(-k_2^2)^{R_1}(-k_3^2)^{N_2}(-k_4^2)^{R_2}(\mb k_2\cdot \mb e_{\mb k_s}^{(h)})(\mb k_4\cdot \mb e_{\mb k_s}^{(h)*})\n\\
    & ~\times\sum_{\mathsf{a},\mathsf{b}=\pm}\mathsf{ab}\int_{-\infty}^0 \di\tau_1\di\tau_2\, (-\tau_1)^{Q_1}(-\tau_2)^{Q_2} \,
     \pd_{\tau_1}^{J_1}G_\mathsf{a} (k_1,\tau_1) \pd_{\tau_1}^{K_1} G_\mathsf{a}(k_2,\tau_1)\n\\
     &~\times \pd_{\tau_2}^{J_2} G_\mathsf{b}(k_3,\tau_2) \pd_{\tau_2}^{K_2} G_\mathsf{b}(k_4,\tau_2)
    D_{\mathsf{ab}}^{(h)}(k_s;\tau_1,\tau_2) ,
\end{align}
where $Q_i\equiv J_i+K_i+2(M_i+N_i+R_i)-2$ with $i=1,2$. Then, using the following equation for arbitrary number of temporal derivatives on the bulk-to-boundary propagator:
\bge
  \pd_{\tau}^{J}\varphi(k,\tau)=\FR{H}{\sqrt{2k^3}}(-\ii k)^{J}(1-J+\ii k\tau)e^{-\ii k\tau},
\ede
we get,
\begin{align}
    \mathcal{T}_{\text{2}}^{(h)}
    =&\lam_{\mathbb{P}_1}\lam_{\mathbb{P}_2}(-1)^{M_1+M_2+N_1+N_2+R_1+R_2}k_1^{M_1+2N_1}k_2^{M_1+2R_1}k_3^{M_2+2N_2}k_4^{M_2+2R_2}\cos^{M_1}\theta_{12}\cos^{M_2}\theta_{34}\n\\
     &~\times (\mb k_2\cdot \mb e_{\mb k_s}^{(h)})(\mb k_4\cdot \mb e_{\mb k_s}^{(h)*})
     \FR{1}{16(k_1k_2k_3k_4)^3} \sum_{\aa,\bb=\pm}\aa\bb(\ii\aa k_1)^{J_1}(\ii\aa k_2)^{K_1}(\ii\bb k_3)^{J_2}(\ii\bb k_4)^{K_2}\n\\
     &~\times\int_{-\infty}^0\di\tau_1\di\tau_2\,(-\tau_1)^{Q_1}(-\tau_2)^{Q_2}(1-J_1-\ii\aa k_1\tau_1)(1-K_1-\ii\aa k_2\tau_1)\n\\
     &~\times(1-J_2-\ii\bb k_3\tau_2)(1-K_2-\ii\bb k_4\tau_2)e^{\ii \aa k_{12}\tau_1+\ii \bb k_{34}\tau_2 }D_{\aa\bb}^{(h)}(k_s;\tau_1,\tau_2).
\end{align}
If we define:
\bge
v_1 = k_1/k_s, \quad v_2 = k_2/k_s, \quad v_3 = k_3/k_s, \quad v_4 = k_4/k_s,
\ede
then, in terms of these momentum ratios and the vector seed integral defined in (\ref{eq_vecIab}), we can write the final answer to this process as:
\begin{align}
\label{eq_T2general}
    \mathcal{T}_{\text{2}}^{(h)}
   =&~\lam_{\mathbb{P}_1}\lam_{\mathbb{P}_2}(-1)^{M_1+M_2+N_1+N_2+R_1+R_2}v_1^{M_1+2N_1+J_1}v_2^{M_1+2R_1+K_1}v_3^{M_2+2N_2+J_2}v_4^{M_2+2R_2+K_2}\n\\
     &~\times \cos^{M_1}\theta_{12}\cos^{M_2}\theta_{34}(\mb k_2\cdot \mb e_{\mb k_s}^{(h)})(\mb k_4\cdot \mb e_{\mb k_s}^{(h)*})
     \FR{-k_s}{16(k_1k_2k_3k_4)^3} \sum_{\aa,\bb=\pm} (\ii\aa)^{J_1+K_1}(\ii\bb)^{J_2+K_2} \n\\
     &~\times\Big\{(1-J_1)(1-K_1)(1-J_2)(1-K_2)\mathcal{I}_{\aa\bb}^{(h)Q_1,Q_2}
     + v_1v_2v_3v_4\mathcal{I}_{\aa\bb}^{(h)Q_1+2,Q_2+2}\n\\
     &~-\aa\bb\big[(1-J_1)v_2+(1-K_1)v_1\big]\big[(1-J_2)v_4+(1-K_2)v_3\big]\mathcal{I}_{\aa\bb}^{(h)Q_1+1,Q_2+1}\n\\
     &~-\ii\aa\big[(1-J_1)v_2+(1-K_1)v_1\big]\Big[(1-J_2)(1-K_2)\mathcal{I}_{\aa\bb}^{(h)Q_1+1,Q_2}-v_3v_4\mathcal{I}_{\aa\bb}^{(h)Q_1+1,Q_2+2}\Big] \n\\
     &~-\ii\bb\big[(1-J_2)v_4+(1-K_2)v_3\big]\Big[(1-J_1)(1-K_1)\mathcal{I}_{\aa\bb}^{(h)Q_1,Q_2+1}-v_1v_2\mathcal{I}_{\aa\bb}^{(h)Q_1+2,Q_2+1}\Big]\n\\
     &~-v_1v_2(1-J_2)(1-K_2)\mathcal{I}_{\aa\bb}^{(h)Q_1+2,Q_2}
       -v_3v_4(1-J_1)(1-K_1)\mathcal{I}_{\aa\bb}^{(h)Q_1,Q_2+2} \Big\}.
\end{align} 

\section{Fully Resolved MB Representation}
\label{app_AlterMB}

In this appendix, we show that the vector seed integral (\ref{eq_vecIab}) can also be computed using the fully resolved MB representation for the Whittaker function (\ref{eq_WhitFullMB}). The resulting expression would be functions of momentum ratios $r_1=k_s/k_{12}$ and $r_{2}=k_s/k_{34}$ instead of $u_{1,2}=2r_{1,2}/(1+r_{1,2})$. With the fully resolved representation, the propagators of massive spin-1 field can be written as
\begin{align}
    D_{\lessgtr}^{(h)}(k_s;\tau_1,\tau_2)=&~\FR{e^{-h\pi\wt\mu}}{2k_s}
    \text{W}_{\mp\ii h\wt\mu,\mp\ii\wt\nu}(\mp2\ii k_s \tau_1)
    \text{W}_{\pm\ii h\wt\mu,\pm\ii\wt\nu}(\pm2\ii k_s \tau_2)\n\\
    =&~e^{-h\pi\wt\mu}\int_{-\ii\infty}^{\ii\infty}\FR{\di s_1}{2\pi\ii}\FR{\di s_2}{2\pi\ii}\, e^{\mp\ii\pi(s_1-s_2)/2}(2k_s)^{-s_{12}}
    (-\tau_1)^{-s_1+1/2}(-\tau_2)^{-s_2+1/2}\n\\
    &\times \Gamma\Big[s_1-\ii\wt\nu,s_1+\ii\wt\nu,s_2-\ii\wt\nu,s_2+\ii\wt\nu\Big]\n\\
    &\times {}_2\wt{\mathrm{F}}_1\left[\bgm s_1-\ii\wt\nu,s_1+\ii\wt\nu\\s_1\pm\ii h\wt\mu+\fr12\edm\middle|\,\fr12\right]
    {}_2\wt{\mathrm{F}}_1\left[\bgm s_2-\ii\wt\nu,s_2+\ii\wt\nu\\s_2\mp\ii h\wt\mu+\fr12\edm\middle|\,\fr12\right],
\end{align}

Plugging the above representation into our definition for the vector seed integral \eqref{eq_vecIab}, we can complete the time integral. It is easy to carry out the opposite-sign integrals:
\begin{align}
    {\mathcal I}_{\pm\mp}^{(h)p_1p_2}=&~ e^{-h\pi\wt\mu}r_1^{3/2+p_1}r_2^{3/2+p_2}e^{\mp\ii(p_1-p_2)\pi/2}\int_{-\ii\infty}^{\ii\infty}\FR{\di s_1}{2\pi\ii}\FR{\di s_2}{2\pi\ii}\,(2r_1)^{-s_1}(2r_2)^{-s_2}\n\\
    &\times \Gamma\Big[p_1+\FR32-s_1,p_2+\FR32-s_2,s_1-\ii\wt\nu,s_1+\ii\wt\nu,s_2-\ii\wt\nu,s_2+\ii\wt\nu\Big]\n\\
    &\times{}_2\wt{\mathrm{F}}_1\left[\bgm s_1-\ii\wt\nu,s_1+\ii\wt\nu\\s_1\pm\ii h\wt\mu+\fr12\edm\middle|\,\fr12\right]
    {}_2\wt{\mathrm{F}}_1\left[\bgm s_2-\ii\wt\nu,s_2+\ii\wt\nu\\s_2\mp\ii h\wt\mu+\fr12\edm\middle|\,\fr12\right].
\end{align}
For the same-sign integrals, we still focus on the case that $k_{12}>k_{34}$. Similar to the scalar case and the vector case, the integral can be divided into a factorized part ${\mathcal I}^{(h)p_1p_2}_{\pm\pm,\text{F},>}$ and a time-ordered part ${\mathcal I}^{(h)p_1p_2}_{\pm\pm,\text{TO},>}$, contributing to the signal and the background, respectively. Using the formulae given in App.\ \ref{app_formulae}, we obtain:
\begin{align}
    {\mathcal I}_{\pm\pm,\text{F},>}^{(h)p_1p_2}=&~ e^{-h\pi\wt\mu}r_1^{3/2+p_1}r_2^{3/2+p_2}e^{\mp\ii(p_1+p_2)\pi/2}\int_{-\ii\infty}^{\ii\infty}\FR{\di s_1}{2\pi\ii}\FR{\di s_2}{2\pi\ii}\,(\mp \ii e^{\pm\ii\pi s_1})(2r_1)^{-s_1}(2r_2)^{-s_2}\n\\
    &\times \Gamma\Big[p_1+\FR32-s_1,p_2+\FR32-s_2,s_1-\ii\wt\nu,s_1+\ii\wt\nu,s_2-\ii\wt\nu,s_2+\ii\wt\nu\Big]\n\\
    &\times{}_2\wt{\mathrm{F}}_1\left[\bgm s_1-\ii\wt\nu,s_1+\ii\wt\nu\\s_1\mp\ii h\wt\mu+\fr12\edm\middle|\,\fr12\right]
    {}_2\wt{\mathrm{F}}_1\left[\bgm s_2-\ii\wt\nu,s_2+\ii\wt\nu\\s_2\pm\ii h\wt\mu+\fr12\edm\middle|\,\fr12\right],
\end{align}
\begin{align}
    {\mathcal I}_{\pm\pm,\text{TO},>}^{(h)p_1p_2}=&~ e^{-h\pi\wt\mu}r_1^{3+p_1+p_2}e^{\mp\ii(p_1+p_2)\pi/2}\int_{-\ii\infty}^{\ii\infty}\FR{\di s_1}{2\pi\ii}\FR{\di s_2}{2\pi\ii}\,\n\\
    &\times \bigg[(\pm\ii e^{\pm\ii\pi s_1})
    {}_2\wt{\mathrm{F}}_1\left[\bgm s_1-\ii\wt\nu,s_1+\ii\wt\nu\\s_1\mp\ii h\wt\mu+\fr12\edm\middle|\,\fr12\right]
    {}_2\wt{\mathrm{F}}_1\left[\bgm s_2-\ii\wt\nu,s_2+\ii\wt\nu\\s_2\pm\ii h\wt\mu+\fr12\edm\middle|\,\fr12\right]\n\\
    &~~+(\mp\ii e^{\pm\ii\pi s_2})
    {}_2\wt{\mathrm{F}}_1\left[\bgm s_1-\ii\wt\nu,s_1+\ii\wt\nu\\s_1\pm\ii h\wt\mu+\fr12\edm\middle|\,\fr12\right]
    {}_2\wt{\mathrm{F}}_1\left[\bgm s_2-\ii\wt\nu,s_2+\ii\wt\nu\\s_2\mp\ii h\wt\mu+\fr12\edm\middle|\,\fr12\right]\bigg]\n\\
    &\times (2r_1)^{-s_{12}} {}_2\wt{\mathrm{F}}_1\left[\bgm p_2+\fr32-s_2,p_1+p_2+3-s_{12}\\p_2+\fr52-s_2\edm\middle|\,-\FR{r_1}{r_2}\right]\n\\
    &\times \Gamma\Big[p_2+\FR32-s_2,p_1+p_2+3-s_{12},s_1-\ii\wt\nu,s_1+\ii\wt\nu,s_2-\ii\wt\nu,s_2+\ii\wt\nu\Big].
\end{align}
Finally, we close the contour from left and pick up poles \eqref{eq_VecPole1} and \eqref{eq_VecPole2} (notice that the contribution of \eqref{eq_VecPole2} vanishes for the time-ordered part), and obtain the (local and nonlocal) signals and the background piece:
\begin{align}
	{\mathcal I}_{\text{L},>}^{(h)p_1p_2}
	=&~ e^{-h\pi\wt\nu}r_1^{3/2+p_1}r_2^{3/2+p_2}\sum_{n_1,n_2=0}^\infty\bigg\{
	\FR{(-1)^{n_{12}}}{n_1!n_2!}
	\Big(e^{\ii(p_1-p_2)\pi/2}-(-1)^{n_1}\ii e^{+\ii(p_1+p_2)\pi/2}e^{\pi\wt\nu}\Big)\n\\
	&\times (2r_1)^{n_1+\ii\wt\nu}(2r_2)^{n_2-\ii\wt\nu}
	\times \Gamma\Big[n_1+p_1+\FR32+\ii\wt\nu,n_2+p_2+\FR32-\ii\wt\nu,-n_1-2\ii\wt\nu,-n_2+2\ii\wt\nu\Big]\n\\
	&\times{}_2\wt{\mathrm{F}}_1\left[\bgm -n_1,-n_1-2\ii\wt\nu\\-n_1-\ii\wt\nu-\ii h\wt\mu+\fr12\edm\middle|\,\fr12\right]
	{}_2\wt{\mathrm{F}}_1\left[\bgm -n_2,-n_2+2\ii\wt\nu\\-n_2+\ii\wt\nu+\ii h\wt\mu+\fr12\edm\middle|\,\fr12\right]\n\\
	&+(\wt\nu\to-\wt\nu)\bigg\}+\text{c.c.},
\end{align}
\begin{align}
	{\mathcal I}_{\text{NL}}^{(h)p_1p_2}
	=&~ e^{-h\pi\wt\nu}r_1^{3/2+p_1}r_2^{3/2+p_2}\sum_{n_1,n_2=0}^\infty\bigg\{
	\FR{(-1)^{n_{12}}}{n_1!n_2!}
	\Big(e^{\ii(p_1-p_2)\pi/2}-(-1)^{n_1}\ii e^{+\ii(p_1+p_2)\pi/2}e^{\pi\wt\nu}\Big)\n\\
	&\times (2r_1)^{n_1+\ii\wt\nu}(2r_2)^{n_1+\ii\wt\nu}
	\times \Gamma\Big[n_1+p_1+\FR32+\ii\wt\nu,n_2+p_2+\FR32+\ii\wt\nu,-n_1-2\ii\wt\nu,-n_2-2\ii\wt\nu\Big]\n\\
	&\times{}_2\wt{\mathrm{F}}_1\left[\bgm -n_1,-n_1-2\ii\wt\nu\\-n_1-\ii\wt\nu-\ii h\wt\mu+\fr12\edm\middle|\,\fr12\right]
	{}_2\wt{\mathrm{F}}_1\left[\bgm -n_2,-n_2-2\ii\wt\nu\\-n_2-\ii\wt\nu+\ii h\wt\mu+\fr12\edm\middle|\,\fr12\right]\n\\
	&+(\wt\nu\to-\wt\nu)\bigg\}+\text{c.c.},
\end{align}
\begin{align}
    {\mathcal I}_{\text{BG},>}^{(h)p_1p_2}=&~ e^{-h\pi\wt\mu}r_1^{3+p_1+p_2}
    \sum_{n_1,n_2=0}^\infty\bigg\{
    \FR{(-1)^{n_{12}}}{n_1!n_2!}\times\ii\Big[(-1)^{n_1}e^{-\ii(p_1+p_2)\pi/2}+(-1)^{n_2}e^{+\ii(p_1+p_2)\pi/2}\Big]e^{\pi\wt\nu}(2r_1)^{n_{12}}\n\\
    &\times 
    {}_2\wt{\mathrm{F}}_1\left[\bgm -n_1,-n_1-2\ii\wt\nu\\-n_1-\ii\wt\nu-\ii h\wt\mu+\fr12\edm\middle|\,\fr12\right]
    {}_2\wt{\mathrm{F}}_1\left[\bgm -n_2,-n_2+2\ii\wt\nu\\-n_2+\ii\wt\nu+\ii h\wt\mu+\fr12\edm\middle|\,\fr12\right]\n\\
        &\times \mathcal F\left[\bgm n_2+p_2+\fr32-\ii\wt\nu,n_{12}+p_1+p_2+3\\ n_2+p_2+\fr52-\ii\wt\nu\edm\middle|\,-\FR{r_1}{r_2}\right]\times \Gamma\Big[-n_1-2\ii\wt\nu,-n_2+2\ii\wt\nu\Big]\n\\
        &+(\wt\nu\to-\wt\nu)\bigg\}+\text{c.c.}.
\end{align}
We have numerically checked that these are identical to our results in Sec.\ \ref{sec_vec_summary}. 

\end{appendix}

\newpage
\bibliography{CosmoCollider} 
\bibliographystyle{utphys}

\end{document}